\documentclass[12pt]{iopart}

\usepackage{graphicx}
\usepackage{braket}
\usepackage{iopams}
\usepackage{tikz}
\usepackage{subfig}
\usepackage[font={small}]{caption}
	
\expandafter\let\csname equation*\endcsname\relax
\expandafter\let\csname endequation*\endcsname\relax
\usepackage{amsmath}
\usepackage{blkarray}
\usepackage{color}	
\usepackage{graphicx}

\newcommand{\joschka}[1]{{\color{black}#1}}
\newcommand{\joschkaEdit}[1]{{\color{black}#1}}

\newcommand{\cnot}{\leavevmode\hbox{\footnotesize{CNOT }}}
\newcommand{\cnoteq}{\leavevmode\hbox{\footnotesize{CNOT}}}

\newcommand{\swap}{\leavevmode\hbox{\footnotesize{SWAP }}}
\newcommand{\swapeq}{\leavevmode\hbox{\footnotesize{SWAP}}}
\newcommand{\CZ}{\leavevmode\hbox{\footnotesize{CZ }}}
\newcommand{\CZeq}{\leavevmode\hbox{\footnotesize{CZ}}}

\DeclareRobustCommand\openone{\leavevmode\hbox{\small1\normalsize\kern-.33em1}}
\makeatletter
\newcommand*\bigcdot{\mathpalette\bigcdot@{.5}}
\newcommand*\bigcdot@[2]{\mathbin{\vcenter{\hbox{\scalebox{#2}{$\m@th#1\bullet$}}}}}
\makeatother

\begin{document}

\title[Protecting quantum memories using coherent parity check codes]{Protecting quantum memories using coherent parity check codes}

\author{Joschka Roffe, David Headley, Nicholas Chancellor, Dominic Horsman \& Viv Kendon}

\address{Joint Quantum Centre (JQC) Durham-Newcastle, Department of Physics, Durham University, South Road, Durham DH1 3LE, United Kingdom}
\ead{joshua.roffe@durham.ac.uk}
\vspace{10pt}
\begin{indented}
\item[]21 May 2018
\end{indented}

	\begin{abstract}
	\noindent Coherent parity check (CPC) codes are a new framework for the construction of quantum error correction codes that encode multiple qubits per logical block. \joschka{CPC codes have a canonical structure involving successive rounds of bit and phase parity checks, supplemented by cross-checks to fix the code distance.} In this paper, we provide a detailed introduction to CPC codes using conventional quantum circuit notation. \joschkaEdit{We demonstrate the implementation of a CPC code on real hardware, by designing a $[[4,2,2]]$ detection code for the IBM 5Q superconducting qubit device. Whilst the individual gate-error rates on the IBM device are too high to realise a fault tolerant quantum detection code, our results show that the syndrome information from a full encode-decode cycle of the $[[4,2,2]]$ CPC code can be used to increase the output state fidelity by post-selection. Following this, we} generalise CPC codes to other quantum technologies by showing that their structure allows them to be efficiently compiled using any experimentally realistic native two-qubit gate. We introduce a three-stage CPC design process for the construction of hardware-optimised quantum memories. As a proof-of-concept example, we apply our design process to an idealised linear seven-qubit ion trap. In the first stage of the process, we use exhaustive search methods to find a large set of $[[7,3,3]]$ codes that saturate the quantum Hamming bound for seven qubits. We then optimise over the discovered set of codes to meet the hardware and layout demands of the ion trap device. We also discuss how the CPC design process will generalise to larger-scale codes and other qubit technologies.
\end{abstract}

\section{Introduction}

\noindent Quantum computing experiments have now matured to the extent to which we can realistically expect to see a medium-scale circuit-model device within the next decade \cite{nqit,IBM}. It is hoped these near-future quantum computers will be sufficient for simple algorithms, possibly beyond what can be solved classically. However, the fulfilment of these aims will usually depend upon the efficacy of the adopted quantum error correction (QEC) code and the ease with which it can be compiled onto the chosen quantum technology platform.

Recently, Chancellor et al.~\cite{cpc1} introduced the \emph{coherent parity check} (CPC) framework as a toolset for the construction of a versatile new class of QEC codes. CPC codes have a canonical structure that allows any sequence of parity checks to be performed on a quantum register without risk of inducing decoherence. This is in contrast to most traditional QEC protocols, where the choice of parity checks is limited to stabilizers of the encoded quantum data. The freedom in the choice of parity checks therefore affords the CPC framework multiple advantages over conventional QEC.

In the original CPC paper \cite{cpc1}, graphical methods based on the \emph{{\sc zx} calculus} \cite{Coecke2011,Aleks_zx} were used to give a construction for re-purposing general classical error correction codes for QEC. This opens the possibility of constructing QEC codes inspired by highly-optimised classical codes, such as \emph{low density parity check} codes \cite{MacKay1996}. Furthermore, as the CPC formalism allows for complete freedom in the choice of parity checks, new CPC codes can be discovered numerically, either via brute-force or more sophisticated search techniques. 

In this work, we demonstrate a further \joschka{feature} of CPC codes with regards to their implementation on physical hardware. \joschka{In a theoretical setting}, QEC codes are usually formulated in terms of \joschka{idealised} controlled-not (\hbox{\footnotesize{CNOT}}) gates. However, the native two-qubit entangling gates provided by various qubit technologies are usually of a different form. Consequently, one of the challenges in realising quantum codes is developing efficient methods by which \joschka{QEC circuits can be realised using} the native interaction \joschka{of the chosen experiment}. Here, we show that the symmetric structure of CPC codes \joschka{enables efficient mapping from the theoretical representation of the code to the hardware-compiled. In particular we show that CPC codes can be implemented with any realistic maximally entangling Clifford native gate, meaning they will be suitable for deployment across a broad range of quantum hardware.}

As a simple first example on real hardware, we implement a $[[n=4, \ k=2, \ d=2]]$ CPC quantum code on the IBM Q five-qubit superconducting device (where we have adopted the usual convention whereby $n$ represents the number of physical qubits, $k$ the number of data qubits and $d$ the code distance) \cite{IBM}. We demonstrate that, for a simple known input state, a version of the $[[4,2,2]]$ circuit can be compiled to accommodate the connectivity constraints of the IBM chip.  By analysing the experimental data using quantum state tomography, we show that the $[[4,2,2]]$ code's syndrome information can be used to improve the fidelity of the output state by post-selection.

There is currently no preferred qubit technology and the first quantum computers will likely be hybrid devices that interface multiple qubit types \cite{nqit_nickerson, Proctor2016}. In order to realise their full potential, these hybrid schemes will require tailor-made QEC strategies. To this end, we outline a three-stage CPC design process for the construction of hardware-optimised QEC memories.

As a proof-of-concept example, we demonstrate the use of the CPC design process in creating a quantum code for a seven qubit linear ion trap. In the first stage of this process, we show that exhaustive search techniques can be used to discover a large set of $[[7,3,3]]$ CPC codes. These $[[7,3,3]]$ codes have the highest possible information density for a non-degenerate QEC code, as dictated by the quantum Hamming bound \cite{GottesmanHammingBound}.

The second stage of the code design process involves implementing strategies to select the best CPC code from the discovered set. For the purposes of the ion trap device, we seek to identify the circuits in which the total two-qubit gate count is minimised. This involves consideration of the additional \swap interactions that must be introduced to mediate interactions between spatially separated qubits.

The final hardware optimisation we consider in the CPC design process is compilation of CPC codes with a device's native two-qubit gate. For the ion trap under consideration, we assume the native interaction is of the form of a maximally entangling symmetrised phase (SP) gate \cite{ballance}. A \cnot interaction can be implemented from an SP gate, but this requires addition of local single-qubit gates which increases the code overhead. As an example of the native gate compilation, we demonstrate that for many of the $[[7,3,3]]$ CPC circuits, constructive simplifications can be applied to reduce the total number of local corrections required.

The $[[7,3,3]]$ CPC codes outlined in this paper should be adaptable to many existing ion trap experiments \cite{Ballance2016, sussex_ions, Debnath2016, Brandl2016}. The ability to encode three data qubits in a seven qubit trap \joschka{would mark an improvement over the current most widley adopted protocol for quantum memories, the surface code, which requires a minimum of $13$ qubits per encoded data qubit \cite{lattice-surgery,Fowler2012}}. \joschka{There have been many proposals for quantum codes promising high encoding densities \cite{Tillich2014,Brown2013,Bravyi2014,Breuckmann2016,Audoux2015}. The CPC construction provides a framework to allow for the automated discovery of high-density codes which are optimised for the requirements of the chosen experiment. Note, however, that the specific CPC code implementations presented in this work are not yet fault tolerant and that making them such will result in additional overhead. As this work covers quantum memories, we do not include discussion of encoded computation}. Steps towards developing fault tolerant CPC gates are outlined in \cite{cpc1}, and this remains an interesting area for future work.

The paper is structured as follows. In section \ref{sec:intro_to_cpc}, we give a detailed introduction to the CPC framework, and explain how it can be used to construct full QEC codes. This is followed, in section \ref{sec:ibm}, by the presentation of experimental results obtained by running a simple CPC detection code on the IBM Q quantum computer. In section \ref{sec:ion_trap_intro}, we provide an overview of the ion trap hardware for quantum computing. In section \ref{sec:any_native_gate}, we demonstrate that the fundamental structure of CPC codes allow them to be efficiently compiled using a wide range of native gates. Section \ref{sec:cpc_code_design} describes the CPC design process and how it can be used to construct hardware-optimised $[[7,3,3]]$ codes for the ion trap device. Finally, in section \ref{sec:outlook}, we discuss possible improvements to the CPC design process and how it might be applied in the discovery of larger quantum codes.

\section{Coherent parity check (CPC) codes} \label{sec:intro_to_cpc}
The signature feature of CPC codes is the ability to implement QEC routines with any sequence of parity checks. This is possible due to a fail-safe code structure that ensures syndrome measurements cannot decohere the register. This freedom in the choice of parity checks gives the CPC framework multiple advantages over traditional QEC techniques. First, it is possible to directly translate the parity checking sequences from classical codes into a CPC code, which allows the derivation of dense QEC codes that encode multiple data qubits per logical block. Second, the CPC framework does not require quantum data to be initially redundantly encoded. Third, the space of possible CPC codes can be searched numerically, meaning code discovery can be automated.

In this section, we outline the tools of the CPC framework, starting with the fundamental CPC gadget. This gadget has a symmetric \textit{encode-error-decode} structure that amounts to an extended measurement of the identity operator. We prove that the CPC gadget is inherently non-disturbing and can be implemented using any parity checking sequence. Following this, we demonstrate how multiple CPC gadgets can be combined to form QEC codes. Finally, we introduce the automated search techniques that will be used in the CPC code design process.

\subsection{Traditional quantum error correction} \label{sec:trad_qec}
Before beginning our presentation of the CPC framework, we briefly outline the key concepts and shortcomings of conventional stabilizer QEC codes. This will provide a point-of-reference with which to compare CPC codes.

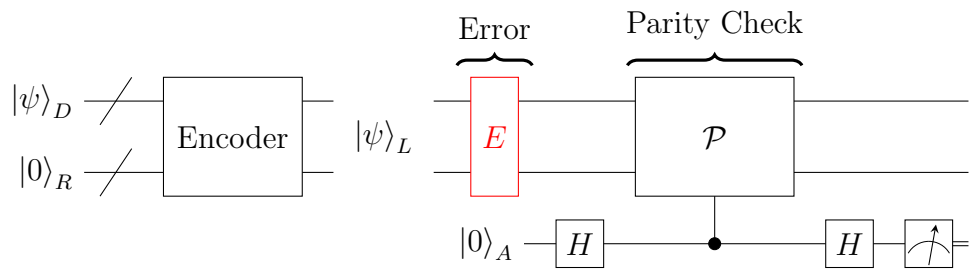
\begin{figure}
\usetikzlibrary{decorations.pathreplacing,decorations.pathmorphing}
\providecommand{\ket}[1]{\left|#1\right\rangle}
\providecommand{\phase}[1]{e^{2{\pi}i\cdot#1}}
\begin{tikzpicture}[scale=1.500000,x=1pt,y=1pt]
\filldraw[color=white] (0.000000, -9.000000) rectangle (223.000000, 45.000000);
\draw[color=black] (0.000000,36.000000) -- (223.000000,36.000000);
\draw[color=black] (0.000000,36.000000) node[left] {$\ket{\psi}_D$};
\draw[color=black] (0.000000,18.000000) -- (223.000000,18.000000);
\draw[color=black] (0.000000,18.000000) node[left] {$\ket{0}_R$};
\draw[color=black] (103.500000,0.000000) -- (213.000000,0.000000);
\draw[color=black] (213.000000,-0.500000) -- (223.000000,-0.500000);
\draw[color=black] (213.000000,0.500000) -- (223.000000,0.500000);
\draw (4.000000, 30.000000) -- (12.000000, 42.000000);
\draw (4.000000, 12.000000) -- (12.000000, 24.000000);
\draw (37.500000,36.000000) -- (37.500000,18.000000);
\begin{scope}
\draw[fill=white] (37.500000, 27.000000) +(-45.000000:24.748737pt and 21.213203pt) -- +(45.000000:24.748737pt and 21.213203pt) -- +(135.000000:24.748737pt and 21.213203pt) -- +(225.000000:24.748737pt and 21.213203pt) -- cycle;
\clip (37.500000, 27.000000) +(-45.000000:24.748737pt and 21.213203pt) -- +(45.000000:24.748737pt and 21.213203pt) -- +(135.000000:24.748737pt and 21.213203pt) -- +(225.000000:24.748737pt and 21.213203pt) -- cycle;
\draw (37.500000, 27.000000) node {{Encoder}};
\end{scope}
\draw[fill=white,color=white] (63.000000, 12.000000) rectangle (88.000000, 42.000000);
\draw (75.500000, 27.000000) node {${\ket{\psi}_L}$};
\begin{scope}[color=red]
\draw (103.500000,36.000000) -- (103.500000,18.000000);
\begin{scope}[color=red]
\begin{scope}
\draw[fill=white] (103.500000, 27.000000) +(-45.000000:8.485281pt and 21.213203pt) -- +(45.000000:8.485281pt and 21.213203pt) -- +(135.000000:8.485281pt and 21.213203pt) -- +(225.000000:8.485281pt and 21.213203pt) -- cycle;
\clip (103.500000, 27.000000) +(-45.000000:8.485281pt and 21.213203pt) -- +(45.000000:8.485281pt and 21.213203pt) -- +(135.000000:8.485281pt and 21.213203pt) -- +(225.000000:8.485281pt and 21.213203pt) -- cycle;
\draw (103.500000, 27.000000) node {{$E$}};
\end{scope}
\end{scope}
\end{scope}
\draw[color=black] (111.000000,0.000000) node[fill=white,left,minimum height=18.000000pt,minimum width=15.000000pt,inner sep=0pt] {\phantom{$\ket{0}_A$}};
\draw[color=black] (111.000000,0.000000) node[left] {$\ket{0}_A$};
\begin{scope}
\draw[fill=white] (125.000000, -0.000000) +(-45.000000:8.485281pt and 8.485281pt) -- +(45.000000:8.485281pt and 8.485281pt) -- +(135.000000:8.485281pt and 8.485281pt) -- +(225.000000:8.485281pt and 8.485281pt) -- cycle;
\clip (125.000000, -0.000000) +(-45.000000:8.485281pt and 8.485281pt) -- +(45.000000:8.485281pt and 8.485281pt) -- +(135.000000:8.485281pt and 8.485281pt) -- +(225.000000:8.485281pt and 8.485281pt) -- cycle;
\draw (125.000000, -0.000000) node {$H$};
\end{scope}
\draw (159.000000,36.000000) -- (159.000000,0.000000);
\begin{scope}
\draw[fill=white] (159.000000, 27.000000) +(-45.000000:28.284271pt and 21.213203pt) -- +(45.000000:28.284271pt and 21.213203pt) -- +(135.000000:28.284271pt and 21.213203pt) -- +(225.000000:28.284271pt and 21.213203pt) -- cycle;
\clip (159.000000, 27.000000) +(-45.000000:28.284271pt and 21.213203pt) -- +(45.000000:28.284271pt and 21.213203pt) -- +(135.000000:28.284271pt and 21.213203pt) -- +(225.000000:28.284271pt and 21.213203pt) -- cycle;
\draw (159.000000, 27.000000) node {{$\mathcal{P}$}};
\end{scope}
\filldraw (159.000000, 0.000000) circle(1.500000pt);
\begin{scope}
\draw[fill=white] (193.000000, -0.000000) +(-45.000000:8.485281pt and 8.485281pt) -- +(45.000000:8.485281pt and 8.485281pt) -- +(135.000000:8.485281pt and 8.485281pt) -- +(225.000000:8.485281pt and 8.485281pt) -- cycle;
\clip (193.000000, -0.000000) +(-45.000000:8.485281pt and 8.485281pt) -- +(45.000000:8.485281pt and 8.485281pt) -- +(135.000000:8.485281pt and 8.485281pt) -- +(225.000000:8.485281pt and 8.485281pt) -- cycle;
\draw (193.000000, -0.000000) node {$H$};
\end{scope}
\draw[fill=white] (207.000000, -6.000000) rectangle (219.000000, 6.000000);
\draw[very thin] (213.000000, 0.600000) arc (90:150:6.000000pt);
\draw[very thin] (213.000000, 0.600000) arc (90:30:6.000000pt);
\draw[->,>=stealth] (213.000000, -5.400000) -- +(80:10.392305pt);
\draw[decorate,decoration={brace,amplitude = 4.000000pt},very thick] (137.000000,45.000000) -- (181.000000,45.000000);
\draw (159.000000, 49.000000) node[text width=144pt,above,text centered] {Parity Check};
\draw[decorate,decoration={brace,amplitude = 4.000000pt},very thick] (94.000000,45.000000) -- (113.000000,45.000000);
\draw (103.500000, 49.000000) node[text width=144pt,above,text centered] {Error};
\end{tikzpicture}
	\caption{Circuit illustrating the structure of a traditional stabilizer code. A quantum data register $\ket{\psi}_D=\ket{\psi_{d_1}\psi_{d_2}...\psi_{d_n}}$ is entangled with redundancy qubits $\ket{0}_R=\ket{0_{r_1}0_{r_2}...0_{r_m}}$ via an encoding operation to create a logical qubit $\ket{\psi}_L$. After encoding, a parity check $\mathcal{P}$ can be performed on the register to determine whether an error has occurred. The result of this parity check is measured via an auxillary qubit $A$, which is prepared in the conjugate basis by Hadamard gates $H$. The slashed wires denote that $\ket{\psi}_D$ and $\ket{0}_R$ are multi-qubit registers. The measurement operator at the end of the wire for qubit $A$ represents a measurement in the computational basis.}
	\label{fig:trad_qec}
\end{figure}

The circuit in figure \ref{fig:trad_qec} shows the basic structure of a traditional stabilizer code. A register of data qubits, $\ket{\psi}_D$, is entangled with a number of blank redundancy qubits, $\ket{0}_R$, via an encoding operation to create a logical qubit $\ket{\psi}_L$. At this stage, the data previously stored solely in $\ket{\psi}_D$ is distributed across the combined Hilbert space of data and redundancy qubits \cite{Shor1995}.

Once the quantum information has been encoded as a logical qubit, errors can be detected by making parity measurements. In practice, this is achieved via the construction shown to the right of the circuit in figure \ref{fig:trad_qec}. A parity check $\mathcal{P}$ is applied to the logical qubit, and the result copied to an auxillary qubit $A$, which is prepared in the conjugate basis by Hadamard gates $H$. Note that a parity check $\mathcal{P}$ is a product of Pauli operators and has eigenvalues $\pm 1$ (for the definition of the Pauli group, consult appendix \ref{app:pauli}). The auxillary qubit is then measured to yield a syndrome. For a well chosen parity check, this syndrome measurement provides information about whether the logical qubit has been subject to an error.

It has been shown that QEC codes based on the above construction can achieve arbitrarily low logical error rates, provided certain threshold conditions are met by qubits at the physical level \cite{Gottesman97}. However, constructing efficient codes with this approach is difficult owing to limitations on the type of parity check that can implemented. In order to ensure that the syndrome measurement of qubit $A$ does not decohere the encoded quantum information, the parity check must stabilize the logical qubit. Formally \cite{devitt2013}, we can write this requirement as follows
\begin{equation}\label{eq:trad_qec_req}
\mathcal{P}\in \mathcal{S},
\end{equation} 
\noindent where the stabilizer $\mathcal{S}=\langle K_i,...,K_n \rangle$ is a sub-group of the Pauli group $\mathcal{G}$ defined by
\begin{equation}
\mathcal{S}=\Big\{ K_i \ | \ K_i\ket{\psi}_L=(+1)\ket{\psi}_L, \ [K_i,K_j]=0, \ K_i\neq-\openone, \ \forall \ (i,j) \Big\}\in \mathcal{G}\rm,
\end{equation}
where $K_{\{i,j\}}$ are the elements of the stabilizer group and $\ket{\psi}_L$ is the logical codeword. The challenges of constructing traditional stabilizer quantum codes are therefore twofold. First, an appropriate encoding operation must be built to create the logical qubit. Second, a compatible set of stabilizer parity checks needs to be discovered so that errors can be checked without compromising the encoded quantum data. As a result of these challenges, the majority of existing QEC codes are limited to the simplest case in which only a single qubit is encoded per logical block. Such $[[n,1,d]]$ codes can be considered quantum analogues of the most basic classical repetition codes, and incur high overheads in terms of the number of redundancy qubits necessary to achieve the desired error suppression rate.

\subsection{The fundamental CPC gadget}

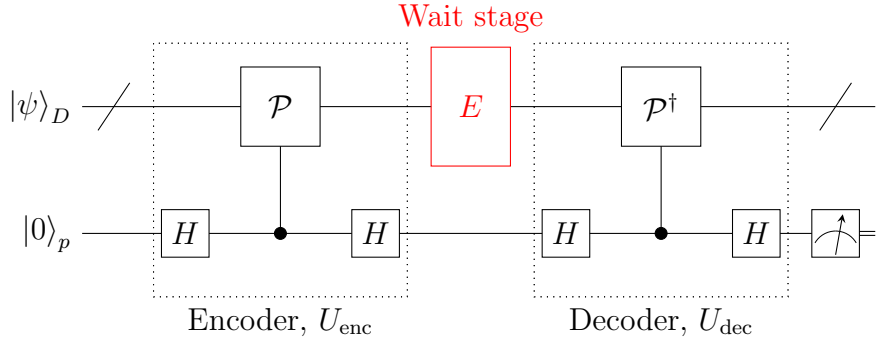
\begin{figure}
	\usetikzlibrary{decorations.pathreplacing,decorations.pathmorphing}
\providecommand{\ket}[1]{\left|#1\right\rangle}
\providecommand{\phase}[1]{e^{2{\pi}i\cdot#1}}
\begin{tikzpicture}[scale=1.500000,x=1pt,y=1pt]
\filldraw[color=white] (0.000000, -16.000000) rectangle (200.000000, 48.000000);
\draw[color=black] (0.000000,32.000000) -- (200.000000,32.000000);
\draw[color=black] (0.000000,32.000000) node[left] {$\ket{\psi}_D$};
\draw[color=black] (0.000000,0.000000) -- (190.000000,0.000000);
\draw[color=black] (190.000000,-0.500000) -- (200.000000,-0.500000);
\draw[color=black] (190.000000,0.500000) -- (200.000000,0.500000);
\draw[color=black] (0.000000,0.000000) node[left] {$\ket{0}_p$};
\draw (4.000000, 26.000000) -- (12.000000, 38.000000);
\begin{scope}
\draw[fill=white] (26.000000, -0.000000) +(-45.000000:8.485281pt and 8.485281pt) -- +(45.000000:8.485281pt and 8.485281pt) -- +(135.000000:8.485281pt and 8.485281pt) -- +(225.000000:8.485281pt and 8.485281pt) -- cycle;
\clip (26.000000, -0.000000) +(-45.000000:8.485281pt and 8.485281pt) -- +(45.000000:8.485281pt and 8.485281pt) -- +(135.000000:8.485281pt and 8.485281pt) -- +(225.000000:8.485281pt and 8.485281pt) -- cycle;
\draw (26.000000, -0.000000) node {$H$};
\end{scope}
\draw (50.000000,32.000000) -- (50.000000,0.000000);
\begin{scope}
\draw[fill=white] (50.000000, 32.000000) +(-45.000000:14.142136pt and 14.142136pt) -- +(45.000000:14.142136pt and 14.142136pt) -- +(135.000000:14.142136pt and 14.142136pt) -- +(225.000000:14.142136pt and 14.142136pt) -- cycle;
\clip (50.000000, 32.000000) +(-45.000000:14.142136pt and 14.142136pt) -- +(45.000000:14.142136pt and 14.142136pt) -- +(135.000000:14.142136pt and 14.142136pt) -- +(225.000000:14.142136pt and 14.142136pt) -- cycle;
\draw (50.000000, 32.000000) node {{$\mathcal{P}$}};
\end{scope}
\filldraw (50.000000, 0.000000) circle(1.500000pt);
\begin{scope}
\draw[fill=white] (74.000000, -0.000000) +(-45.000000:8.485281pt and 8.485281pt) -- +(45.000000:8.485281pt and 8.485281pt) -- +(135.000000:8.485281pt and 8.485281pt) -- +(225.000000:8.485281pt and 8.485281pt) -- cycle;
\clip (74.000000, -0.000000) +(-45.000000:8.485281pt and 8.485281pt) -- +(45.000000:8.485281pt and 8.485281pt) -- +(135.000000:8.485281pt and 8.485281pt) -- +(225.000000:8.485281pt and 8.485281pt) -- cycle;
\draw (74.000000, -0.000000) node {$H$};
\end{scope}
\begin{scope}[color=red]
\draw (98.000000, 48.000000) node[text width=144pt,above,text centered] {{Wait stage}};
\begin{scope}[color=red]
\begin{scope}
\draw[fill=white] (98.000000, 32.000000) +(-45.000000:14.142136pt and 21.213203pt) -- +(45.000000:14.142136pt and 21.213203pt) -- +(135.000000:14.142136pt and 21.213203pt) -- +(225.000000:14.142136pt and 21.213203pt) -- cycle;
\clip (98.000000, 32.000000) +(-45.000000:14.142136pt and 21.213203pt) -- +(45.000000:14.142136pt and 21.213203pt) -- +(135.000000:14.142136pt and 21.213203pt) -- +(225.000000:14.142136pt and 21.213203pt) -- cycle;
\draw (98.000000, 32.000000) node {{$E$}};
\end{scope}
\end{scope}
\end{scope}
\begin{scope}
\draw[fill=white] (122.000000, -0.000000) +(-45.000000:8.485281pt and 8.485281pt) -- +(45.000000:8.485281pt and 8.485281pt) -- +(135.000000:8.485281pt and 8.485281pt) -- +(225.000000:8.485281pt and 8.485281pt) -- cycle;
\clip (122.000000, -0.000000) +(-45.000000:8.485281pt and 8.485281pt) -- +(45.000000:8.485281pt and 8.485281pt) -- +(135.000000:8.485281pt and 8.485281pt) -- +(225.000000:8.485281pt and 8.485281pt) -- cycle;
\draw (122.000000, -0.000000) node {$H$};
\end{scope}
\draw (146.000000,32.000000) -- (146.000000,0.000000);
\begin{scope}
\draw[fill=white] (146.000000, 32.000000) +(-45.000000:14.142136pt and 14.142136pt) -- +(45.000000:14.142136pt and 14.142136pt) -- +(135.000000:14.142136pt and 14.142136pt) -- +(225.000000:14.142136pt and 14.142136pt) -- cycle;
\clip (146.000000, 32.000000) +(-45.000000:14.142136pt and 14.142136pt) -- +(45.000000:14.142136pt and 14.142136pt) -- +(135.000000:14.142136pt and 14.142136pt) -- +(225.000000:14.142136pt and 14.142136pt) -- cycle;
\draw (146.000000, 32.000000) node {{$\mathcal{P^\dagger}$}};
\end{scope}
\filldraw (146.000000, 0.000000) circle(1.500000pt);
\begin{scope}
\draw[fill=white] (170.000000, -0.000000) +(-45.000000:8.485281pt and 8.485281pt) -- +(45.000000:8.485281pt and 8.485281pt) -- +(135.000000:8.485281pt and 8.485281pt) -- +(225.000000:8.485281pt and 8.485281pt) -- cycle;
\clip (170.000000, -0.000000) +(-45.000000:8.485281pt and 8.485281pt) -- +(45.000000:8.485281pt and 8.485281pt) -- +(135.000000:8.485281pt and 8.485281pt) -- +(225.000000:8.485281pt and 8.485281pt) -- cycle;
\draw (170.000000, -0.000000) node {$H$};
\end{scope}
\draw (186.000000, 26.000000) -- (194.000000, 38.000000);
\draw[fill=white] (184.000000, -6.000000) rectangle (196.000000, 6.000000);
\draw[very thin] (190.000000, 0.600000) arc (90:150:6.000000pt);
\draw[very thin] (190.000000, 0.600000) arc (90:30:6.000000pt);
\draw[->,>=stealth] (190.000000, -5.400000) -- +(80:10.392305pt);
\draw[draw opacity=1.000000,fill opacity=0.200000,color=black,dotted] (18.000000,48.000000) rectangle (82.000000,-16.000000);
\draw (50.000000, -16.000000) node[text width=144pt,below,text centered,color=black] {{Encoder, $U_{\rm enc}$}};
\draw[draw opacity=1.000000,fill opacity=0.200000,color=black,dotted] (18.000000,48.000000) rectangle (82.000000,-16.000000);
\draw[draw opacity=1.000000,fill opacity=0.200000,color=black,dotted] (114.000000,48.000000) rectangle (178.000000,-16.000000);
\draw (146.000000, -16.000000) node[text width=144pt,below,text centered,color=black] {{Decoder,  $U_{\rm dec}$}};
\draw[draw opacity=1.000000,fill opacity=0.200000,color=black,dotted] (114.000000,48.000000) rectangle (178.000000,-16.000000);
\end{tikzpicture}
	\caption{The fundamental CPC gadget illustrating the symmetric \textit{encode-error-decode} structure. The parity qubit $p$ is prepared in the conjugate basis by Hadamard gates, $H$. Encode stage: a parity check $\mathcal{P}$, controlled by the parity qubit, is applied to the multi-qubit register $\ket{\psi}_D=\ket{\psi_{d_1}\psi_{d_2}...\psi_{d_n}}$ and the result is copied to the parity qubit. The parity qubit is kept coherent throughout the wait stage, during which an error $E$ can occur on the register. Decode stage: the register is disentangled from the parity qubit via the application of the unitary inverse of the first parity check $\mathcal{P}^\dagger$. The final syndrome measurement of qubit $p$ tells us whether the results of the two parity checks differ. For appropriately chosen parity checks, this information can be used to detect errors. The slashed wire denotes that $\ket{\psi}_D$ is a multi-qubit register.}
	\label{fig:fund_gadget}
\end{figure}

The fundamental CPC gadget, shown in figure \ref{fig:fund_gadget}, is the building block upon which all CPC codes are based \cite{cpc1}. The basic premise behind the CPC gadget is that the parity of the quantum register is never explicitly measured. Instead, parity information is stored coherently as quantum data and compared over time. This is made possible by the gadget's symmetric \textit{encode-error-decode} structure.

The CPC gadget takes a multi-qubit register $\ket{\psi}_D$ and a parity qubit $p$, prepared in the state $\ket{0}_p$, as its input. The action of the encode stage of the gadget, labelled $U_{\rm enc}$ in figure \ref{fig:fund_gadget}, is to apply the parity operator $\mathcal{P}$ to the register and record the outcome in parity qubit $p$. Rather than measuring the syndrome immediately, the parity qubit is kept \joschka{coherent} during a wait stage in which the register is potentially subject to an error $E$. Note that we are not yet considering errors that occur on the parity qubit. In section \ref{sec:422}, we outline how multiple CPC gadgets can be combined to allow for error detection on the combined system of register and parity qubits.

Following the wait stage, the parity qubit is disentangled from the register via a decoder operation, labelled $U_{\rm dec}$ in figure \ref{fig:fund_gadget}, which is the unitary inverse of the encoder. The encoder applies the parity operator $\mathcal{P}$ to the register and the decoder applies its inverse $\mathcal{P}^\dagger$. The final syndrome measurement of parity qubit $p$ tells us whether the results of these two parity checks differ. For an appropriately chosen parity check, this syndrome information can indicate whether an error occurred during the wait stage.

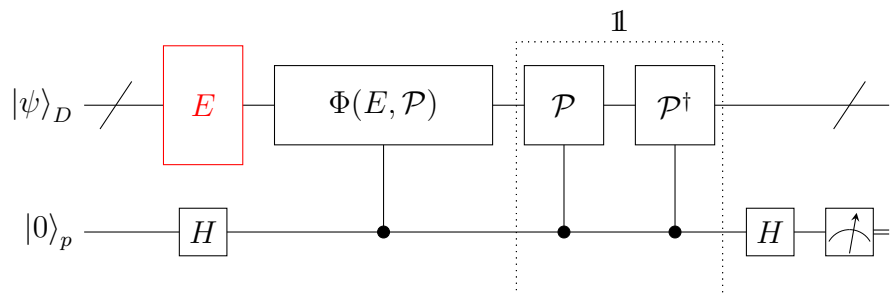
\begin{figure}
\usetikzlibrary{decorations.pathreplacing,decorations.pathmorphing}
\providecommand{\ket}[1]{\left|#1\right\rangle}
\providecommand{\phase}[1]{e^{2{\pi}i\cdot#1}}
\begin{tikzpicture}[scale=1.500000,x=1pt,y=1pt]
\filldraw[color=white] (0.000000, -16.000000) rectangle (203.000000, 48.000000);
\draw[color=black] (0.000000,32.000000) -- (203.000000,32.000000);
\draw[color=black] (0.000000,32.000000) node[left] {$\ket{\psi}_D$};
\draw[color=black] (0.000000,0.000000) -- (193.000000,0.000000);
\draw[color=black] (193.000000,-0.500000) -- (203.000000,-0.500000);
\draw[color=black] (193.000000,0.500000) -- (203.000000,0.500000);
\draw[color=black] (0.000000,0.000000) node[left] {$\ket{0}_p$};
\draw (4.000000, 26.000000) -- (12.000000, 38.000000);
\begin{scope}
\draw[fill=white] (30.000000, -0.000000) +(-45.000000:8.485281pt and 8.485281pt) -- +(45.000000:8.485281pt and 8.485281pt) -- +(135.000000:8.485281pt and 8.485281pt) -- +(225.000000:8.485281pt and 8.485281pt) -- cycle;
\clip (30.000000, -0.000000) +(-45.000000:8.485281pt and 8.485281pt) -- +(45.000000:8.485281pt and 8.485281pt) -- +(135.000000:8.485281pt and 8.485281pt) -- +(225.000000:8.485281pt and 8.485281pt) -- cycle;
\draw (30.000000, -0.000000) node {$H$};
\end{scope}
\begin{scope}[color=red]
\begin{scope}[color=red]
\begin{scope}
\draw[fill=white] (30.000000, 32.000000) +(-45.000000:14.142136pt and 21.213203pt) -- +(45.000000:14.142136pt and 21.213203pt) -- +(135.000000:14.142136pt and 21.213203pt) -- +(225.000000:14.142136pt and 21.213203pt) -- cycle;
\clip (30.000000, 32.000000) +(-45.000000:14.142136pt and 21.213203pt) -- +(45.000000:14.142136pt and 21.213203pt) -- +(135.000000:14.142136pt and 21.213203pt) -- +(225.000000:14.142136pt and 21.213203pt) -- cycle;
\draw (30.000000, 32.000000) node {{$E$}};
\end{scope}
\end{scope}
\end{scope}
\draw (75.500000,32.000000) -- (75.500000,0.000000);
\begin{scope}
\draw[fill=white] (75.500000, 32.000000) +(-45.000000:38.890873pt and 14.142136pt) -- +(45.000000:38.890873pt and 14.142136pt) -- +(135.000000:38.890873pt and 14.142136pt) -- +(225.000000:38.890873pt and 14.142136pt) -- cycle;
\clip (75.500000, 32.000000) +(-45.000000:38.890873pt and 14.142136pt) -- +(45.000000:38.890873pt and 14.142136pt) -- +(135.000000:38.890873pt and 14.142136pt) -- +(225.000000:38.890873pt and 14.142136pt) -- cycle;
\draw (75.500000, 32.000000) node {{$\Phi(E, \mathcal{P})$}};
\end{scope}
\filldraw (75.500000, 0.000000) circle(1.500000pt);
\draw (121.000000,32.000000) -- (121.000000,0.000000);
\begin{scope}
\draw[fill=white] (121.000000, 32.000000) +(-45.000000:14.142136pt and 14.142136pt) -- +(45.000000:14.142136pt and 14.142136pt) -- +(135.000000:14.142136pt and 14.142136pt) -- +(225.000000:14.142136pt and 14.142136pt) -- cycle;
\clip (121.000000, 32.000000) +(-45.000000:14.142136pt and 14.142136pt) -- +(45.000000:14.142136pt and 14.142136pt) -- +(135.000000:14.142136pt and 14.142136pt) -- +(225.000000:14.142136pt and 14.142136pt) -- cycle;
\draw (121.000000, 32.000000) node {{$\mathcal{P}$}};
\end{scope}
\filldraw (121.000000, 0.000000) circle(1.500000pt);
\draw (149.000000,32.000000) -- (149.000000,0.000000);
\begin{scope}
\draw[fill=white] (149.000000, 32.000000) +(-45.000000:14.142136pt and 14.142136pt) -- +(45.000000:14.142136pt and 14.142136pt) -- +(135.000000:14.142136pt and 14.142136pt) -- +(225.000000:14.142136pt and 14.142136pt) -- cycle;
\clip (149.000000, 32.000000) +(-45.000000:14.142136pt and 14.142136pt) -- +(45.000000:14.142136pt and 14.142136pt) -- +(135.000000:14.142136pt and 14.142136pt) -- +(225.000000:14.142136pt and 14.142136pt) -- cycle;
\draw (149.000000, 32.000000) node {{$\mathcal{P^\dagger}$}};
\end{scope}
\filldraw (149.000000, 0.000000) circle(1.500000pt);
\begin{scope}
\draw[fill=white] (173.000000, -0.000000) +(-45.000000:8.485281pt and 8.485281pt) -- +(45.000000:8.485281pt and 8.485281pt) -- +(135.000000:8.485281pt and 8.485281pt) -- +(225.000000:8.485281pt and 8.485281pt) -- cycle;
\clip (173.000000, -0.000000) +(-45.000000:8.485281pt and 8.485281pt) -- +(45.000000:8.485281pt and 8.485281pt) -- +(135.000000:8.485281pt and 8.485281pt) -- +(225.000000:8.485281pt and 8.485281pt) -- cycle;
\draw (173.000000, -0.000000) node {$H$};
\end{scope}
\draw (189.000000, 26.000000) -- (197.000000, 38.000000);
\draw[fill=white] (187.000000, -6.000000) rectangle (199.000000, 6.000000);
\draw[very thin] (193.000000, 0.600000) arc (90:150:6.000000pt);
\draw[very thin] (193.000000, 0.600000) arc (90:30:6.000000pt);
\draw[->,>=stealth] (193.000000, -5.400000) -- +(80:10.392305pt);
\draw[draw opacity=1.000000,fill opacity=0.200000,color=black,dotted] (109.000000,48.000000) rectangle (161.000000,-16.000000);
\draw (135.000000, 48.000000) node[text width=144pt,above,text centered,color=black] {{$\openone$}};
\draw[draw opacity=1.000000,fill opacity=0.200000,color=black,dotted] (109.000000,48.000000) rectangle (161.000000,-16.000000);
\end{tikzpicture}
	\caption{To prove the non-disturbing nature of the fundamental CPC gadget, it is useful to rearrange the circuit by moving the error operator $E$ through the first parity check $\mathcal{P}$. Following this rewrite, the controlled parity check operators are adjacent and cancel. In this form, the CPC gadget can be viewed as a measurement of the $\Phi(E,\mathcal{P})=\pm \openone_D$ operator on the data register. The value of the final syndrome measurement depends only upon whether $E$ commutes with $\mathcal{P}$. The slashed wire denotes that $\ket{\psi}_D$ is a multi-qubit register.}
	\label{fig:id_gadget}
\end{figure}

To prove its error detection capabilities, it is convenient to rearrange the circuit for the CPC gadget into the form shown in figure \ref{fig:id_gadget}. This rewrite is achieved by moving the error operator $E$ through the parity check operator $\mathcal{P}$. Both the error gate and the parity check gate are Pauli group operations. A property of the Pauli group is that its elements either commute or anti-commute with one another. Consequently, the effect of pushing the error operator to the front of the circuit is to introduce a global phase $\Phi(E,\mathcal{P})$ on the register which is controlled by the parity qubit. This global phase is dependent upon both the parity check and the error operator, and is defined as follows,
\begin{eqnarray}\label{eq:global_phase}
\Phi(E,\mathcal{P})=
\begin{cases}(+1)\openone_D,   & \text{if} \ [E,\mathcal{P}]=0\\(-1) \openone_D,   & \text{if} \ [E,\mathcal{P}]\neq0\rm,
\end{cases}
\end{eqnarray}
where $\openone_D$ is the identity operator on the data register and the commutator is given by $[E,\mathcal{P}]=E\bigcdot\mathcal{P}-\mathcal{P}\bigcdot E$. Note that, after the rewrite, the controlled parity-check operators are adjacent to each other and cancel. The full mathematical action of the CPC circuit $U_{\rm CPC}$, can now be expressed as follows,
\begin{equation}
U_{\rm CPC} \ket{\psi}_D \ket{0}_p=\left(\openone+\Phi(E,\mathcal{P})\right)E\ket{\psi}_D\ket{0}_p+\left(\openone-\Phi(E,\mathcal{P})\right)E\ket{\psi}_D\ket{1}_p.
\end{equation} 
Using the definition of the global phase operator $\Phi(E,\mathcal{P})$ given in equation (\ref{eq:global_phase}), the output of the CPC gadget simplifies to
\begin{eqnarray}\label{eq:CPC_output}
U_{\rm CPC} \ket{\psi}_D \ket{0}_p=
\begin{cases}E\ket{\psi}_D\ket{0}_p,   & \text{if} \ [E,\mathcal{P}]=0\\E\ket{\psi}_D\ket{1}_p,   & \text{if} \ [E,\mathcal{P}]\neq0\rm.
\end{cases}
\end{eqnarray}
From the above we can see that eventual syndrome measurement of parity qubit $p$ depends only upon whether $\mathcal{P}$ commutes with $E$. If no error occurs during the wait stage, then $E=\openone_D$ and the syndrome is measured deterministically as `$0$'. Likewise, if an error does occur, but it commutes with the parity operator, $[E,\mathcal{P}]=0$, then the syndrome is also `$0$'. Finally, if the error anti-commutes with the parity check, $[E,P]\neq 0$, then the syndrome is measured as `$1$'. A quantum error detection protocol can therefore be constructed from the CPC gadget by selecting a parity check that anti-commutes with the error to be identified. In the following subsections, we will show that CPC gadgets can be combined to create full QEC codes which can detect and localise multiple error types simultaneously.

The CPC gadget can be thought of as an extended measurement of the $\pm \openone_D$ operator on the data register, where the sign depends upon the commutation relation between $\mathcal{P}$ and $E$. As the $\pm \openone_D$ operator is trivially non-disturbing for all quantum states, there is no need for CPC codes to encode quantum information as logical qubits. Furthermore, it is clear from the output of the CPC gadget in equation (\ref{eq:CPC_output}), that the quantum data register is completely disentangled from the parity qubit prior to syndrome measurement. As a result, the only requirement on the parity checks is that they are Pauli group operators
\begin{equation}
\mathcal{P}\in \mathcal{G}\rm.
\end{equation}
Recall from equation (\ref{eq:trad_qec_req}), that for traditional codes, the choice of parity checks is limited to the set of stabilizers of the encoded logical qubits. The CPC framework lifts this restriction.

\joschka{It should be noted that, as the encoders and decoders consists entirely of Clifford operations, CPC codes form a class of stabilizer codes. A detailed explanation of the correspondence between CPC codes and stabilizer codes can be found in Chancellor et al \cite{cpc1}. The specific strength of the CPC framework lies in the fact that the symmetric \textit{encode-error-decode} structure provides a general method for creating a stabilizer code using any sequence of parity checks.}

\subsection{A CPC gadget for detecting bit-flips} \label{sec:prop_rules}

\begin{figure}
	\usetikzlibrary{decorations.pathreplacing,decorations.pathmorphing}
\providecommand{\ket}[1]{\left|#1\right\rangle}
\providecommand{\phase}[1]{e^{2{\pi}i\cdot#1}}
\begin{tikzpicture}[scale=1.500000,x=1pt,y=1pt]
\filldraw[color=white] (0.000000, -11.000000) rectangle (178.000000, 55.000000);
\draw[color=black] (0.000000,44.000000) -- (178.000000,44.000000);
\draw[color=black] (0.000000,22.000000) -- (178.000000,22.000000);
\draw[color=black] (0.000000,33.000000) node[left] {$\ket{\psi_{AB}}$};
\draw[color=black] (0.000000,0.000000) -- (170.000000,0.000000);
\draw[color=black] (170.000000,-0.500000) -- (178.000000,-0.500000);
\draw[color=black] (170.000000,0.500000) -- (178.000000,0.500000);
\draw[color=black] (0.000000,0.000000) node[left] {$\ket{0}_p$};
\begin{scope}
\draw[fill=white] (12.000000, -0.000000) +(-45.000000:8.485281pt and 8.485281pt) -- +(45.000000:8.485281pt and 8.485281pt) -- +(135.000000:8.485281pt and 8.485281pt) -- +(225.000000:8.485281pt and 8.485281pt) -- cycle;
\clip (12.000000, -0.000000) +(-45.000000:8.485281pt and 8.485281pt) -- +(45.000000:8.485281pt and 8.485281pt) -- +(135.000000:8.485281pt and 8.485281pt) -- +(225.000000:8.485281pt and 8.485281pt) -- cycle;
\draw (12.000000, -0.000000) node {$H$};
\end{scope}
\draw (28.000000,44.000000) -- (28.000000,0.000000);
\begin{scope}
\draw[fill=white] (28.000000, 44.000000) +(-45.000000:8.485281pt and 8.485281pt) -- +(45.000000:8.485281pt and 8.485281pt) -- +(135.000000:8.485281pt and 8.485281pt) -- +(225.000000:8.485281pt and 8.485281pt) -- cycle;
\clip (28.000000, 44.000000) +(-45.000000:8.485281pt and 8.485281pt) -- +(45.000000:8.485281pt and 8.485281pt) -- +(135.000000:8.485281pt and 8.485281pt) -- +(225.000000:8.485281pt and 8.485281pt) -- cycle;
\draw (28.000000, 44.000000) node {$Z$};
\end{scope}
\filldraw (28.000000, 0.000000) circle(1.500000pt);
\draw (44.000000,22.000000) -- (44.000000,0.000000);
\begin{scope}
\draw[fill=white] (44.000000, 22.000000) +(-45.000000:8.485281pt and 8.485281pt) -- +(45.000000:8.485281pt and 8.485281pt) -- +(135.000000:8.485281pt and 8.485281pt) -- +(225.000000:8.485281pt and 8.485281pt) -- cycle;
\clip (44.000000, 22.000000) +(-45.000000:8.485281pt and 8.485281pt) -- +(45.000000:8.485281pt and 8.485281pt) -- +(135.000000:8.485281pt and 8.485281pt) -- +(225.000000:8.485281pt and 8.485281pt) -- cycle;
\draw (44.000000, 22.000000) node {$Z$};
\end{scope}
\filldraw (44.000000, 0.000000) circle(1.500000pt);
\begin{scope}
\draw[fill=white] (60.000000, -0.000000) +(-45.000000:8.485281pt and 8.485281pt) -- +(45.000000:8.485281pt and 8.485281pt) -- +(135.000000:8.485281pt and 8.485281pt) -- +(225.000000:8.485281pt and 8.485281pt) -- cycle;
\clip (60.000000, -0.000000) +(-45.000000:8.485281pt and 8.485281pt) -- +(45.000000:8.485281pt and 8.485281pt) -- +(135.000000:8.485281pt and 8.485281pt) -- +(225.000000:8.485281pt and 8.485281pt) -- cycle;
\draw (60.000000, -0.000000) node {$H$};
\end{scope}
\begin{scope}[color=red]
\draw (76.000000,44.000000) -- (76.000000,22.000000);
\begin{scope}[color=red]
\begin{scope}
\draw[fill=white] (76.000000, 33.000000) +(-45.000000:8.485281pt and 24.041631pt) -- +(45.000000:8.485281pt and 24.041631pt) -- +(135.000000:8.485281pt and 24.041631pt) -- +(225.000000:8.485281pt and 24.041631pt) -- cycle;
\clip (76.000000, 33.000000) +(-45.000000:8.485281pt and 24.041631pt) -- +(45.000000:8.485281pt and 24.041631pt) -- +(135.000000:8.485281pt and 24.041631pt) -- +(225.000000:8.485281pt and 24.041631pt) -- cycle;
\draw (76.000000, 33.000000) node {$E$};
\end{scope}
\end{scope}
\end{scope}
\begin{scope}
\draw[fill=white] (92.000000, -0.000000) +(-45.000000:8.485281pt and 8.485281pt) -- +(45.000000:8.485281pt and 8.485281pt) -- +(135.000000:8.485281pt and 8.485281pt) -- +(225.000000:8.485281pt and 8.485281pt) -- cycle;
\clip (92.000000, -0.000000) +(-45.000000:8.485281pt and 8.485281pt) -- +(45.000000:8.485281pt and 8.485281pt) -- +(135.000000:8.485281pt and 8.485281pt) -- +(225.000000:8.485281pt and 8.485281pt) -- cycle;
\draw (92.000000, -0.000000) node {$H$};
\end{scope}
\draw (108.000000,22.000000) -- (108.000000,0.000000);
\begin{scope}
\draw[fill=white] (108.000000, 22.000000) +(-45.000000:8.485281pt and 8.485281pt) -- +(45.000000:8.485281pt and 8.485281pt) -- +(135.000000:8.485281pt and 8.485281pt) -- +(225.000000:8.485281pt and 8.485281pt) -- cycle;
\clip (108.000000, 22.000000) +(-45.000000:8.485281pt and 8.485281pt) -- +(45.000000:8.485281pt and 8.485281pt) -- +(135.000000:8.485281pt and 8.485281pt) -- +(225.000000:8.485281pt and 8.485281pt) -- cycle;
\draw (108.000000, 22.000000) node {$Z$};
\end{scope}
\filldraw (108.000000, 0.000000) circle(1.500000pt);
\draw (124.000000,44.000000) -- (124.000000,0.000000);
\begin{scope}
\draw[fill=white] (124.000000, 44.000000) +(-45.000000:8.485281pt and 8.485281pt) -- +(45.000000:8.485281pt and 8.485281pt) -- +(135.000000:8.485281pt and 8.485281pt) -- +(225.000000:8.485281pt and 8.485281pt) -- cycle;
\clip (124.000000, 44.000000) +(-45.000000:8.485281pt and 8.485281pt) -- +(45.000000:8.485281pt and 8.485281pt) -- +(135.000000:8.485281pt and 8.485281pt) -- +(225.000000:8.485281pt and 8.485281pt) -- cycle;
\draw (124.000000, 44.000000) node {$Z$};
\end{scope}
\filldraw (124.000000, 0.000000) circle(1.500000pt);
\begin{scope}
\draw[fill=white] (140.000000, -0.000000) +(-45.000000:8.485281pt and 8.485281pt) -- +(45.000000:8.485281pt and 8.485281pt) -- +(135.000000:8.485281pt and 8.485281pt) -- +(225.000000:8.485281pt and 8.485281pt) -- cycle;
\clip (140.000000, -0.000000) +(-45.000000:8.485281pt and 8.485281pt) -- +(45.000000:8.485281pt and 8.485281pt) -- +(135.000000:8.485281pt and 8.485281pt) -- +(225.000000:8.485281pt and 8.485281pt) -- cycle;
\draw (140.000000, -0.000000) node {$H$};
\end{scope}
\draw[fill=white] (164.000000, -6.000000) rectangle (176.000000, 6.000000);
\draw[very thin] (170.000000, 0.600000) arc (90:150:6.000000pt);
\draw[very thin] (170.000000, 0.600000) arc (90:30:6.000000pt);
\draw[->,>=stealth] (170.000000, -5.400000) -- +(80:10.392305pt);
\draw[draw opacity=1.000000,fill opacity=0.200000,color=black,dotted] (5.000000,55.000000) rectangle (67.000000,-11.000000);
\draw (36.000000, -11.000000) node[text width=144pt,below,text centered,color=black] {Encoder};
\draw[draw opacity=1.000000,fill opacity=0.200000,color=black,dotted] (5.000000,55.000000) rectangle (67.000000,-11.000000);
\draw[draw opacity=1.000000,fill opacity=0.200000,color=black,dotted] (85.000000,55.000000) rectangle (147.000000,-11.000000);
\draw (116.000000, -11.000000) node[text width=144pt,below,text centered,color=black] {Decoder};
\draw[draw opacity=1.000000,fill opacity=0.200000,color=black,dotted] (85.000000,55.000000) rectangle (147.000000,-11.000000);
\end{tikzpicture}
	\caption{A CPC gadget for detecting single bit-flips on a two-qubit data register $\ket{\psi_{AB}}$. The gadget returns a `$0$' syndrome measurement if there is no error and a `$1$' if a bit-flip occurs during the wait stage.}
	\label{bitFlips}
\end{figure}
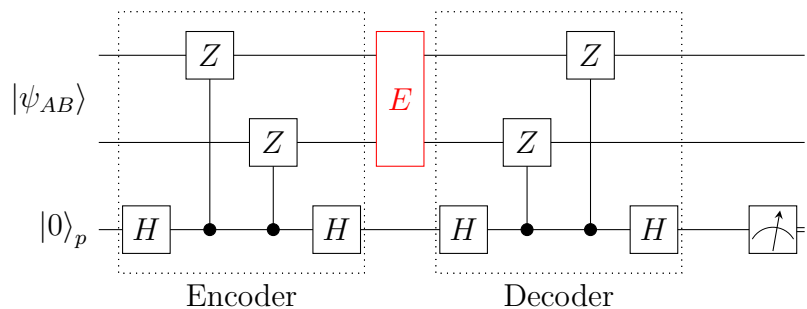 

We now provide specific examples of CPC gadgets to detect bit-flips and phase-flips on a two-qubit data register $\ket{\psi_{AB}}$. Following this, we describe how the two types of CPC gadget can be combined to create a $[[4,2,2]]$ detection code. 

\begin{figure}
	\includegraphics[width=0.8\textwidth]{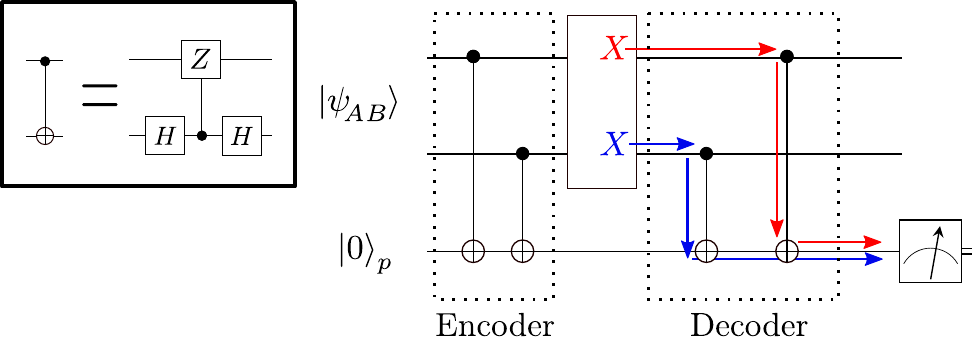}
	\caption{Right: the bit-flip CPC gadget rewritten using the circuit rewrite rule to the left. Expressing the gadget in this form allows for easy visualisation of the propagation of errors from the wait stage to the parity check measurement. The red and blue arrows show the two possible bit-flip propagation pathways.}
	\label{bitFlipsCNOT}
\end{figure}

In order to design a CPC gadget that will detect single bit-flips on the register $\ket{\psi_{AB}}$, we need a parity check that anti-commutes with the errors in the set $\mathcal{E}_X=\{X_A,X_B\}$. Setting $\mathcal{P}_{AB}=Z_AZ_B$ satisfies this requirement to give the bit-flip CPC gadget depicted in figure \ref{bitFlips}. Note that $X$ and $Z$ are Pauli operators which are defined in appendix \ref{app:pauli}. It is useful to rewrite the circuit in figure \ref{bitFlips} in terms of \cnot gates using the gate substitution defined by the following matrix equation
\begin{equation}
\CZeq^{q_1}_{q_2} = \left( \openone_{q_1} \otimes H_{q_2} \right)\bigcdot \cnoteq^{q_1}_{q_2} \bigcdot \left( \openone_{q_1} \otimes H_{q_2} \right)\rm,
\end{equation}
\noindent where \CZ is a controlled-Z gate and $q_1$ and $q_2$ are the input qubits. The resultant circuit is shown in figure \ref{bitFlipsCNOT}.  In this form, the operation of the CPC gadget can easily be visualised by considering the propagation of errors through the decoder. A \cnot gate will propagate a bit-flip error from the control qubit $q_1$ to the target $q_2$ as follows,
\begin{equation}\label{eq:cnot_prop}
\cnoteq^{q_1}_{q_2} \bigcdot \left(X_{q_1}\otimes \openone_{q_2}\right) \bigcdot  \cnoteq^{q_1}_{q_2} = X_{q_1} \otimes X_{q_2}\rm .
\end{equation}

\noindent Implementing the above propagation rule, the red and blue arrows in figure \ref{bitFlipsCNOT} depict the possible detection pathways for bit-errors from the wait stage to the parity check qubit.

\subsection{A CPC gadget for detecting phase-flips} \label{sec:cpg_prop}

\begin{figure}
	\includegraphics[width=0.8\textwidth]{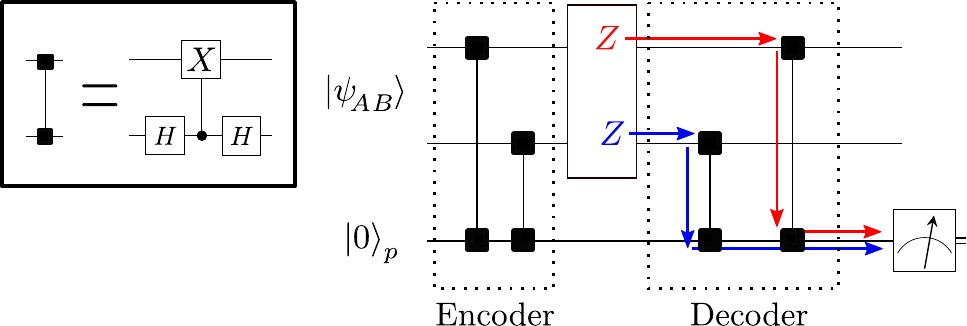}
	\caption{A phase-flip CPC gadget. Phase flips on the register $\ket{\psi_A\psi_B}$ can be detected by setting $\mathcal{P}=X_AX_B$. In the above-right, we have expressed the phase-flip gadget in terms of conjugate-propagator gates, which are defined in the box to the left. The conjugate-propagator gates are symmetric gates which are designed to copy $Z$ errors from one qubit to another. The red and blue arrows depict the possible propagation pathways for $Z$ errors from the wait stage to the parity check measurement.}  
	\label{phaseFlip}
\end{figure}

A CPC gadget that detects errors from the set $\mathcal{E}_Z=\{Z_A,Z_B\}$ can be obtained using a parity check of the form $\mathcal{P}_{AB}=X_AX_B$. Figure \ref{phaseFlip} depicts the phase-flip CPC gadget expressed in terms of the conjugate-propagator gate $\Lambda^{q_1}_{q_2}$ given by
\begin{equation}
\Lambda^{q_1}_{q_2}=\left(\openone_{q_1}\otimes H_{q_2}\right) \bigcdot \cnoteq^{q_2}_{q_1} \bigcdot \left(\openone_{q_1}\otimes H_{q_2}\right) .
\end{equation}
\noindent The conjugate-propagator gate is a symmetric two-qubit operator with the following propagation rule for $Z$-errors  
\begin{equation}
\Lambda^{q_1}_{q_2} \bigcdot \left(Z_{q_1}\otimes \openone_{q_2}\right) \bigcdot \Lambda^{q_1}_{q_2}  \rightarrow Z_{q_1}\otimes X_{q_2}.
\end{equation}

\noindent Phase-flip errors in the wait stage are copied to the parity qubit via a conjugate-propagator gate which converts the $Z$-error to an $X$-error that can be detected in the computational basis. Figure \ref{phaseFlip} depicts the possible error propagation pathways for errors in the phase-flip CPC gadget.

\subsection{The [[4,2,2]] error detection code}\label{sec:422}

\begin{figure}
	\includegraphics[width=0.8\textwidth]{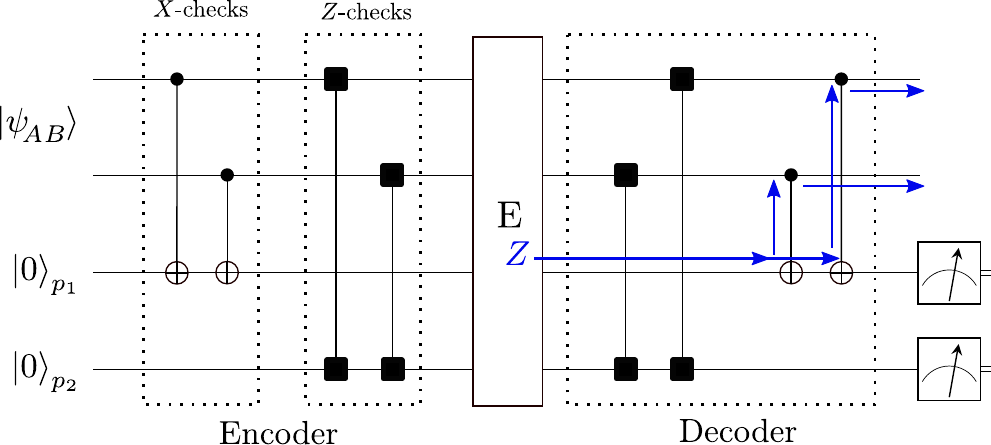}
	
	\caption{The circuit formed by combining the bit-flip CPC gadget to the phase-flip CPC gadget. This circuit can detect both $X$ and $Z$ errors on qubits $A$, $B$ and $p_2$. However, a phase-flip error on qubit $p_1$ will propagate errors to the register without triggering a syndrome, as shown by the blue arrows. This propagation loophole can be closed through the addition of cross-check operators.}
	\label{fig:422_circ1}
\end{figure}

\begin{figure}
	\includegraphics[width=0.9\textwidth]{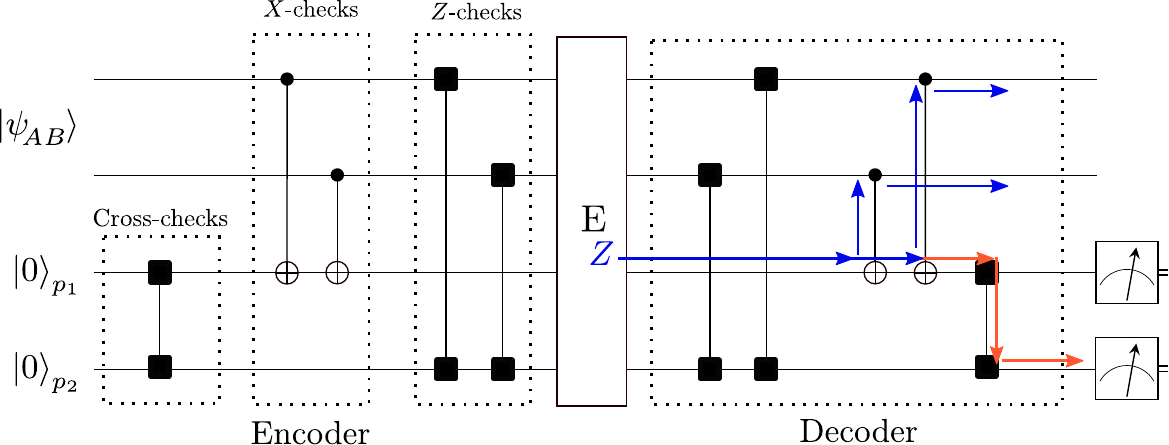}
	
	\caption{The $[[4,4,2]]$ CPC quantum detection code, formed by combining the bit-flip and phase-flip gadgets. The addition of cross-check operators ensures that errors do not propagate from the parity qubits to the register in an undetectable way.}
	\label{fig:422_circ2}
\end{figure}

\begin{table}[]
	\centering
	\begin{tabular}{|l|l|}
		\hline
		\textbf{Error}             & \textbf{Syndrome} \\ \hline
		$I$                        & $0_{p_1}0_{p_2}$    \\ \hline
		$X_A, X_B, X_{p_1}, Z_{p_2}$ & $1_{p_1}0_{p_2}$    \\ \hline
		$Z_A, Z_B, Z_{p_1}, X_{p_2}$ & $0_{p_1}1_{p_2}$    \\ \hline
		$Y_A, Y_B, Y_{p_1}, Y_{p_2}$ & $1_{p_1}1_{p_2}$    \\ \hline
	\end{tabular}
	\caption{The syndrome table for the $[[4,2,2]]$ quantum error detection code. If no errors occur, the code returns a `$00$' syndrome. If a single $X$, $Y$ or $Z$ error occurs on any of the four qubits, a non-zero syndrome is returned.}
	\label{tab:422synd}
\end{table} 

We now show how the bit-flip and phase-flip CPC gadgets can be combined to form a full quantum error detection code. Figure \ref{fig:422_circ1} shows the CPC circuit formed by combining the bit-flip gadget with the phase flip-gadget. By considering the error propagation rules outlined in the previous subsections, it can be verified that this circuit will detect errors which occur on the register qubits $\ket{\psi_{AB}}$, but not errors which occur the parity qubits ${p_1}$ and $p_2$. We now show how the code can be modified to enable error detection across all four of the qubits.

The blue arrows in figure \ref{fig:422_circ1} show that a phase-flip error on the first parity qubit $p_1$ will propagate errors to the register in an undetectable way. Fortunately, a detection pathway can be created by applying a conjugate-propagator gate between the parity qubits at the end of the decoder (from now on, we will refer to these additional gates as `cross-checks'). As shown by the orange arrows in figure \ref{fig:422_circ2}, this cross-check propagates the phase-error to the parity-check qubit $p_2$ and converts it to an $X$-error that can be picked up by a computational basis measurement. With the addition of the cross-check, the circuit becomes a fully functional [[4,2,2]] quantum error detection code. The single-qubit error syndromes are given in table \ref{tab:422synd}, and demonstrate the code can detect the occurrence of $X$, $Y$ and $Z$ errors on any of the 4 qubits.

As the [[4,2,2]] code is a detection code, the syndromes do not give us enough information to pinpoint which qubit the error occurred on. The construction of full error correcting CPC codes, that can both identify and localise errors, will be outlined in the next section.

\subsection{The canonical form of CPC codes}\label{sec:can_form}

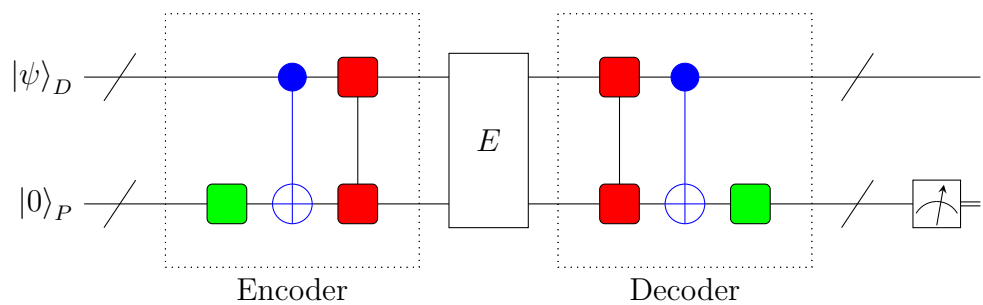
\begin{figure}
	\usetikzlibrary{decorations.pathreplacing,decorations.pathmorphing}
\providecommand{\ket}[1]{\left|#1\right\rangle}
\providecommand{\phase}[1]{e^{2{\pi}i\cdot#1}}
\begin{tikzpicture}[scale=1.500000,x=1pt,y=1pt]
\filldraw[color=white] (0.000000, -16.000000) rectangle (226.000000, 48.000000);
\draw[color=black] (0.000000,32.000000) -- (226.000000,32.000000);
\draw[color=black] (0.000000,32.000000) node[left] {$\ket{\psi}_D$};
\draw[color=black] (0.000000,0.000000) -- (215.000000,0.000000);
\draw[color=black] (215.000000,-0.500000) -- (226.000000,-0.500000);
\draw[color=black] (215.000000,0.500000) -- (226.000000,0.500000);
\draw[color=black] (0.000000,0.000000) node[left] {$\ket{0}_P$};
\draw (5.000000, 26.000000) -- (13.000000, 38.000000);
\draw (5.000000, -6.000000) -- (13.000000, 6.000000);
\begin{scope}
\begin{scope}[shift={(36.000000,0.000000)}]
\draw[fill=green, rounded corners=2] (-5, -5) rectangle (5, 5) {};
\end{scope}
\end{scope}
\begin{scope}[color=blue]
\draw (52.500000,32.000000) -- (52.500000,0.000000);
\filldraw (52.500000, 32.000000) circle(3.500000pt);
\begin{scope}
\draw[fill=white] (52.500000, 0.000000) circle(5.000000pt);
\clip (52.500000, 0.000000) circle(5.000000pt);
\draw (47.500000, 0.000000) -- (57.500000, 0.000000);
\draw (52.500000, -5.000000) -- (52.500000, 5.000000);
\end{scope}
\end{scope}
\draw (69.000000,32.000000) -- (69.000000,0.000000);
\begin{scope}
\begin{scope}[shift={(69.000000,32.000000)}]
\draw[fill=red, rounded corners=2] (-5, -5) rectangle (5, 5) {};
\end{scope}
\end{scope}
\begin{scope}
\begin{scope}[shift={(69.000000,0.000000)}]
\draw[fill=red, rounded corners=2] (-5, -5) rectangle (5, 5) {};
\end{scope}
\end{scope}
\draw (102.000000,32.000000) -- (102.000000,0.000000);
\begin{scope}
\draw[fill=white] (102.000000, 16.000000) +(-45.000000:14.142136pt and 31.112698pt) -- +(45.000000:14.142136pt and 31.112698pt) -- +(135.000000:14.142136pt and 31.112698pt) -- +(225.000000:14.142136pt and 31.112698pt) -- cycle;
\clip (102.000000, 16.000000) +(-45.000000:14.142136pt and 31.112698pt) -- +(45.000000:14.142136pt and 31.112698pt) -- +(135.000000:14.142136pt and 31.112698pt) -- +(225.000000:14.142136pt and 31.112698pt) -- cycle;
\draw (102.000000, 16.000000) node {{$E$}};
\end{scope}
\draw (135.000000,32.000000) -- (135.000000,0.000000);
\begin{scope}
\begin{scope}[shift={(135.000000,32.000000)}]
\draw[fill=red, rounded corners=2] (-5, -5) rectangle (5, 5) {};
\end{scope}
\end{scope}
\begin{scope}
\begin{scope}[shift={(135.000000,0.000000)}]
\draw[fill=red, rounded corners=2] (-5, -5) rectangle (5, 5) {};
\end{scope}
\end{scope}
\begin{scope}[color=blue]
\draw (151.500000,32.000000) -- (151.500000,0.000000);
\filldraw (151.500000, 32.000000) circle(3.500000pt);
\begin{scope}
\draw[fill=white] (151.500000, 0.000000) circle(5.000000pt);
\clip (151.500000, 0.000000) circle(5.000000pt);
\draw (146.500000, 0.000000) -- (156.500000, 0.000000);
\draw (151.500000, -5.000000) -- (151.500000, 5.000000);
\end{scope}
\end{scope}
\begin{scope}
\begin{scope}[shift={(168.000000,0.000000)}]
\draw[fill=green, rounded corners=2] (-5, -5) rectangle (5, 5) {};
\end{scope}
\end{scope}
\draw (191.000000, 26.000000) -- (199.000000, 38.000000);
\draw (191.000000, -6.000000) -- (199.000000, 6.000000);
\draw[fill=white] (209.000000, -6.000000) rectangle (221.000000, 6.000000);
\draw[very thin] (215.000000, 0.600000) arc (90:150:6.000000pt);
\draw[very thin] (215.000000, 0.600000) arc (90:30:6.000000pt);
\draw[->,>=stealth] (215.000000, -5.400000) -- +(80:10.392305pt);
\draw[draw opacity=1.000000,fill opacity=0.200000,color=black,dotted] (20.500000,48.000000) rectangle (84.500000,-16.000000);
\draw (52.500000, -16.000000) node[text width=144pt,below,text centered,color=black] {{Encoder}};
\draw[draw opacity=1.000000,fill opacity=0.200000,color=black,dotted] (20.500000,48.000000) rectangle (84.500000,-16.000000);
\draw[draw opacity=1.000000,fill opacity=0.200000,color=black,dotted] (119.500000,48.000000) rectangle (183.500000,-16.000000);
\draw (151.500000, -16.000000) node[text width=144pt,below,text centered,color=black] {{Decoder}};
\draw[draw opacity=1.000000,fill opacity=0.200000,color=black,dotted] (119.500000,48.000000) rectangle (183.500000,-16.000000);
\end{tikzpicture}
	\caption{The canonical form of CPC codes, showing the symmetric \textit{encode-error-decode} structure. In an $[[n,k,d]]$ CPC code the qubits are split into two distinct types: $k$ data qubits, $\ket{\psi}_D=\ket{\psi_{D_1}\psi_{D_2}...\psi_{D_k}}$, and $n-k$ parity qubits, $\ket{0}_P=\ket{0_{p_1}0_{p_2}...0_{p_{n-k}}  }$. The encoder involves successive rounds of cross-checks (green), bit-checks (blue) and phase-checks (red). The decoder is simply the unitary inverse of the encoder.}
	\label{fig:can_qec}
\end{figure} 

The [[4,2,2]] quantum error detection code illustrates the basic principles behind the operation of a CPC code. The encoder is constructed by combining a bit-flip CPC gadget with a phase-flip CPC gadget. Under this canonical ordering, errors on the parity qubits are identifiable via the addition of the cross-check operators. A compact way of representing CPC codes is in terms of adjacency matrices which describe the connectivity between the register and parity qubits. For example, the adjacency matrices for the [[4,2,2]] code are
	\begin{equation} \label{eq:adj_mat}
	m_b=\begin{blockarray}{cccccc}
\ & \text{\scriptsize $[p1]$} & \text{\scriptsize $[p2]$}  \\
\begin{block}{c(ccccc)}
\text{\scriptsize{$[A]$}} & 1 & 0  \\
\text{\scriptsize{$[B]$}} & 1 & 0   \\
\end{block}
\end{blockarray},
 \ \
	m_p=\begin{blockarray}{cccccc}
\ & \text{\scriptsize $[p1]$} & \text{\scriptsize $[p2]$}  \\
\begin{block}{c(ccccc)}
\text{\scriptsize{$[A]$}} & 0 & 1  \\
\text{\scriptsize{$[B]$}} & 0 & 1   \\
\end{block}
\end{blockarray},
 \ \
m_c=\begin{blockarray}{cccccc}
\ & \text{\scriptsize $[p1]$} & \text{\scriptsize $[p2]$}  \\
\begin{block}{c(ccccc)}
\text{\scriptsize{$[p1]$}} & 0 & 1  \\
\text{\scriptsize{$[p2]$}} & 0 & 0   \\
\end{block}
\end{blockarray},
	\end{equation}		
\noindent where $m_b$ represents the bit-checks, $m_p$ the phase-checks and $m_c$ the cross-checks. For the bit-flip and phase-flip adjacency matrices, $m_b$ and $m_p$, the rows refer to the data qubits and the columns the parity qubits. Looking at the bit-flip matrix, we can see that both register qubits connect to parity qubit $p_1$ via \cnot gates in accordance with the circuit in figure \ref{fig:422_circ2}. Likewise, matrix $m_P$ tells us that both register qubits are connected to parity qubit $p_2$ via conjugate-propagator gates. Finally, from matrix $m_c$, we see that there is a single cross-check between parity qubits $p_1$ and $p_2$. \joschka{The cross-checks result in a matrix that is always symmetric.  We follow \cite{cpc1} in representing this as an upper triangular matrix, so that the number of non-zero entries corresponds to the number of two-qubit gates.}

We are now in a position to extend the CPC framework to enable the description of more general codes. The canonical form of an [[n,k,d]] CPC code is shown in figure \ref{fig:can_qec}. Such codes have $k$ data qubits, $\ket{\psi}_D=\ket{\psi_{D_1}\psi_{D_2}...\psi_{D_k}}$, and $m=n-k$ parity qubits, $\ket{0}_P=\ket{0_{p_1}0_{p_2}...0_{p_{m}}  }$. As with the detection schemes described previously, the encode stage of a general CPC code involves successive rounds of cross-checks, bit-checks and then phase-checks. The sequence of gates within each stage of the encoder can be compactly described in terms of adjacency matrices of the form   

	\begin{equation}
\begin{split}
	m_b=\begin{blockarray}{cccccc}
\ & \text{\scriptsize $[p_1]$} & \text{\scriptsize $[p_2]$} & \text{\scriptsize [...]}  & \text{\scriptsize $[p_m]$}\\
\begin{block}{c(ccccc)}
\text{\scriptsize{$[D_1]$}} & b_{11} & b_{12} & ... & b_{1} \\
\text{\scriptsize{$[D_2]$}} & b_{21} & b_{22} & ... & b_{2m} \\
\text{\scriptsize{$[...]$}} & ... & ... & ... & ... \\
\text{\scriptsize{$[D_k]$}} & b_{k1} & b_{k2} & ... & b_{km} \\
\end{block}
\end{blockarray} \ \rm,
\quad
	m_p=\begin{blockarray}{cccccc}
\ & \text{\scriptsize $[p_1]$} & \text{\scriptsize $[p_2]$} & \text{\scriptsize [...]}  & \text{\scriptsize $[p_m]$}\\
\begin{block}{c(ccccc)}
\text{\scriptsize{$[D_1]$}} & h_{11} & h_{12} & ... & h_{1m} \\
\text{\scriptsize{$[D_2]$}} & h_{21} & h_{22} & ... & h_{2m} \\
\text{\scriptsize{$[...]$}} & ... & ... & ... & ... \\
\text{\scriptsize{$[D_k]$}} & h_{k1} & h_{k2} & ... & h_{km} \\
\end{block}
\end{blockarray} \ \rm,
 \\
	m_c=\begin{blockarray}{cccccc}
\ & \text{\scriptsize $[p_1]$} & \text{\scriptsize $[p_2]$} & \text{\scriptsize [...]}  & \text{\scriptsize $[p_m]$}\\
\begin{block}{c(ccccc)}
\text{\scriptsize{$[p_1]$}} & 0 & c_{12} & ... & c_{1m} \\
\text{\scriptsize{$[p_2]$}} & 0 & 0 & ... & c_{2m} \\
\text{\scriptsize{$[...]$}} & ... & ... & ... & ... \\
\text{\scriptsize{$[p_{(m-1)}]$}} & 0 & 0 & ... & c_{(m-1)m} \\
\text{\scriptsize{$[p_{m}]$}} & 0 & 0 & ... & 0 \\
\end{block}
\end{blockarray} \ \rm,
\end{split}
\end{equation}

\noindent where $b_{xy}$, $h_{xy}$, $c_{xy}$ are binary values. \joschka{As mentioned previously}, for simplicity, the cross-check matrix $m_c$ is always represented as an upper triangular matrix.

\subsection{Numerical CPC code discovery}

We have now outlined the canonical structure of CPC codes, and shown how they can be represented in terms of three adjacency matrices. The CPC framework removes the need to start out by redundantly encoding quantum data, and allows QEC protocols to be implemented with any parity check. As such, the CPC framework essentially reduces the task of deriving QEC codes to a classical decoding problem.   

A new $[[n,k,d=?]]$ CPC circuit can be generated simply by selecting a random instance of the adjacency matrices for an $n$-qubit code with $k$ data qubits. The symmetric \textit{encode-error-decode} structure of the CPC code will ensure that the random sequence of parity checks this set of adjacency matrices represents does not decohere the register. The only task necessary to verify whether the circuit represents a working CPC code is to measure the code distance $d$. This can be done by testing the circuits with all of the errors in the chosen error model. If each error produces a unique syndrome, then the code distance is $d\geq 3$, and the circuit represents a working CPC code.

In this paper we only consider quantum memories. As a result, the codes under consideration are Clifford circuits. The code distance can therefore be efficiently verified \joschka{for small-distance codes} using a stabilizer simulator such as \cite{Aaronson2004, Anders2006}. Alternatively, we have developed an algorithm specifically for calculating the syndromes of CPC codes, which is based on error propagation rules outlined in sections \ref{sec:prop_rules} and \ref{sec:cpg_prop}. This algorithm is described in appendix \ref{app:synd_tab}, and can be implemented in less than $200$ lines of Python code.

\section{Implementation of the $[[4,2,2]]$ code on the IBM 5Q device}\label{sec:ibm}

\begin{figure}
	\centering
	\includegraphics[width=0.3\textwidth]{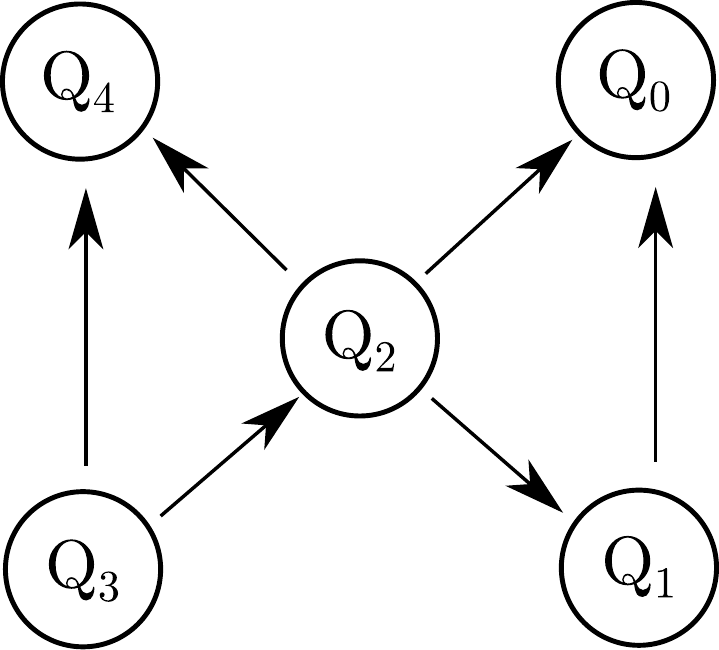}
\caption{The connectivity map of the IBMQX4 version of the IBM 5Q quantum computer illustrating the `bowtie' layout. The arrows indicate the allowed \cnot operations and their preferred directions.}
	\label{fig:ibmqx4}
\end{figure}

As a simple first experimental example of a CPC code, we now consider the compilation and execution of a $[[4,2,2]]$ quantum detection code on a superconducting qubit device. The IBM 5Q is a small-scale quantum computer, built and maintained by IBM Quantum \cite{IBM}. The device has five programmable superconducting transmon qubits, and is accessible to the public via the Internet. In \cite{Takita2017}, the IBM 5Q was shown to allow fault tolerant preparation of codewords for a $[[4,2,2]]$ code. It has also been demonstrated, in \cite{Vuillot2017}, that certain $[[4,2,2]]$ encoded operations on the IBM 5Q have a lower error rate than the equivalent operation on the device's raw qubits. Here, we implement a complete encode-decode cycle of a $[[4,2,2]]$ CPC quantum memory using the IBM 5Q. Our aim is to demonstrate that the fidelity of the code's output state can be improved by post-selection.

\subsection{Experimental overview and conditions for success}

Our experiment on the IBM 5Q encodes a single input state $\ket{\psi_{AB}}=\ket{+_A 0_B}$ using a $[[4,2,2]]$ CPC quantum memory of the type described in section \ref{sec:422}. The $\ket{+_A 0_B}$ state is an easy-to-prepare quantum state that is susceptible to both bit- and phase-flip errors, and therefore provides a suitable test of the $[[4,2,2]]$ CPC code as a quantum memory.

Ultimately, the condition for success for a quantum code is to test whether the encoded protocol has a lower logical error rate than the equivalent circuit before encoding. In the case of quantum memory, the circuit that is encoded is simply an extended identity operation. The usefulness of the $[[4,2,2]]$ code could therefore be assessed by comparing the fidelity of the encoded output to the equivalent output of an unprotected two-qubit data register. \joschkaEdit{However, the gate error rates on the IBM 5Q hardware are too high for such a comparison to yield a positive result. This problem is compounded by the fact that the IBM hardware limits the experiment to a single encode-decode cycle, meaning certain regions of the $[[4,2,2]]$ circuit -- before the encoder and after the decoder -- are left unprotected.} Consequently, the aim of the experiment presented here is restricted to demonstrating that, whilst not suppressing the logical error rate, the $[[4,2,2]]$ CPC code does detect errors. We now describe the method by which this is achieved.

\joschka{The compiled $[[4,2,2]]$ CPC code is run multiple times with the input state $\ket{\psi_{AB}}=\ket{+_A 0_B}$ on the IBM 5Q hardware. At the end of each CPC code cycle,} the parity qubits are measured to provide a syndrome \joschka{designed to} indicate if an error has occurred. \joschka{An approximation to the output state of the register is reconstructed from the experimental data using quantum state tomography. The quality of this output is quantified by calculating its fidelity relative to the input state $\ket{+_A 0_B}$. In this experiment we compare the output fidelity of the $[[4,2,2]]$ protocol before and after post-selection. In the former, the syndrome information is ignored, whereas in the latter it is used to determine which experimental runs are discarded during post-selection. The condition for success is that the post-selection should improve the output fidelity. If this is the case, it will demonstrate that the $[[4,2,2]]$ CPC code is detecting errors and produces useful syndrome information.} 

\begin{figure*}[t]

	\subfloat[]{%
		\providecommand{\ket}[1]{\left|#1\right\rangle}
\providecommand{\phase}[1]{e^{2{\pi}i\cdot#1}}
\begin{tikzpicture}[scale=1.350000,x=1pt,y=1pt]
\filldraw[color=white] (0.000000, -7.000000) rectangle (71.000000, 21.000000);
\draw[color=black] (0.000000,14.000000) -- (71.000000,14.000000);
\draw[color=black] (0.000000,0.000000) -- (71.000000,0.000000);
\draw (5.000000,14.000000) -- (5.000000,0.000000);
\filldraw (5.000000, 14.000000) circle(1.500000pt);
\begin{scope}
\draw[fill=white] (5.000000, 0.000000) circle(3.000000pt);
\clip (5.000000, 0.000000) circle(3.000000pt);
\draw (2.000000, 0.000000) -- (8.000000, 0.000000);
\draw (5.000000, -3.000000) -- (5.000000, 3.000000);
\end{scope}
\draw[fill=white,color=white] (12.000000, -6.000000) rectangle (27.000000, 20.000000);
\draw (19.500000, 7.000000) node {$=$};
\begin{scope}
\draw[fill=white] (37.000000, 14.000000) +(-45.000000:8.485281pt and 8.485281pt) -- +(45.000000:8.485281pt and 8.485281pt) -- +(135.000000:8.485281pt and 8.485281pt) -- +(225.000000:8.485281pt and 8.485281pt) -- cycle;
\clip (37.000000, 14.000000) +(-45.000000:8.485281pt and 8.485281pt) -- +(45.000000:8.485281pt and 8.485281pt) -- +(135.000000:8.485281pt and 8.485281pt) -- +(225.000000:8.485281pt and 8.485281pt) -- cycle;
\draw (37.000000, 14.000000) node {$H$};
\end{scope}
\begin{scope}
\draw[fill=white] (37.000000, -0.000000) +(-45.000000:8.485281pt and 8.485281pt) -- +(45.000000:8.485281pt and 8.485281pt) -- +(135.000000:8.485281pt and 8.485281pt) -- +(225.000000:8.485281pt and 8.485281pt) -- cycle;
\clip (37.000000, -0.000000) +(-45.000000:8.485281pt and 8.485281pt) -- +(45.000000:8.485281pt and 8.485281pt) -- +(135.000000:8.485281pt and 8.485281pt) -- +(225.000000:8.485281pt and 8.485281pt) -- cycle;
\draw (37.000000, -0.000000) node {$H$};
\end{scope}
\draw (50.000000,14.000000) -- (50.000000,0.000000);
\filldraw (50.000000, 0.000000) circle(1.500000pt);
\begin{scope}
\draw[fill=white] (50.000000, 14.000000) circle(3.000000pt);
\clip (50.000000, 14.000000) circle(3.000000pt);
\draw (47.000000, 14.000000) -- (53.000000, 14.000000);
\draw (50.000000, 11.000000) -- (50.000000, 17.000000);
\end{scope}
\begin{scope}
\draw[fill=white] (63.000000, 14.000000) +(-45.000000:8.485281pt and 8.485281pt) -- +(45.000000:8.485281pt and 8.485281pt) -- +(135.000000:8.485281pt and 8.485281pt) -- +(225.000000:8.485281pt and 8.485281pt) -- cycle;
\clip (63.000000, 14.000000) +(-45.000000:8.485281pt and 8.485281pt) -- +(45.000000:8.485281pt and 8.485281pt) -- +(135.000000:8.485281pt and 8.485281pt) -- +(225.000000:8.485281pt and 8.485281pt) -- cycle;
\draw (63.000000, 14.000000) node {$H$};
\end{scope}
\begin{scope}
\draw[fill=white] (63.000000, -0.000000) +(-45.000000:8.485281pt and 8.485281pt) -- +(45.000000:8.485281pt and 8.485281pt) -- +(135.000000:8.485281pt and 8.485281pt) -- +(225.000000:8.485281pt and 8.485281pt) -- cycle;
\clip (63.000000, -0.000000) +(-45.000000:8.485281pt and 8.485281pt) -- +(45.000000:8.485281pt and 8.485281pt) -- +(135.000000:8.485281pt and 8.485281pt) -- +(225.000000:8.485281pt and 8.485281pt) -- cycle;
\draw (63.000000, -0.000000) node {$H$};
\end{scope}
\end{tikzpicture}%
		\label{}%
	}\qquad
	\subfloat[]{%
		\providecommand{\ket}[1]{\left|#1\right\rangle}
\providecommand{\phase}[1]{e^{2{\pi}i\cdot#1}}
\begin{tikzpicture}[scale=1.350000,x=1pt,y=1pt]
\filldraw[color=white] (0.000000, -7.000000) rectangle (71.000000, 21.000000);
\draw[color=black] (0.000000,14.000000) -- (71.000000,14.000000);
\draw[color=black] (0.000000,0.000000) -- (71.000000,0.000000);
\draw (5.000000,14.000000) -- (5.000000,0.000000);
\begin{scope}
\begin{scope}[shift={(5.000000,14.000000)}]
\draw[fill=black, rounded corners=2] (-3, -3) rectangle (3, 3) {};
\end{scope}
\end{scope}
\begin{scope}
\begin{scope}[shift={(5.000000,0.000000)}]
\draw[fill=black, rounded corners=2] (-3, -3) rectangle (3, 3) {};
\end{scope}
\end{scope}
\draw[fill=white,color=white] (12.000000, -6.000000) rectangle (27.000000, 20.000000);
\draw (19.500000, 7.000000) node {$=$};
\begin{scope}
\draw[fill=white] (37.000000, 14.000000) +(-45.000000:8.485281pt and 8.485281pt) -- +(45.000000:8.485281pt and 8.485281pt) -- +(135.000000:8.485281pt and 8.485281pt) -- +(225.000000:8.485281pt and 8.485281pt) -- cycle;
\clip (37.000000, 14.000000) +(-45.000000:8.485281pt and 8.485281pt) -- +(45.000000:8.485281pt and 8.485281pt) -- +(135.000000:8.485281pt and 8.485281pt) -- +(225.000000:8.485281pt and 8.485281pt) -- cycle;
\draw (37.000000, 14.000000) node {$H$};
\end{scope}
\draw (50.000000,14.000000) -- (50.000000,0.000000);
\filldraw (50.000000, 14.000000) circle(1.500000pt);
\begin{scope}
\draw[fill=white] (50.000000, 0.000000) circle(3.000000pt);
\clip (50.000000, 0.000000) circle(3.000000pt);
\draw (47.000000, 0.000000) -- (53.000000, 0.000000);
\draw (50.000000, -3.000000) -- (50.000000, 3.000000);
\end{scope}
\begin{scope}
\draw[fill=white] (63.000000, 14.000000) +(-45.000000:8.485281pt and 8.485281pt) -- +(45.000000:8.485281pt and 8.485281pt) -- +(135.000000:8.485281pt and 8.485281pt) -- +(225.000000:8.485281pt and 8.485281pt) -- cycle;
\clip (63.000000, 14.000000) +(-45.000000:8.485281pt and 8.485281pt) -- +(45.000000:8.485281pt and 8.485281pt) -- +(135.000000:8.485281pt and 8.485281pt) -- +(225.000000:8.485281pt and 8.485281pt) -- cycle;
\draw (63.000000, 14.000000) node {$H$};
\end{scope}
\end{tikzpicture}%
		\label{}%
	}\qquad
	\subfloat[]{%
		\providecommand{\ket}[1]{\left|#1\right\rangle}
\providecommand{\phase}[1]{e^{2{\pi}i\cdot#1}}
\begin{tikzpicture}[scale=1.350000,x=1pt,y=1pt]
\filldraw[color=white] (0.000000, -7.000000) rectangle (91.000000, 21.000000);
\draw[color=black] (0.000000,14.000000) -- (91.000000,14.000000);
\draw[color=black] (0.000000,0.000000) -- (91.000000,0.000000);
\draw (5.000000,14.000000) -- (5.000000,0.000000);
\begin{scope}
\draw (2.878680, 11.878680) -- (7.121320, 16.121320);
\draw (2.878680, 16.121320) -- (7.121320, 11.878680);
\end{scope}
\begin{scope}
\draw (2.878680, -2.121320) -- (7.121320, 2.121320);
\draw (2.878680, 2.121320) -- (7.121320, -2.121320);
\end{scope}
\draw[fill=white,color=white] (12.000000, -6.000000) rectangle (27.000000, 20.000000);
\draw (19.500000, 7.000000) node {$=$};
\draw (34.000000,14.000000) -- (34.000000,0.000000);
\filldraw (34.000000, 14.000000) circle(1.500000pt);
\begin{scope}
\draw[fill=white] (34.000000, 0.000000) circle(3.000000pt);
\clip (34.000000, 0.000000) circle(3.000000pt);
\draw (31.000000, 0.000000) -- (37.000000, 0.000000);
\draw (34.000000, -3.000000) -- (34.000000, 3.000000);
\end{scope}
\begin{scope}
\draw[fill=white] (47.000000, 14.000000) +(-45.000000:8.485281pt and 8.485281pt) -- +(45.000000:8.485281pt and 8.485281pt) -- +(135.000000:8.485281pt and 8.485281pt) -- +(225.000000:8.485281pt and 8.485281pt) -- cycle;
\clip (47.000000, 14.000000) +(-45.000000:8.485281pt and 8.485281pt) -- +(45.000000:8.485281pt and 8.485281pt) -- +(135.000000:8.485281pt and 8.485281pt) -- +(225.000000:8.485281pt and 8.485281pt) -- cycle;
\draw (47.000000, 14.000000) node {$H$};
\end{scope}
\begin{scope}
\draw[fill=white] (47.000000, -0.000000) +(-45.000000:8.485281pt and 8.485281pt) -- +(45.000000:8.485281pt and 8.485281pt) -- +(135.000000:8.485281pt and 8.485281pt) -- +(225.000000:8.485281pt and 8.485281pt) -- cycle;
\clip (47.000000, -0.000000) +(-45.000000:8.485281pt and 8.485281pt) -- +(45.000000:8.485281pt and 8.485281pt) -- +(135.000000:8.485281pt and 8.485281pt) -- +(225.000000:8.485281pt and 8.485281pt) -- cycle;
\draw (47.000000, -0.000000) node {$H$};
\end{scope}
\draw (60.000000,14.000000) -- (60.000000,0.000000);
\filldraw (60.000000, 14.000000) circle(1.500000pt);
\begin{scope}
\draw[fill=white] (60.000000, 0.000000) circle(3.000000pt);
\clip (60.000000, 0.000000) circle(3.000000pt);
\draw (57.000000, 0.000000) -- (63.000000, 0.000000);
\draw (60.000000, -3.000000) -- (60.000000, 3.000000);
\end{scope}
\begin{scope}
\draw[fill=white] (73.000000, 14.000000) +(-45.000000:8.485281pt and 8.485281pt) -- +(45.000000:8.485281pt and 8.485281pt) -- +(135.000000:8.485281pt and 8.485281pt) -- +(225.000000:8.485281pt and 8.485281pt) -- cycle;
\clip (73.000000, 14.000000) +(-45.000000:8.485281pt and 8.485281pt) -- +(45.000000:8.485281pt and 8.485281pt) -- +(135.000000:8.485281pt and 8.485281pt) -- +(225.000000:8.485281pt and 8.485281pt) -- cycle;
\draw (73.000000, 14.000000) node {$H$};
\end{scope}
\begin{scope}
\draw[fill=white] (73.000000, -0.000000) +(-45.000000:8.485281pt and 8.485281pt) -- +(45.000000:8.485281pt and 8.485281pt) -- +(135.000000:8.485281pt and 8.485281pt) -- +(225.000000:8.485281pt and 8.485281pt) -- cycle;
\clip (73.000000, -0.000000) +(-45.000000:8.485281pt and 8.485281pt) -- +(45.000000:8.485281pt and 8.485281pt) -- +(135.000000:8.485281pt and 8.485281pt) -- +(225.000000:8.485281pt and 8.485281pt) -- cycle;
\draw (73.000000, -0.000000) node {$H$};
\end{scope}
\draw (86.000000,14.000000) -- (86.000000,0.000000);
\filldraw (86.000000, 14.000000) circle(1.500000pt);
\begin{scope}
\draw[fill=white] (86.000000, 0.000000) circle(3.000000pt);
\clip (86.000000, 0.000000) circle(3.000000pt);
\draw (83.000000, 0.000000) -- (89.000000, 0.000000);
\draw (86.000000, -3.000000) -- (86.000000, 3.000000);
\end{scope}
\end{tikzpicture}%
		\label{}%
	}
	
	\caption{(a) The direction of a \cnot operation on the IBMQX4 can be reversed via the addition of Hadamard gates to the inputs and outputs. (b) The conjugate propagator gate expressed in terms of a \cnot gate. (c) Realisation of a \swap gate using three \cnot operations.}
	
	\label{fig:ibmqx_gates}
\end{figure*}
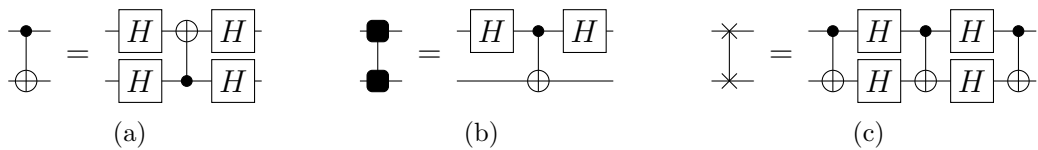

\begin{figure*}[]
	
	\subfloat[The \texttt{$[[$}4,2,2\texttt{$]]$} CPC code with a $\ket{\psi_{AB}}=\ket{+_A0_B}$ input mapped onto the IMBQX4 chip. The red conjugate propagator gates are not possible according to the connectivity map for the IBMQX4 shown in figure \ref{fig:ibmqx4}.]{
		\providecommand{\ket}[1]{\left|#1\right\rangle}
\providecommand{\phase}[1]{e^{2{\pi}i\cdot#1}}
\begin{tikzpicture}[scale=1.500000,x=1pt,y=1pt]
\filldraw[color=white] (0.000000, -9.000000) rectangle (253.000000, 63.000000);
\draw[color=black] (0.000000,54.000000) -- (253.000000,54.000000);
\draw[color=black] (0.000000,54.000000) node[left] {${A: \ \ket{+}_{Q3}}$};
\draw[color=black] (0.000000,36.000000) -- (253.000000,36.000000);
\draw[color=black] (0.000000,36.000000) node[left] {${B: \ \ket{0}_{Q0}}$};
\draw[color=black] (0.000000,18.000000) -- (240.500000,18.000000);
\draw[color=black] (240.500000,17.500000) -- (253.000000,17.500000);
\draw[color=black] (240.500000,18.500000) -- (253.000000,18.500000);
\draw[color=black] (0.000000,18.000000) node[left] {${p_1: \ \ket{0}_{Q2}}$};
\draw[color=black] (0.000000,0.000000) -- (240.500000,0.000000);
\draw[color=black] (240.500000,-0.500000) -- (253.000000,-0.500000);
\draw[color=black] (240.500000,0.500000) -- (253.000000,0.500000);
\draw[color=black] (0.000000,0.000000) node[left] {${p_2: \ \ket{0}_{Q1}}$};
\draw (22.500000,18.000000) -- (22.500000,0.000000);
\begin{scope}
\begin{scope}[shift={(22.500000,18.000000)}]
\draw[fill=black, rounded corners=2] (-3, -3) rectangle (3, 3) {};
\end{scope}
\end{scope}
\begin{scope}
\begin{scope}[shift={(22.500000,0.000000)}]
\draw[fill=black, rounded corners=2] (-3, -3) rectangle (3, 3) {};
\end{scope}
\end{scope}
\draw (41.500000,36.000000) -- (41.500000,18.000000);
\filldraw (41.500000, 36.000000) circle(1.500000pt);
\begin{scope}
\draw[fill=white] (41.500000, 18.000000) circle(3.000000pt);
\clip (41.500000, 18.000000) circle(3.000000pt);
\draw (38.500000, 18.000000) -- (44.500000, 18.000000);
\draw (41.500000, 15.000000) -- (41.500000, 21.000000);
\end{scope}
\draw (60.500000,54.000000) -- (60.500000,18.000000);
\filldraw (60.500000, 54.000000) circle(1.500000pt);
\begin{scope}
\draw[fill=white] (60.500000, 18.000000) circle(3.000000pt);
\clip (60.500000, 18.000000) circle(3.000000pt);
\draw (57.500000, 18.000000) -- (63.500000, 18.000000);
\draw (60.500000, 15.000000) -- (60.500000, 21.000000);
\end{scope}
\draw (79.500000,36.000000) -- (79.500000,0.000000);
\begin{scope}
\begin{scope}[shift={(79.500000,36.000000)}]
\draw[fill=black, rounded corners=2] (-3, -3) rectangle (3, 3) {};
\end{scope}
\end{scope}
\begin{scope}
\begin{scope}[shift={(79.500000,0.000000)}]
\draw[fill=black, rounded corners=2] (-3, -3) rectangle (3, 3) {};
\end{scope}
\end{scope}
\draw (98.500000,54.000000) -- (98.500000,0.000000);
\begin{scope}
\begin{scope}[shift={(98.500000,54.000000)}]
\draw[fill=red, rounded corners=2] (-    3, -3) rectangle (3, 3) {};
\end{scope}
\end{scope}
\begin{scope}
\begin{scope}[shift={(98.500000,0.000000)}]
\draw[fill=red, rounded corners=2] (-    3, -3) rectangle (3, 3) {};
\end{scope}
\end{scope}
\draw (120.500000,54.000000) -- (120.500000,0.000000);
\begin{scope}
\draw[fill=white] (120.500000, 27.000000) +(-45.000000:8.485281pt and 46.669048pt) -- +(45.000000:8.485281pt and 46.669048pt) -- +(135.000000:8.485281pt and 46.669048pt) -- +(225.000000:8.485281pt and 46.669048pt) -- cycle;
\clip (120.500000, 27.000000) +(-45.000000:8.485281pt and 46.669048pt) -- +(45.000000:8.485281pt and 46.669048pt) -- +(135.000000:8.485281pt and 46.669048pt) -- +(225.000000:8.485281pt and 46.669048pt) -- cycle;
\draw (120.500000, 27.000000) node {$E$};
\end{scope}
\draw (142.500000,54.000000) -- (142.500000,0.000000);
\begin{scope}
\begin{scope}[shift={(142.500000,54.000000)}]
\draw[fill=red, rounded corners=2] (-    3, -3) rectangle (3, 3) {};
\end{scope}
\end{scope}
\begin{scope}
\begin{scope}[shift={(142.500000,0.000000)}]
\draw[fill=red, rounded corners=2] (-    3, -3) rectangle (3, 3) {};
\end{scope}
\end{scope}
\draw (161.500000,36.000000) -- (161.500000,0.000000);
\begin{scope}
\begin{scope}[shift={(161.500000,36.000000)}]
\draw[fill=black, rounded corners=2] (-3, -3) rectangle (3, 3) {};
\end{scope}
\end{scope}
\begin{scope}
\begin{scope}[shift={(161.500000,0.000000)}]
\draw[fill=black, rounded corners=2] (-3, -3) rectangle (3, 3) {};
\end{scope}
\end{scope}
\draw (180.500000,54.000000) -- (180.500000,18.000000);
\filldraw (180.500000, 54.000000) circle(1.500000pt);
\begin{scope}
\draw[fill=white] (180.500000, 18.000000) circle(3.000000pt);
\clip (180.500000, 18.000000) circle(3.000000pt);
\draw (177.500000, 18.000000) -- (183.500000, 18.000000);
\draw (180.500000, 15.000000) -- (180.500000, 21.000000);
\end{scope}
\draw (199.500000,36.000000) -- (199.500000,18.000000);
\filldraw (199.500000, 36.000000) circle(1.500000pt);
\begin{scope}
\draw[fill=white] (199.500000, 18.000000) circle(3.000000pt);
\clip (199.500000, 18.000000) circle(3.000000pt);
\draw (196.500000, 18.000000) -- (202.500000, 18.000000);
\draw (199.500000, 15.000000) -- (199.500000, 21.000000);
\end{scope}
\draw (218.500000,18.000000) -- (218.500000,0.000000);
\begin{scope}
\begin{scope}[shift={(218.500000,18.000000)}]
\draw[fill=black, rounded corners=2] (-3, -3) rectangle (3, 3) {};
\end{scope}
\end{scope}
\begin{scope}
\begin{scope}[shift={(218.500000,0.000000)}]
\draw[fill=black, rounded corners=2] (-3, -3) rectangle (3, 3) {};
\end{scope}
\end{scope}
\draw[fill=white] (234.500000, 12.000000) rectangle (246.500000, 24.000000);
\draw[very thin] (240.500000, 18.600000) arc (90:150:6.000000pt);
\draw[very thin] (240.500000, 18.600000) arc (90:30:6.000000pt);
\draw[->,>=stealth] (240.500000, 12.600000) -- +(80:10.392305pt);
\draw[fill=white] (234.500000, -6.000000) rectangle (246.500000, 6.000000);
\draw[very thin] (240.500000, 0.600000) arc (90:150:6.000000pt);
\draw[very thin] (240.500000, 0.600000) arc (90:30:6.000000pt);
\draw[->,>=stealth] (240.500000, -5.400000) -- +(80:10.392305pt);
\end{tikzpicture}%
		\label{}%
	}

	\subfloat[A modified version of the \texttt{$[[$}4,2,2\texttt{$]]$} circuit in which the order of gates in the encoder and decoder has be rearranged. In this new form, the circuit can be simplified by noting that the action of the gates marked in green is the identity.]{
		\providecommand{\ket}[1]{\left|#1\right\rangle}
\providecommand{\phase}[1]{e^{2{\pi}i\cdot#1}}
\begin{tikzpicture}[scale=1.500000,x=1pt,y=1pt]
\filldraw[color=white] (0.000000, -9.000000) rectangle (253.000000, 63.000000);
\draw[color=black] (0.000000,54.000000) -- (253.000000,54.000000);
\draw[color=black] (0.000000,54.000000) node[left] {${A: \ \ket{+}_{Q3}}$};
\draw[color=black] (0.000000,36.000000) -- (253.000000,36.000000);
\draw[color=black] (0.000000,36.000000) node[left] {${B: \ \ket{0}_{Q0}}$};
\draw[color=black] (0.000000,18.000000) -- (240.500000,18.000000);
\draw[color=black] (240.500000,17.500000) -- (253.000000,17.500000);
\draw[color=black] (240.500000,18.500000) -- (253.000000,18.500000);
\draw[color=black] (0.000000,18.000000) node[left] {${p_1: \ \ket{0}_{Q2}}$};
\draw[color=black] (0.000000,0.000000) -- (240.500000,0.000000);
\draw[color=black] (240.500000,-0.500000) -- (253.000000,-0.500000);
\draw[color=black] (240.500000,0.500000) -- (253.000000,0.500000);
\draw[color=black] (0.000000,0.000000) node[left] {${p_2: \ \ket{0}_{Q1}}$};
\draw (22.500000,18.000000) -- (22.500000,0.000000);
\begin{scope}
\begin{scope}[shift={(22.500000,18.000000)}]
\draw[fill=black, rounded corners=2] (-3, -3) rectangle (3, 3) {};
\end{scope}
\end{scope}
\begin{scope}
\begin{scope}[shift={(22.500000,0.000000)}]
\draw[fill=black, rounded corners=2] (-3, -3) rectangle (3, 3) {};
\end{scope}
\end{scope}
\draw (41.500000,54.000000) -- (41.500000,0.000000);
\begin{scope}
\begin{scope}[shift={(41.500000,54.000000)}]
\draw[fill=green, rounded corners=2] (    -    3, -3) rectangle (3, 3) {};
\end{scope}
\end{scope}
\begin{scope}
\begin{scope}[shift={(41.500000,0.000000)}]
\draw[fill=green, rounded corners=2] (    -    3, -3) rectangle (3, 3) {};
\end{scope}
\end{scope}
\begin{scope}[color=green]
\draw (60.500000,36.000000) -- (60.500000,18.000000);
\filldraw (60.500000, 36.000000) circle(1.500000pt);
\begin{scope}
\draw[fill=white] (60.500000, 18.000000) circle(3.000000pt);
\clip (60.500000, 18.000000) circle(3.000000pt);
\draw (57.500000, 18.000000) -- (63.500000, 18.000000);
\draw (60.500000, 15.000000) -- (60.500000, 21.000000);
\end{scope}
\end{scope}
\draw (79.500000,36.000000) -- (79.500000,0.000000);
\begin{scope}
\begin{scope}[shift={(79.500000,36.000000)}]
\draw[fill=black, rounded corners=2] (-3, -3) rectangle (3, 3) {};
\end{scope}
\end{scope}
\begin{scope}
\begin{scope}[shift={(79.500000,0.000000)}]
\draw[fill=black, rounded corners=2] (-3, -3) rectangle (3, 3) {};
\end{scope}
\end{scope}
\draw (98.500000,54.000000) -- (98.500000,18.000000);
\filldraw (98.500000, 54.000000) circle(1.500000pt);
\begin{scope}
\draw[fill=white] (98.500000, 18.000000) circle(3.000000pt);
\clip (98.500000, 18.000000) circle(3.000000pt);
\draw (95.500000, 18.000000) -- (101.500000, 18.000000);
\draw (98.500000, 15.000000) -- (98.500000, 21.000000);
\end{scope}
\draw (120.500000,54.000000) -- (120.500000,0.000000);
\begin{scope}
\draw[fill=white] (120.500000, 27.000000) +(-45.000000:8.485281pt and 46.669048pt) -- +(45.000000:8.485281pt and 46.669048pt) -- +(135.000000:8.485281pt and 46.669048pt) -- +(225.000000:8.485281pt and 46.669048pt) -- cycle;
\clip (120.500000, 27.000000) +(-45.000000:8.485281pt and 46.669048pt) -- +(45.000000:8.485281pt and 46.669048pt) -- +(135.000000:8.485281pt and 46.669048pt) -- +(225.000000:8.485281pt and 46.669048pt) -- cycle;
\draw (120.500000, 27.000000) node {$E$};
\end{scope}
\draw (142.500000,54.000000) -- (142.500000,18.000000);
\filldraw (142.500000, 54.000000) circle(1.500000pt);
\begin{scope}
\draw[fill=white] (142.500000, 18.000000) circle(3.000000pt);
\clip (142.500000, 18.000000) circle(3.000000pt);
\draw (139.500000, 18.000000) -- (145.500000, 18.000000);
\draw (142.500000, 15.000000) -- (142.500000, 21.000000);
\end{scope}
\draw (161.500000,36.000000) -- (161.500000,0.000000);
\begin{scope}
\begin{scope}[shift={(161.500000,36.000000)}]
\draw[fill=black, rounded corners=2] (-3, -3) rectangle (3, 3) {};
\end{scope}
\end{scope}
\begin{scope}
\begin{scope}[shift={(161.500000,0.000000)}]
\draw[fill=black, rounded corners=2] (-3, -3) rectangle (3, 3) {};
\end{scope}
\end{scope}
\draw (180.500000,36.000000) -- (180.500000,18.000000);
\filldraw (180.500000, 36.000000) circle(1.500000pt);
\begin{scope}
\draw[fill=white] (180.500000, 18.000000) circle(3.000000pt);
\clip (180.500000, 18.000000) circle(3.000000pt);
\draw (177.500000, 18.000000) -- (183.500000, 18.000000);
\draw (180.500000, 15.000000) -- (180.500000, 21.000000);
\end{scope}
\draw (199.500000,54.000000) -- (199.500000,0.000000);
\begin{scope}
\begin{scope}[shift={(199.500000,54.000000)}]
\draw[fill=red, rounded corners=2] (-    3, -3) rectangle (3, 3) {};
\end{scope}
\end{scope}
\begin{scope}
\begin{scope}[shift={(199.500000,0.000000)}]
\draw[fill=red, rounded corners=2] (-    3, -3) rectangle (3, 3) {};
\end{scope}
\end{scope}
\draw (218.500000,18.000000) -- (218.500000,0.000000);
\begin{scope}
\begin{scope}[shift={(218.500000,18.000000)}]
\draw[fill=black, rounded corners=2] (-3, -3) rectangle (3, 3) {};
\end{scope}
\end{scope}
\begin{scope}
\begin{scope}[shift={(218.500000,0.000000)}]
\draw[fill=black, rounded corners=2] (-3, -3) rectangle (3, 3) {};
\end{scope}
\end{scope}
\draw[fill=white] (234.500000, 12.000000) rectangle (246.500000, 24.000000);
\draw[very thin] (240.500000, 18.600000) arc (90:150:6.000000pt);
\draw[very thin] (240.500000, 18.600000) arc (90:30:6.000000pt);
\draw[->,>=stealth] (240.500000, 12.600000) -- +(80:10.392305pt);
\draw[fill=white] (234.500000, -6.000000) rectangle (246.500000, 6.000000);
\draw[very thin] (240.500000, 0.600000) arc (90:150:6.000000pt);
\draw[very thin] (240.500000, 0.600000) arc (90:30:6.000000pt);
\draw[->,>=stealth] (240.500000, -5.400000) -- +(80:10.392305pt);
\end{tikzpicture}%
		\label{}%
	}

	\subfloat[A \swap gate can be added to the \texttt{$[[$}4,2,2\texttt{$]]$} circuit to exchange the $p_1$ and $p_2$ parity qubits. This allows the `illegal' operation marked in red in the decoder to be performed via a nearest-neighbour interaction.]{
		\providecommand{\ket}[1]{\left|#1\right\rangle}
\providecommand{\phase}[1]{e^{2{\pi}i\cdot#1}}
\begin{tikzpicture}[scale=1.400000,x=1pt,y=1pt]
\filldraw[color=white] (0.000000, -9.000000) rectangle (270.000000, 63.000000);
\draw[color=black] (0.000000,54.000000) -- (270.000000,54.000000);
\draw[color=black] (0.000000,54.000000) node[left] {${A: \ \ket{+}_{Q3}}$};
\draw[color=black] (0.000000,36.000000) -- (270.000000,36.000000);
\draw[color=black] (0.000000,36.000000) node[left] {${B: \ \ket{0}_{Q0}}$};
\draw[color=black] (0.000000,18.000000) -- (256.000000,18.000000);
\draw[color=black] (256.000000,17.500000) -- (270.000000,17.500000);
\draw[color=black] (256.000000,18.500000) -- (270.000000,18.500000);
\draw[color=black] (0.000000,18.000000) node[left] {${p_1: \ \ket{0}_{Q2}}$};
\draw[color=black] (0.000000,0.000000) -- (256.000000,0.000000);
\draw[color=black] (256.000000,-0.500000) -- (270.000000,-0.500000);
\draw[color=black] (256.000000,0.500000) -- (270.000000,0.500000);
\draw[color=black] (0.000000,0.000000) node[left] {${p_2: \ \ket{0}_{Q1}}$};
\draw (27.000000,18.000000) -- (27.000000,0.000000);
\begin{scope}
\begin{scope}[shift={(27.000000,18.000000)}]
\draw[fill=black, rounded corners=2] (-3, -3) rectangle (3, 3) {};
\end{scope}
\end{scope}
\begin{scope}
\begin{scope}[shift={(27.000000,0.000000)}]
\draw[fill=black, rounded corners=2] (-3, -3) rectangle (3, 3) {};
\end{scope}
\end{scope}
\draw (49.000000,36.000000) -- (49.000000,0.000000);
\begin{scope}
\begin{scope}[shift={(49.000000,36.000000)}]
\draw[fill=black, rounded corners=2] (-3, -3) rectangle (3, 3) {};
\end{scope}
\end{scope}
\begin{scope}
\begin{scope}[shift={(49.000000,0.000000)}]
\draw[fill=black, rounded corners=2] (-3, -3) rectangle (3, 3) {};
\end{scope}
\end{scope}
\draw (71.000000,54.000000) -- (71.000000,18.000000);
\filldraw (71.000000, 54.000000) circle(1.500000pt);
\begin{scope}
\draw[fill=white] (71.000000, 18.000000) circle(3.000000pt);
\clip (71.000000, 18.000000) circle(3.000000pt);
\draw (68.000000, 18.000000) -- (74.000000, 18.000000);
\draw (71.000000, 15.000000) -- (71.000000, 21.000000);
\end{scope}
\draw (96.000000,54.000000) -- (96.000000,0.000000);
\begin{scope}
\draw[fill=white] (96.000000, 27.000000) +(-45.000000:8.485281pt and 46.669048pt) -- +(45.000000:8.485281pt and 46.669048pt) -- +(135.000000:8.485281pt and 46.669048pt) -- +(225.000000:8.485281pt and 46.669048pt) -- cycle;
\clip (96.000000, 27.000000) +(-45.000000:8.485281pt and 46.669048pt) -- +(45.000000:8.485281pt and 46.669048pt) -- +(135.000000:8.485281pt and 46.669048pt) -- +(225.000000:8.485281pt and 46.669048pt) -- cycle;
\draw (96.000000, 27.000000) node {$E$};
\end{scope}
\draw (121.000000,54.000000) -- (121.000000,18.000000);
\filldraw (121.000000, 54.000000) circle(1.500000pt);
\begin{scope}
\draw[fill=white] (121.000000, 18.000000) circle(3.000000pt);
\clip (121.000000, 18.000000) circle(3.000000pt);
\draw (118.000000, 18.000000) -- (124.000000, 18.000000);
\draw (121.000000, 15.000000) -- (121.000000, 21.000000);
\end{scope}
\draw (143.000000,36.000000) -- (143.000000,0.000000);
\begin{scope}
\begin{scope}[shift={(143.000000,36.000000)}]
\draw[fill=black, rounded corners=2] (-3, -3) rectangle (3, 3) {};
\end{scope}
\end{scope}
\begin{scope}
\begin{scope}[shift={(143.000000,0.000000)}]
\draw[fill=black, rounded corners=2] (-3, -3) rectangle (3, 3) {};
\end{scope}
\end{scope}
\draw (165.000000,36.000000) -- (165.000000,18.000000);
\filldraw (165.000000, 36.000000) circle(1.500000pt);
\begin{scope}
\draw[fill=white] (165.000000, 18.000000) circle(3.000000pt);
\clip (165.000000, 18.000000) circle(3.000000pt);
\draw (162.000000, 18.000000) -- (168.000000, 18.000000);
\draw (165.000000, 15.000000) -- (165.000000, 21.000000);
\end{scope}
\draw (187.000000,18.000000) -- (187.000000,0.000000);
\begin{scope}
\draw (184.878680, 15.878680) -- (189.121320, 20.121320);
\draw (184.878680, 20.121320) -- (189.121320, 15.878680);
\end{scope}
\begin{scope}
\draw (184.878680, -2.121320) -- (189.121320, 2.121320);
\draw (184.878680, 2.121320) -- (189.121320, -2.121320);
\end{scope}
\draw (209.000000,54.000000) -- (209.000000,18.000000);
\begin{scope}
\begin{scope}[shift={(209.000000,54.000000)}]
\draw[fill=red, rounded corners=2] (-    3, -3) rectangle (3, 3) {};
\end{scope}
\end{scope}
\begin{scope}
\begin{scope}[shift={(209.000000,18.000000)}]
\draw[fill=red, rounded corners=2] (-    3, -3) rectangle (3, 3) {};
\end{scope}
\end{scope}
\draw (231.000000,18.000000) -- (231.000000,0.000000);
\begin{scope}
\begin{scope}[shift={(231.000000,18.000000)}]
\draw[fill=black, rounded corners=2] (-3, -3) rectangle (3, 3) {};
\end{scope}
\end{scope}
\begin{scope}
\begin{scope}[shift={(231.000000,0.000000)}]
\draw[fill=black, rounded corners=2] (-3, -3) rectangle (3, 3) {};
\end{scope}
\end{scope}
\draw[fill=white] (250.000000, 12.000000) rectangle (262.000000, 24.000000);
\draw[very thin] (256.000000, 18.600000) arc (90:150:6.000000pt);
\draw[very thin] (256.000000, 18.600000) arc (90:30:6.000000pt);
\draw[->,>=stealth] (256.000000, 12.600000) -- +(80:10.392305pt);
\draw[fill=white] (250.000000, -6.000000) rectangle (262.000000, 6.000000);
\draw[very thin] (256.000000, 0.600000) arc (90:150:6.000000pt);
\draw[very thin] (256.000000, 0.600000) arc (90:30:6.000000pt);
\draw[->,>=stealth] (256.000000, -5.400000) -- +(80:10.392305pt);
\end{tikzpicture}
		\label{}
	}

	\subfloat[When running the \texttt{$[[$}4,2,2\texttt{$]]$} code on a real device, it can no longer be assumed that errors occur only in the wait-stage between the encoder and decoder. For the \texttt{$[[$}4,2,2\texttt{$]]$} circuit in question, a single-qubit fault $E$ after any gate will not propagate a multi-qubit error to the register without triggering a syndrome.]{%
		\input{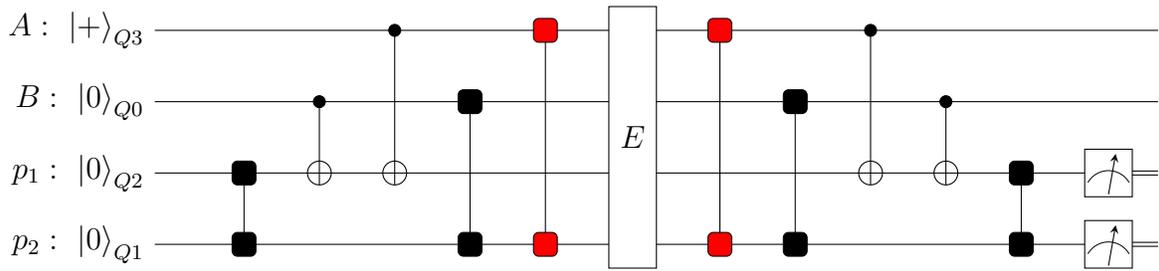}
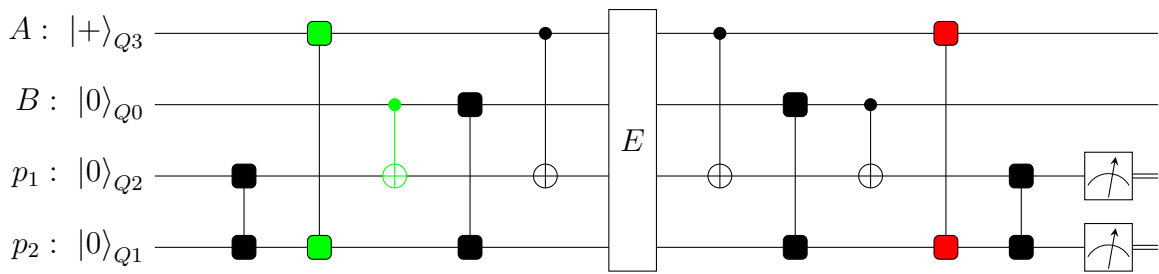
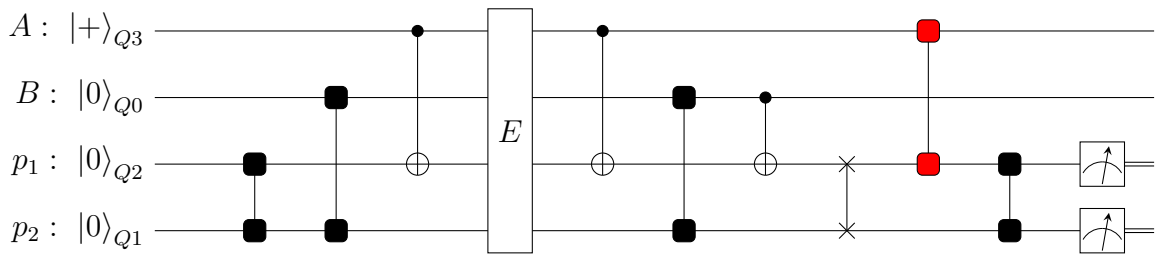
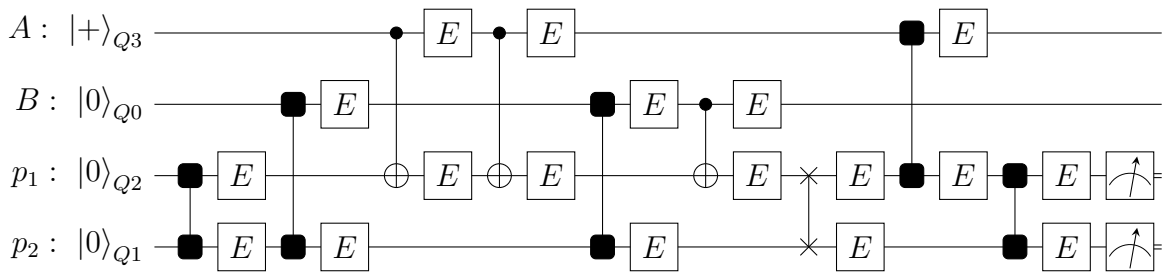%
		\label{}%
	}

	\caption{Steps for compiling the \texttt{$[[$}4,2,2\texttt{$]]$} CPC quantum memory onto the IBMQX4 chip.}
	
	\label{fig:ibmqx4_circuits}
\end{figure*} 

\subsection{Compiling a $[[4,2,2]]$ CPC circuit onto the IBM 5Q}

Our experiment is run on the IBMQX4 version of the IBM 5Q, the technical details for which can be found in \cite{ibmqx4}. Figure \ref{fig:ibmqx4} depicts the `bow tie' layout of the chip. The arrows represent the allowed \cnot operations between qubits. The direction of the arrow indicates the preferred \cnot direction, but the operation can be reversed via the circuit transformation shown in figure \ref{fig:ibmqx_gates}a.

The $[[4,2,2]]$ code, as depicted in figure \ref{fig:422_circ2}, has two data qubits $\{A,B\}$ and two parity qubits $\{p_1,p_2\}$. In this experiment, the code qubits are mapped onto the physical qubits of the IBMQX4 device as follows: $\{A\rightarrow Q_3, \ B\rightarrow Q_0, \ p_1 \rightarrow Q_2, \ p_2 \rightarrow Q_1 \}$. The input state becomes $\ket{\psi_{AB}}=\ket{+}_{Q_3}\otimes \ket{0}_{Q_0}$, and the resultant circuit is shown in figure \ref{fig:ibmqx4_circuits}a. The two conjugate propagator gates marked in red are not possible on the IBMQX4, as there is no connectivity between qubits $Q_3$ and $Q_1$ (see figure \ref{fig:ibmqx4}). The $[[4,2,2]]$ CPC circuit must therefore be modified to accommodate this hardware constraint.

The first step in compiling the $[[4,2,2]]$ circuit for the IBMQX4 is to rearrange the gates into the order shown in figure \ref{fig:ibmqx4_circuits}b. This is a departure from the canonical form of CPC codes outlined in section \ref{sec:can_form}. However, it can easily be checked that the modified circuit remains a functional $[[4,2,2]]$ CPC code capable of detecting single $X$- and $Z$-errors on any of the qubits during the wait-stage.

In the rearranged form of the circuit in figure \ref{fig:ibmqx4_circuits}b, and when the input state is $\ket{\psi_{AB}}=\ket{+}_{Q_3}\otimes \ket{0}_{Q_0}$, it can be seen that the action of the gates highlighted in green is the identity. The green gates can therefore be omitted from the circuit without affecting the function of the quantum memory. Following this simplification, the only operation that remains prohibited by the IMBQX4's connectivity constraints is the red conjugate propagator gate between $Q_1$ and $Q_3$ in the decoder. One way of resolving this problem is to perform a \swap operation between $Q_2$ and $Q_1$, as shown in figure \ref{fig:ibmqx4_circuits}c. The \swap gate exchanges the positions of the $p_1$ and $p_2$ parity check qubits, enabling the red conjugate propagator gate to be performed via a nearest-neighbour interaction. A \swap gate is achieved via the application of three \cnot gates (see figure \ref{fig:ibmqx_gates}c), and is therefore an expensive operation that should be used sparingly. In section \ref{sec:cpc_code_design}, we explore how the CPC code design process can be used to minimise the \swap gate count when compiling larger codes onto quantum hardware.       

\subsection{A note on fault tolerance for the $[[4,2,2]]$ circuit} \label{sec:fault_tolerance}

So far, we have considered a simplified model of CPC code operation in which it is assumed errors only occur during the wait-stage between the encoder and the decoder. However, we have observed that the error rates for \cnot operations and readout on the IBMQX4 are of the order $10^{-2}$ (daily calibration data can be obtained from the IBM Q website \cite{IBM}). This realistically means that any quantum code must be designed to detect errors that occur at any point in the circuit. To this end, figure \ref{fig:ibmqx4_circuits}d shows the IMBQX4-compiled $[[4,2,2]]$ circuit under a more general error model.

Fault tolerant circuit construction ordinarily necessitates the introduction of additional qubits \cite{ft_shor,ft_steane,flag}. However, in this particular instance of the $[[4,2,2]]$ CPC code with a known $\ket{+}_{Q_3}\otimes\ket{0}_{Q_0}$ input, it can be verified that a single fault at any of the locations marked on figure \ref{fig:ibmqx4_circuits}d will not propagate a multi-qubit error to the register without triggering a syndrome. The circuit can therefore be considered to have been hardened against single-qubit errors in the encode and decode stages of the circuit. It should be noted, however, that this does not extend the circuit to full fault tolerance when implemented on the IBMQX4 chip. State preparation and measurement (SPAM) errors are not accounted for, nor is the $[[4,2,2]]$ code capable of detecting correlated two-qubit errors that might occur after a \cnot gate. Another issue is that the circuit allows certain single-qubit errors to propagate to the register in an undetectable way. It is not currently possible to measure, then reset a qubit on the IBMQX4 via the public API. As a result, our implementation is restricted to a single encode-decode cycle, meaning the undetected single-qubit errors will reduce the output fidelity. \joschkaEdit{However, as outlined in \cite{cpc1}, CPC codes can be expressed in terms of stabilizer codes. Adopting this approach allows CPC codes to be implemented using existing syndrome extraction techniques, and enables errors to be decoded over multiple cycles}. Assuming access to hardware that allows qubit reset, CPC codes implemented in this way would be tolerant of the single-qubit errors that propagate to the register.

\subsection{Experimental data reconstruction methods}

The IBM Quantum Information Software Kit (QISKIT) \cite{qiskit} was used to prepare the $[[4,2,2]]$ experiment for quantum state tomography on the output qubits $Q_0$ and $Q_3$. QISKIT quantum tomography tools were used to create a set of nine circuits from the original $[[4,2,2]]$ circuit (depicted in figure \ref{fig:ibmqx4_circuits}c), each of which was designed to measure the output qubits $Q_0$ and $Q_3$ in a different measurement basis from the list $\{XX,\ XY,\ XZ,\ YX,\ YY,\ YZ,\ ZX,\ ZY,\ ZZ\}$. These quantum tomography circuits were then run multiple times to create a distribution of results that could be used reconstruct an approximation to the density matrix of the output state. The QISKIT method used for state reconstruction from the experimental data was the fast maximum likelihood method for quantum tomography, a description of which can be found in \cite{Smolin2012}.

The quantum tomography circuits for the $[[4,2,2]]$ memory were run in batches of $8192$ shots. After each batch, the QISKIT maximum likelihood method was used to reconstruct the density matrix $\rho_{dd}$ of the directly decoded output before post-selection. The syndrome qubits were then inspected to determine which of the shots in the batch should be discarded during post-selection. State reconstruction was then performed again on the reduced set to obtain a post-selected density matrix $\rho_{ps}$. The quality of the directly decoded and post-selected output state for each batch was quantified by calculating the fidelity, $F(\rho)=\big(\text{Tr}\left[\sqrt{\rho^{1/2} \sigma \rho^{1/2}}\ \right]\big)^2$, where $\sigma$ is the target density matrix. For the chosen input state $\ket{\psi_{AB}}=\ket{+}_{Q_3}\otimes \ket{0}_{Q_0}$, the target density matrix is given by
\begin{equation} \label{eq:tomo_target}
\sigma=\left(
\begin{array}{cccc}
\frac{1}{2} & 0 & \frac{1}{2} & 0 \\
0 & 0 & 0 & 0 \\
\frac{1}{2} & 0 & \frac{1}{2} & 0 \\
0 & 0 & 0 & 0 \\
\end{array}
\right)\rm.
\end{equation}
The purity of the density matrices, defined by $P(\rho)=\text{Tr}\left[ \rho^2 \right]$, was also calculated to provide a coherence measure for the output states. 

\subsection{Experimental results}

\begin{figure}
	\centering
	\input{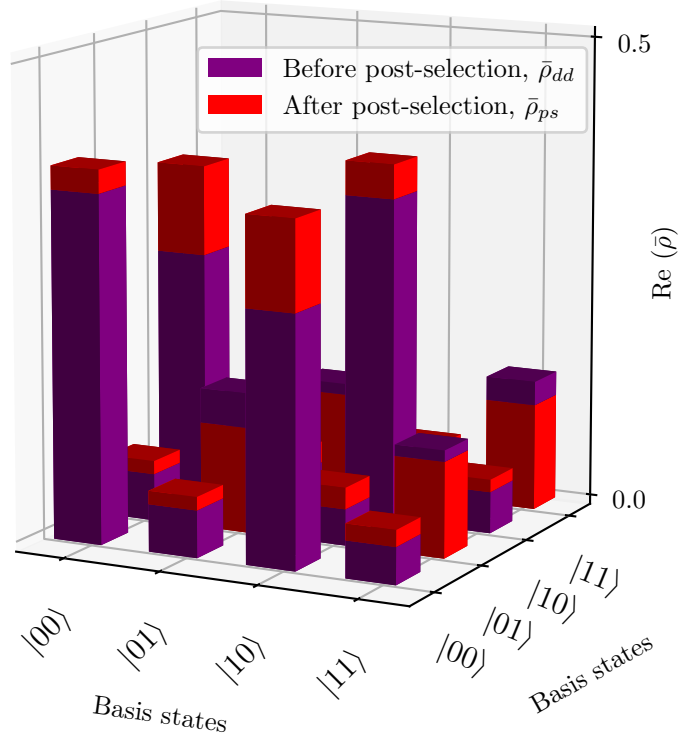}
	\caption{Plot of the real components of the density matrices $\bar{\rho}_{dd}$ and $\bar{\rho}_{ps}$ corresponding to the experimental output state before and after post-selection.}
	\label{fig:tomo_city}
\end{figure}

\begin{figure}
	\centering
	\input{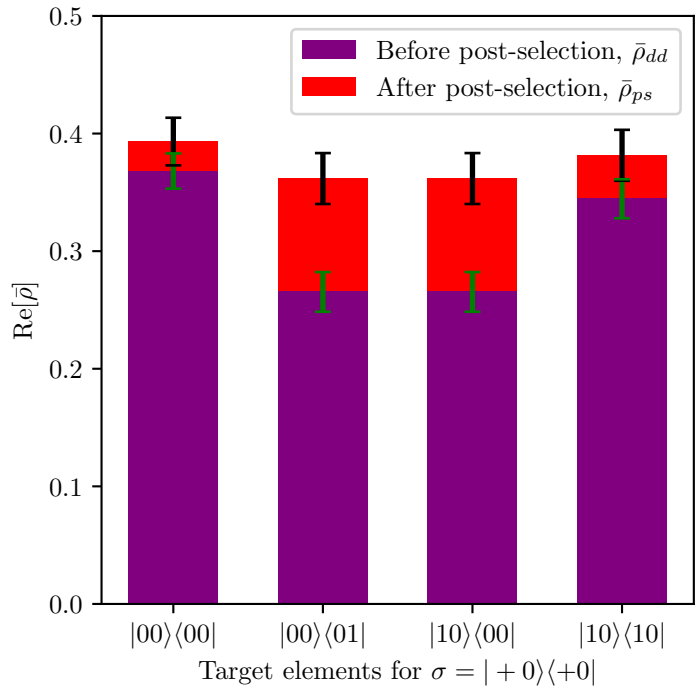}
	\caption{Plot of the target elements for $\rho_{dd}$ and $\rho_{ps}$. The target elements are the non-zero elements in the density matrix $\sigma$ given in equation (\ref{eq:tomo_target}).}
	\label{fig:tomo_target}
\end{figure}

The $[[4,2,2]]$ CPC quantum memory circuit, depicted in figure \ref{fig:ibmqx4_circuits}c, was run on the IBMQX4 device between the 25\textsuperscript{th} and 27\textsuperscript{th} November 2017. A summary of the experimental results for state purity, fidelity and yield can be found in table \ref{tab:tomo_results}. Error bars were calculated as one standard deviation of a single run value consisting of 8192 experimental executions of the quantum tomography circuit set. The standard error of the mean over all $154$ runs was too small to be visible on our plots. Calibration data for the device on each of the three days of the experiment can be found in appendix \ref{sec:ibm_calib}.

A total of $154$ batches of $8192$ shots were run over the course of the experiment. Figure \ref{fig:422_circ1} shows a plot of the real components of the elements of $\bar{\rho}_{dd}$ and $\bar{\rho}_{ps}$ averaged across the $154$ batches. It is immediately clear that the post-selected density matrix $\bar{\rho}_{ps}$ better preserves the four target elements, which we identify as the non-zero elements in the target state $\sigma$ given by equation (\ref{eq:tomo_target}). The bar-chart in figure \ref{fig:tomo_target} shows these target elements in isolation, from which it is apparent that post-selection has the biggest impact in preserving the strength of the off-diagonal coherences. This can also be seen when comparing the purity values, shown in table \ref{tab:tomo_results}, for $\bar{\rho}_{dd}$ and $\bar{\rho}_{ps}$. The directly-decoded density matrix $\bar{\rho}_{dd}$ has a purity of $P(\bar{\rho}_{dd})=0.52\pm0.02$, implying it represents a near-fully mixed classical ensemble with a purity of $0.5$. In contrast, the post-selected density matrix $\bar{\rho}_{ps}$ has a purity of $P(\bar{\rho}_{ps})=0.74\pm 0.03$, suggesting it has undergone only partial decoherence.

The fidelities of $\bar{\rho}_{dd}$ and $\bar{\rho}_{ps}$ relative to the target state are $F(\bar{\rho}_{dd})=0.62\pm0.03$ and $F(\bar{\rho}_{ps})=0.75\pm 0.04$ respectively. The fidelity of the post-selected state is therefore greater than the directly-decoded state with a confidence level of three standard deviations. From this we can conclude that the $[[4,2,2]]$ quantum memory produces useful syndrome information for protecting a $\ket{\psi_{AB}}=\ket{+_A0_B}$ state. A consideration, however, is that the average yield (the proportion of results retained after post-selection) was $(54\pm2)\%$ averaged over the $154$ batches.
 
\begin{table}[]
	\centering
	
	\begin{tabular}{|l|l|l|l|}
		\hline
	 & \textbf{Purity, } $P(\rho)$ & \textbf{Fidelity}, $F(\rho)$     &\textbf{Yield}  \\
	 \hline
	\textbf{Before post-selection}, $\bar{\rho}_{dd}$  & $0.52\pm 0.02$& $ 0.62 \pm0.03 $    &$100\%$  \\
	\textbf{After post-selection}, $\bar{\rho}_{ps}$  & $0.74\pm 0.03$& $ 0.75 \pm0.04 $    &$(54\pm2)\%$ \\ \hline

		\multicolumn{4}{|l|}{no. runs: $154$ batches of $8192$ shots}                       \\ \hline

	\end{tabular}

\caption{Quality metrics for the reconstructed density matrices before and after post-selection. The fidelity is calculated relative the target density matrix $\sigma$ which is defined in equation (\ref{eq:tomo_target}). The yield is the proportion of shots per batch that are retained during the post-selection process. The errors are calculated as one standard deviation of a single run value consisting of 8192 experimental shots.}
\label{tab:tomo_results}
\end{table}

\subsection{Summary of IBM 5Q experiment}

The results of our experiment with the IBMQX4 device show that the syndrome information produced by a $[[4,2,2]]$ CPC quantum memory can be used to improve the fidelity and purity of the code output. The $[[4,2,2]]$ CPC code is one the simplest quantum memories, and as such, it was possible to compile the circuit for the IBMQX4 device by inspection. In the following sections, we outline a CPC design process that provides automated methods for compiling and optimising more complex CPC codes onto quantum hardware. As an example, we demonstrate the utility of the CPC design process in the compilation of a custom quantum memory for an idealised seven-qubit ion trap device. 
 
\section{Overview of ion trap hardware for quantum computing} \label{sec:ion_trap_intro}

Ion traps are considered one the leading platforms for quantum computation. Ion-based qubits have long coherence times, and can be read out with near $100\%$ efficiency \cite{near_perfect_measurement}. It has also been proposed that multiple ion-trap cells could be networked via auxillary qubit systems to create larger hybrid quantum computers \cite{nqit_nickerson}. In such a hybrid networked architecture, good QEC codes will be vital to ensure the quantum data in each ion trap is protected.

In this paper, we provide an illustrative example of how the CPC design process can be used to create a bespoke QEC code for a specific ion trap device. We consider a linear ion-trap with seven application qubits. This scheme has been chosen because several existing ion trap experiments have a similar size and layout \cite{Ballance2016, sussex_ions, Debnath2016, Brandl2016}.

We assume that arbitrary single-qubit operations can be performed on any of the ions in the register. It is in principle possible to implement interactions between spatially separated qubits, for example, by exploiting the collective vibrational modes of the ions as a quantum bus \cite{Cirac1995}. In practice, however, the fidelity of two-qubit interactions decreases with separation \cite{Tan2015}. For this reason, in our idealised model ion trap, two-qubit gates are limited to nearest-neighbour interactions.

Under nearest-neighbour constraints, interactions between spatially separated qubits are achieved by performing \swap operations to move quantum information around the trap. These \swap operations can be realised either by physically shuttling qubits between zones of the trap \cite{Monroe2014}, or by synthesising \swap gates from \cnot interactions \cite{Tan2015}. In the CPC design process, we show how CPC codes can be compiled with \swap gates to allow for implementation with only nearest neighbour interactions. 

We assume that our idealised ion trap has a two-qubit entangling gate that gives rise to a unitary of the form
\begin{equation} \label{eq:entangling_unitary}
U=\exp{\left(-\text{i} \dfrac{\pi}{2}\left[Z\otimes Z\right]t\right)}=e^{\text{i}\pi t/2}\left(\begin{matrix}
1&0&0&0\\
0&e^{-\text{i}\pi t}&0&0\\
0&0&e^{-\text{i}\pi t}&0\\
0&0&0&1
\end{matrix}\right)\rm.
\end{equation}
where $t$ is a tuning parameter. Such interactions can be realised via geometric phase gate procedures \cite{Cirac1995,Leibfried2003,Srensen1999}. In this paper, we consider the symmetrised phase (SP) gate, which is one of the simplest possible maximally entangling gates that arises from the above ion trap unitary \cite{ballance}. \joschkaEdit{The SP native gate is realised by setting the tuning parameter in equation (\ref{eq:entangling_unitary}) to $t_{\text{SP}}=1/2$.} Up to a global phase, the gate can then be described as a matrix, $F$, of the form
\begin{equation}\label{eq:sp_native_gate}
F_{q_2}^{q_1}=
\left(
\begin{matrix}
1	&0			&0			&0\\
0	&\text{i}	&0			&0\\
0	&0			&\text{i}	&0\\
0	&0			&0			&1
\end{matrix}
\right)\rm,
\end{equation}
where $q_1$ and $q_2$ are the input qubits to the gate. In section \ref{sec:any_native_gate}, we explicitly show how a $[[4,2,2]]$ detection code can be efficiently compiled with the SP native gate of equation (\ref{eq:sp_native_gate}). Building on this example, we then demonstrate how efficient compilation is in principle possible for any experimentally realistic maximally entangling native gate.   

\section{Compiling CPC codes with any realistic maximally entangling Clifford gate}\label{sec:any_native_gate}

In our discussion of the CPC framework so far, quantum codes have been expressed in terms of \cnot and conjugate-propagator gates. This allows for intuitive  visualisation of the propagation of errors through the decoder, and simplifies the calculation of syndrome tables via the techniques described in sections \ref{sec:prop_rules} and \ref{sec:cpg_prop}. However, in practice, the native two-qubit entangling interaction of a given experiment will be of a different form. As a result, when compiling a CPC code, additional operations are required to allow \cnot and conjugate-propagator gates to be synthesised from the native interaction. If the native interaction is maximally entangling, this will involve the addition of single-qubit corrections. In this section, we show that the symmetric \textit{encode-error-decode} structure of the CPC framework enables efficient QEC code compilation with a broad range of native gates.

\subsection{Compiling the $[[4,2,2]]$ CPC detection with an ion trap native gate}\label{sec:gp_nat_gate_simp}

\begin{figure*}[t]

	\subfloat[]{%
		\providecommand{\ket}[1]{\left|#1\right\rangle}
\providecommand{\phase}[1]{e^{2{\pi}i\cdot#1}}
\begin{tikzpicture}[scale=1.250000,x=1pt,y=1pt]
\filldraw[color=white] (0.000000, -8.000000) rectangle (88.000000, 24.000000);
\draw[color=black] (0.000000,16.000000) -- (88.000000,16.000000);
\draw[color=black] (0.000000,0.000000) -- (88.000000,0.000000);
\draw (5.000000,16.000000) -- (5.000000,0.000000);
\filldraw (5.000000, 16.000000) circle(1.500000pt);
\begin{scope}
\draw[fill=white] (5.000000, 0.000000) circle(3.000000pt);
\clip (5.000000, 0.000000) circle(3.000000pt);
\draw (2.000000, 0.000000) -- (8.000000, 0.000000);
\draw (5.000000, -3.000000) -- (5.000000, 3.000000);
\end{scope}
\draw[fill=white,color=white] (12.000000, -6.000000) rectangle (27.000000, 22.000000);
\draw (19.500000, 8.000000) node {$=$};
\begin{scope}
\draw[fill=white] (37.000000, -0.000000) +(-45.000000:8.485281pt and 8.485281pt) -- +(45.000000:8.485281pt and 8.485281pt) -- +(135.000000:8.485281pt and 8.485281pt) -- +(225.000000:8.485281pt and 8.485281pt) -- cycle;
\clip (37.000000, -0.000000) +(-45.000000:8.485281pt and 8.485281pt) -- +(45.000000:8.485281pt and 8.485281pt) -- +(135.000000:8.485281pt and 8.485281pt) -- +(225.000000:8.485281pt and 8.485281pt) -- cycle;
\draw (37.000000, -0.000000) node {$H$};
\end{scope}
\draw (50.500000,16.000000) -- (50.500000,0.000000);
\begin{scope}
\draw[fill=blue] (50.500000, 16.000000) +(-54.000000:3.500000pt) -- +(18.000000:3.500000pt) -- +(90.000000:3.500000pt) -- +(162.000000:3.500000pt) -- +(234.000000:3.500000pt) -- cycle;
\clip (50.500000, 16.000000) +(-54.000000:3.500000pt) -- +(18.000000:3.500000pt) -- +(90.000000:3.500000pt) -- +(162.000000:3.500000pt) -- +(234.000000:3.500000pt) -- cycle;
\end{scope}
\begin{scope}
\draw[fill=blue] (50.500000, 0.000000) +(-54.000000:3.500000pt) -- +(18.000000:3.500000pt) -- +(90.000000:3.500000pt) -- +(162.000000:3.500000pt) -- +(234.000000:3.500000pt) -- cycle;
\clip (50.500000, 0.000000) +(-54.000000:3.500000pt) -- +(18.000000:3.500000pt) -- +(90.000000:3.500000pt) -- +(162.000000:3.500000pt) -- +(234.000000:3.500000pt) -- cycle;
\end{scope}
\begin{scope}
\draw[fill=white] (64.000000, 16.000000) +(-45.000000:8.485281pt and 8.485281pt) -- +(45.000000:8.485281pt and 8.485281pt) -- +(135.000000:8.485281pt and 8.485281pt) -- +(225.000000:8.485281pt and 8.485281pt) -- cycle;
\clip (64.000000, 16.000000) +(-45.000000:8.485281pt and 8.485281pt) -- +(45.000000:8.485281pt and 8.485281pt) -- +(135.000000:8.485281pt and 8.485281pt) -- +(225.000000:8.485281pt and 8.485281pt) -- cycle;
\draw (64.000000, 16.000000) node {$P$};
\end{scope}
\begin{scope}
\draw[fill=white] (64.000000, -0.000000) +(-45.000000:8.485281pt and 8.485281pt) -- +(45.000000:8.485281pt and 8.485281pt) -- +(135.000000:8.485281pt and 8.485281pt) -- +(225.000000:8.485281pt and 8.485281pt) -- cycle;
\clip (64.000000, -0.000000) +(-45.000000:8.485281pt and 8.485281pt) -- +(45.000000:8.485281pt and 8.485281pt) -- +(135.000000:8.485281pt and 8.485281pt) -- +(225.000000:8.485281pt and 8.485281pt) -- cycle;
\draw (64.000000, -0.000000) node {$P$};
\end{scope}
\begin{scope}
\draw[fill=white] (80.000000, -0.000000) +(-45.000000:8.485281pt and 8.485281pt) -- +(45.000000:8.485281pt and 8.485281pt) -- +(135.000000:8.485281pt and 8.485281pt) -- +(225.000000:8.485281pt and 8.485281pt) -- cycle;
\clip (80.000000, -0.000000) +(-45.000000:8.485281pt and 8.485281pt) -- +(45.000000:8.485281pt and 8.485281pt) -- +(135.000000:8.485281pt and 8.485281pt) -- +(225.000000:8.485281pt and 8.485281pt) -- cycle;
\draw (80.000000, -0.000000) node {$H$};
\end{scope}
\end{tikzpicture}%
		\label{}%
	}\qquad
	\subfloat[]{%
		\providecommand{\ket}[1]{\left|#1\right\rangle}
\providecommand{\phase}[1]{e^{2{\pi}i\cdot#1}}
\begin{tikzpicture}[scale=1.250000,x=1pt,y=1pt]
\filldraw[color=white] (0.000000, -8.000000) rectangle (87.000000, 24.000000);
\draw[color=black] (0.000000,16.000000) -- (87.000000,16.000000);
\draw[color=black] (0.000000,0.000000) -- (87.000000,0.000000);
\draw (5.000000,16.000000) -- (5.000000,0.000000);
\begin{scope}
\begin{scope}[shift={(5.000000,16.000000)}]
\draw[fill=black, rounded corners=2] (-3, -3) rectangle (3, 3) {};
\end{scope}
\end{scope}
\begin{scope}
\begin{scope}[shift={(5.000000,0.000000)}]
\draw[fill=black, rounded corners=2] (-3, -3) rectangle (3, 3) {};
\end{scope}
\end{scope}
\draw[fill=white,color=white] (12.000000, -6.000000) rectangle (27.000000, 22.000000);
\draw (19.500000, 8.000000) node {$=$};
\begin{scope}
\draw[fill=white] (37.000000, 16.000000) +(-45.000000:8.485281pt and 8.485281pt) -- +(45.000000:8.485281pt and 8.485281pt) -- +(135.000000:8.485281pt and 8.485281pt) -- +(225.000000:8.485281pt and 8.485281pt) -- cycle;
\clip (37.000000, 16.000000) +(-45.000000:8.485281pt and 8.485281pt) -- +(45.000000:8.485281pt and 8.485281pt) -- +(135.000000:8.485281pt and 8.485281pt) -- +(225.000000:8.485281pt and 8.485281pt) -- cycle;
\draw (37.000000, 16.000000) node {$H$};
\end{scope}
\begin{scope}
\draw[fill=white] (37.000000, -0.000000) +(-45.000000:8.485281pt and 8.485281pt) -- +(45.000000:8.485281pt and 8.485281pt) -- +(135.000000:8.485281pt and 8.485281pt) -- +(225.000000:8.485281pt and 8.485281pt) -- cycle;
\clip (37.000000, -0.000000) +(-45.000000:8.485281pt and 8.485281pt) -- +(45.000000:8.485281pt and 8.485281pt) -- +(135.000000:8.485281pt and 8.485281pt) -- +(225.000000:8.485281pt and 8.485281pt) -- cycle;
\draw (37.000000, -0.000000) node {$H$};
\end{scope}
\draw (50.000000,16.000000) -- (50.000000,0.000000);
\begin{scope}
\draw[fill=blue] (50.000000, 16.000000) +(-54.000000:3.000000pt) -- +(18.000000:3.000000pt) -- +(90.000000:3.000000pt) -- +(162.000000:3.000000pt) -- +(234.000000:3.000000pt) -- cycle;
\clip (50.000000, 16.000000) +(-54.000000:3.000000pt) -- +(18.000000:3.000000pt) -- +(90.000000:3.000000pt) -- +(162.000000:3.000000pt) -- +(234.000000:3.000000pt) -- cycle;
\end{scope}
\begin{scope}
\draw[fill=blue] (50.000000, 0.000000) +(-54.000000:3.000000pt) -- +(18.000000:3.000000pt) -- +(90.000000:3.000000pt) -- +(162.000000:3.000000pt) -- +(234.000000:3.000000pt) -- cycle;
\clip (50.000000, 0.000000) +(-54.000000:3.000000pt) -- +(18.000000:3.000000pt) -- +(90.000000:3.000000pt) -- +(162.000000:3.000000pt) -- +(234.000000:3.000000pt) -- cycle;
\end{scope}
\begin{scope}
\draw[fill=white] (63.000000, 16.000000) +(-45.000000:8.485281pt and 8.485281pt) -- +(45.000000:8.485281pt and 8.485281pt) -- +(135.000000:8.485281pt and 8.485281pt) -- +(225.000000:8.485281pt and 8.485281pt) -- cycle;
\clip (63.000000, 16.000000) +(-45.000000:8.485281pt and 8.485281pt) -- +(45.000000:8.485281pt and 8.485281pt) -- +(135.000000:8.485281pt and 8.485281pt) -- +(225.000000:8.485281pt and 8.485281pt) -- cycle;
\draw (63.000000, 16.000000) node {$\Huge P$};
\end{scope}
\begin{scope}
\draw[fill=white] (63.000000, -0.000000) +(-45.000000:8.485281pt and 8.485281pt) -- +(45.000000:8.485281pt and 8.485281pt) -- +(135.000000:8.485281pt and 8.485281pt) -- +(225.000000:8.485281pt and 8.485281pt) -- cycle;
\clip (63.000000, -0.000000) +(-45.000000:8.485281pt and 8.485281pt) -- +(45.000000:8.485281pt and 8.485281pt) -- +(135.000000:8.485281pt and 8.485281pt) -- +(225.000000:8.485281pt and 8.485281pt) -- cycle;
\draw (63.000000, -0.000000) node {$\Huge P$};
\end{scope}
\begin{scope}
\draw[fill=white] (79.000000, 16.000000) +(-45.000000:8.485281pt and 8.485281pt) -- +(45.000000:8.485281pt and 8.485281pt) -- +(135.000000:8.485281pt and 8.485281pt) -- +(225.000000:8.485281pt and 8.485281pt) -- cycle;
\clip (79.000000, 16.000000) +(-45.000000:8.485281pt and 8.485281pt) -- +(45.000000:8.485281pt and 8.485281pt) -- +(135.000000:8.485281pt and 8.485281pt) -- +(225.000000:8.485281pt and 8.485281pt) -- cycle;
\draw (79.000000, 16.000000) node {$H$};
\end{scope}
\begin{scope}
\draw[fill=white] (79.000000, -0.000000) +(-45.000000:8.485281pt and 8.485281pt) -- +(45.000000:8.485281pt and 8.485281pt) -- +(135.000000:8.485281pt and 8.485281pt) -- +(225.000000:8.485281pt and 8.485281pt) -- cycle;
\clip (79.000000, -0.000000) +(-45.000000:8.485281pt and 8.485281pt) -- +(45.000000:8.485281pt and 8.485281pt) -- +(135.000000:8.485281pt and 8.485281pt) -- +(225.000000:8.485281pt and 8.485281pt) -- cycle;
\draw (79.000000, -0.000000) node {$H$};
\end{scope}
\end{tikzpicture}%
		\label{}%
	}\qquad
	\subfloat[]{%
		\providecommand{\ket}[1]{\left|#1\right\rangle}
\providecommand{\phase}[1]{e^{2{\pi}i\cdot#1}}
\begin{tikzpicture}[scale=1.250000,x=1pt,y=1pt]
\filldraw[color=white] (0.000000, -8.000000) rectangle (73.000000, 24.000000);
\draw[color=black] (0.000000,16.000000) -- (73.000000,16.000000);
\draw[color=black] (0.000000,0.000000) -- (73.000000,0.000000);
\begin{scope}
\draw[fill=white] (8.000000, 16.000000) +(-45.000000:8.485281pt and 8.485281pt) -- +(45.000000:8.485281pt and 8.485281pt) -- +(135.000000:8.485281pt and 8.485281pt) -- +(225.000000:8.485281pt and 8.485281pt) -- cycle;
\clip (8.000000, 16.000000) +(-45.000000:8.485281pt and 8.485281pt) -- +(45.000000:8.485281pt and 8.485281pt) -- +(135.000000:8.485281pt and 8.485281pt) -- +(225.000000:8.485281pt and 8.485281pt) -- cycle;
\draw (8.000000, 16.000000) node {$P$};
\end{scope}
\draw (21.500000,16.000000) -- (21.500000,0.000000);
\begin{scope}
\draw[fill=blue] (21.500000, 16.000000) +(-54.000000:3.500000pt) -- +(18.000000:3.500000pt) -- +(90.000000:3.500000pt) -- +(162.000000:3.500000pt) -- +(234.000000:3.500000pt) -- cycle;
\clip (21.500000, 16.000000) +(-54.000000:3.500000pt) -- +(18.000000:3.500000pt) -- +(90.000000:3.500000pt) -- +(162.000000:3.500000pt) -- +(234.000000:3.500000pt) -- cycle;
\end{scope}
\begin{scope}
\draw[fill=blue] (21.500000, 0.000000) +(-54.000000:3.500000pt) -- +(18.000000:3.500000pt) -- +(90.000000:3.500000pt) -- +(162.000000:3.500000pt) -- +(234.000000:3.500000pt) -- cycle;
\clip (21.500000, 0.000000) +(-54.000000:3.500000pt) -- +(18.000000:3.500000pt) -- +(90.000000:3.500000pt) -- +(162.000000:3.500000pt) -- +(234.000000:3.500000pt) -- cycle;
\end{scope}
\draw[fill=white,color=white] (29.000000, -6.000000) rectangle (44.000000, 22.000000);
\draw (36.500000, 8.000000) node {$=$};
\draw (51.500000,16.000000) -- (51.500000,0.000000);
\begin{scope}
\draw[fill=blue] (51.500000, 16.000000) +(-54.000000:3.500000pt) -- +(18.000000:3.500000pt) -- +(90.000000:3.500000pt) -- +(162.000000:3.500000pt) -- +(234.000000:3.500000pt) -- cycle;
\clip (51.500000, 16.000000) +(-54.000000:3.500000pt) -- +(18.000000:3.500000pt) -- +(90.000000:3.500000pt) -- +(162.000000:3.500000pt) -- +(234.000000:3.500000pt) -- cycle;
\end{scope}
\begin{scope}
\draw[fill=blue] (51.500000, 0.000000) +(-54.000000:3.500000pt) -- +(18.000000:3.500000pt) -- +(90.000000:3.500000pt) -- +(162.000000:3.500000pt) -- +(234.000000:3.500000pt) -- cycle;
\clip (51.500000, 0.000000) +(-54.000000:3.500000pt) -- +(18.000000:3.500000pt) -- +(90.000000:3.500000pt) -- +(162.000000:3.500000pt) -- +(234.000000:3.500000pt) -- cycle;
\end{scope}
\begin{scope}
\draw[fill=white] (65.000000, 16.000000) +(-45.000000:8.485281pt and 8.485281pt) -- +(45.000000:8.485281pt and 8.485281pt) -- +(135.000000:8.485281pt and 8.485281pt) -- +(225.000000:8.485281pt and 8.485281pt) -- cycle;
\clip (65.000000, 16.000000) +(-45.000000:8.485281pt and 8.485281pt) -- +(45.000000:8.485281pt and 8.485281pt) -- +(135.000000:8.485281pt and 8.485281pt) -- +(225.000000:8.485281pt and 8.485281pt) -- cycle;
\draw (65.000000, 16.000000) node {$P$};
\end{scope}
\end{tikzpicture}%
		\label{}%
	}
	
	\caption{(a) A \cnot gate expressed in terms of the SP native gate\joschka{, which is represented by the connected blue pentagons. The matrix form of the SP native gate is given in equation (\ref{eq:sp_native_gate}).} (b) A conjugate-propagator gate expressed in terms of the SP native gate. (c) Both the phase gate $P$ and the $SP$ native gate are represented as diagonal matrices in the computational basis. As a result, the $P$ gate can be moved freely through the SP native gate.}
	
	\label{fig:nat_gate_defs}
\end{figure*}
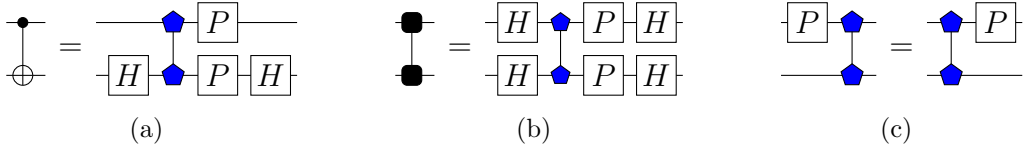

Here we show that the $[[4,2,2]]$ CPC detection code, introduced in section \ref{sec:422}, can be efficiently compiled with an ion trap native gate. For the purposes of this example, we adopt an ion trap with a SP native gate as introduced in equation (\ref{eq:sp_native_gate}) in section \ref{sec:ion_trap_intro}. The SP native gate can be transformed into a \cnot via the application of local unitary operations to its inputs and outputs. A possible mapping, in matrix equation form, is given by
\begin{equation}\label{eq:cnot_map}
\cnoteq^{q_1}_{q_2}=\left(I_{q_1}\otimes H_{q_2}\right)\bigcdot \left(P_{q_1}\otimes P_{q_2}\right)\bigcdot F^{q_1}_{q_2} \bigcdot \left(I_{q_1}\otimes H_{q_2}\right)\rm,
\end{equation} 
where $F^{q_1}_{q_2}$ is the matrix representation of the SP gate defined in equation (\ref{eq:entangling_unitary}), and $P$ is a phase gate defined as $P=\text{diag}\left(1,-\text{i}\right)$. Realising a \cnot gate on ion trap hardware, via the above mapping, requires the application of the native gate combined with four single-qubit gates, as shown in figure \ref{fig:nat_gate_defs}a. Likewise, figure \ref{fig:nat_gate_defs}b shows how the conjugate-propagator gate can be constructed from the native gate via the addition of six single-qubit operations. We will see that, when the native gates are compiled into a CPC circuit, constructive simplifications become possible to reduce the total number of single-qubit gates required.

\begin{figure*}[]
	
	\subfloat[The un-compiled \texttt{$[[$}4,2,2\texttt{$]]$} CPC error detection code expressed in terms of \cnot and conjugate propagator gates.]{
		\providecommand{\ket}[1]{\left|#1\right\rangle}
\providecommand{\phase}[1]{e^{2{\pi}i\cdot#1}}
\begin{tikzpicture}[scale=1.000000,x=1pt,y=1pt]
\filldraw[color=white] (0.000000, -10.000000) rectangle (396.000000, 70.000000);
\draw[color=black] (0.000000,60.000000) -- (396.000000,60.000000);
\draw[color=black] (0.000000,40.000000) -- (396.000000,40.000000);
\draw[color=black] (0.000000,50.000000) node[left] {$\ket{\psi_{AB}}$};
\draw[color=black] (0.000000,20.000000) -- (377.000000,20.000000);
\draw[color=black] (377.000000,19.500000) -- (396.000000,19.500000);
\draw[color=black] (377.000000,20.500000) -- (396.000000,20.500000);
\draw[color=black] (0.000000,20.000000) node[left] {$\ket{0}_{p_1}$};
\draw[color=black] (0.000000,0.000000) -- (377.000000,0.000000);
\draw[color=black] (377.000000,-0.500000) -- (396.000000,-0.500000);
\draw[color=black] (377.000000,0.500000) -- (396.000000,0.500000);
\draw[color=black] (0.000000,0.000000) node[left] {$\ket{0}_{p_2}$};
\draw (16.000000,20.000000) -- (16.000000,0.000000);
\begin{scope}
\begin{scope}[shift={(16.000000,20.000000)}]
\draw[fill=black, rounded corners=2] (-3, -3) rectangle (3, 3) {};
\end{scope}
\end{scope}
\begin{scope}
\begin{scope}[shift={(16.000000,0.000000)}]
\draw[fill=black, rounded corners=2] (-3, -3) rectangle (3, 3) {};
\end{scope}
\end{scope}
\draw (342.000000,20.000000) -- (342.000000,0.000000);
\begin{scope}
\begin{scope}[shift={(342.000000,20.000000)}]
\draw[fill=black, rounded corners=2] (-3, -3) rectangle (3, 3) {};
\end{scope}
\end{scope}
\begin{scope}
\begin{scope}[shift={(342.000000,0.000000)}]
\draw[fill=black, rounded corners=2] (-3, -3) rectangle (3, 3) {};
\end{scope}
\end{scope}
\draw (48.000000,60.000000) -- (48.000000,20.000000);
\filldraw (48.000000, 60.000000) circle(1.500000pt);
\begin{scope}
\draw[fill=white] (48.000000, 20.000000) circle(3.000000pt);
\clip (48.000000, 20.000000) circle(3.000000pt);
\draw (45.000000, 20.000000) -- (51.000000, 20.000000);
\draw (48.000000, 17.000000) -- (48.000000, 23.000000);
\end{scope}
\draw (310.000000,60.000000) -- (310.000000,20.000000);
\filldraw (310.000000, 60.000000) circle(1.500000pt);
\begin{scope}
\draw[fill=white] (310.000000, 20.000000) circle(3.000000pt);
\clip (310.000000, 20.000000) circle(3.000000pt);
\draw (307.000000, 20.000000) -- (313.000000, 20.000000);
\draw (310.000000, 17.000000) -- (310.000000, 23.000000);
\end{scope}
\draw (80.000000,40.000000) -- (80.000000,20.000000);
\filldraw (80.000000, 40.000000) circle(1.500000pt);
\begin{scope}
\draw[fill=white] (80.000000, 20.000000) circle(3.000000pt);
\clip (80.000000, 20.000000) circle(3.000000pt);
\draw (77.000000, 20.000000) -- (83.000000, 20.000000);
\draw (80.000000, 17.000000) -- (80.000000, 23.000000);
\end{scope}
\draw (278.000000,40.000000) -- (278.000000,20.000000);
\filldraw (278.000000, 40.000000) circle(1.500000pt);
\begin{scope}
\draw[fill=white] (278.000000, 20.000000) circle(3.000000pt);
\clip (278.000000, 20.000000) circle(3.000000pt);
\draw (275.000000, 20.000000) -- (281.000000, 20.000000);
\draw (278.000000, 17.000000) -- (278.000000, 23.000000);
\end{scope}
\draw (112.000000,60.000000) -- (112.000000,0.000000);
\begin{scope}
\begin{scope}[shift={(112.000000,60.000000)}]
\draw[fill=black, rounded corners=2] (-3, -3) rectangle (3, 3) {};
\end{scope}
\end{scope}
\begin{scope}
\begin{scope}[shift={(112.000000,0.000000)}]
\draw[fill=black, rounded corners=2] (-3, -3) rectangle (3, 3) {};
\end{scope}
\end{scope}
\draw (246.000000,60.000000) -- (246.000000,0.000000);
\begin{scope}
\begin{scope}[shift={(246.000000,60.000000)}]
\draw[fill=black, rounded corners=2] (-3, -3) rectangle (3, 3) {};
\end{scope}
\end{scope}
\begin{scope}
\begin{scope}[shift={(246.000000,0.000000)}]
\draw[fill=black, rounded corners=2] (-3, -3) rectangle (3, 3) {};
\end{scope}
\end{scope}
\draw (144.000000,40.000000) -- (144.000000,0.000000);
\begin{scope}
\begin{scope}[shift={(144.000000,40.000000)}]
\draw[fill=black, rounded corners=2] (-3, -3) rectangle (3, 3) {};
\end{scope}
\end{scope}
\begin{scope}
\begin{scope}[shift={(144.000000,0.000000)}]
\draw[fill=black, rounded corners=2] (-3, -3) rectangle (3, 3) {};
\end{scope}
\end{scope}
\draw (214.000000,40.000000) -- (214.000000,0.000000);
\begin{scope}
\begin{scope}[shift={(214.000000,40.000000)}]
\draw[fill=black, rounded corners=2] (-3, -3) rectangle (3, 3) {};
\end{scope}
\end{scope}
\begin{scope}
\begin{scope}[shift={(214.000000,0.000000)}]
\draw[fill=black, rounded corners=2] (-3, -3) rectangle (3, 3) {};
\end{scope}
\end{scope}
\draw (179.000000,60.000000) -- (179.000000,0.000000);
\begin{scope}
\draw[fill=white] (179.000000, 30.000000) +(-45.000000:8.485281pt and 50.911688pt) -- +(45.000000:8.485281pt and 50.911688pt) -- +(135.000000:8.485281pt and 50.911688pt) -- +(225.000000:8.485281pt and 50.911688pt) -- cycle;
\clip (179.000000, 30.000000) +(-45.000000:8.485281pt and 50.911688pt) -- +(45.000000:8.485281pt and 50.911688pt) -- +(135.000000:8.485281pt and 50.911688pt) -- +(225.000000:8.485281pt and 50.911688pt) -- cycle;
\draw (179.000000, 30.000000) node {$E$};
\end{scope}
\draw[fill=white] (371.000000, 14.000000) rectangle (383.000000, 26.000000);
\draw[very thin] (377.000000, 20.600000) arc (90:150:6.000000pt);
\draw[very thin] (377.000000, 20.600000) arc (90:30:6.000000pt);
\draw[->,>=stealth] (377.000000, 14.600000) -- +(80:10.392305pt);
\draw[fill=white] (371.000000, -6.000000) rectangle (383.000000, 6.000000);
\draw[very thin] (377.000000, 0.600000) arc (90:150:6.000000pt);
\draw[very thin] (377.000000, 0.600000) arc (90:30:6.000000pt);
\draw[->,>=stealth] (377.000000, -5.400000) -- +(80:10.392305pt);
\end{tikzpicture}%
		\label{}%
	}

	\subfloat[The compiled circuit prior to simplification. Note that parts of the decoder have been hidden to save space. The pairs of Hadamards, labelled red, cancel to the identity. The blue $H$ gates can also be discarded without affecting the operation of the code.]{
		\input{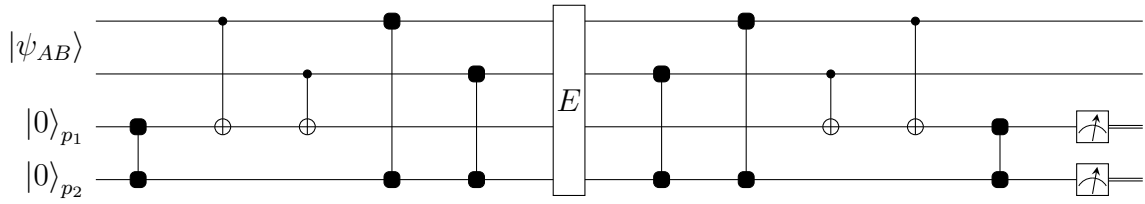}%
		\label{fsfsdfsd}%
		
	}

	\subfloat[The circuit following removal of the $H$ gates. The pairs of red $P$ gates combine to form $Z$ gates.]{%
		\input{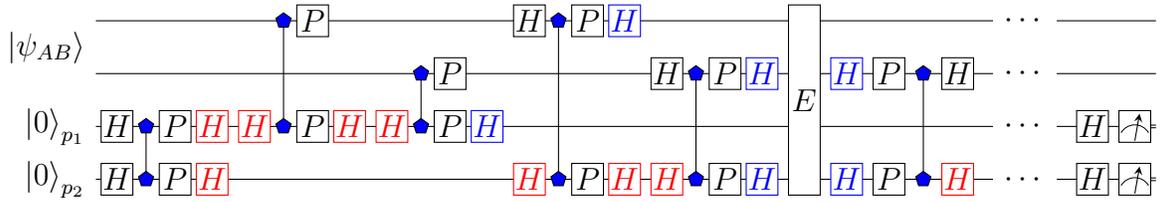}%
		\label{}%
	}

	\subfloat[The $Z$ gates and blue $P$ gates can be moved freely through the SP gates to the centre of the circuit. Due to their symmetry about the error window, the $P$ and $Z$ gates can be omitted from the code. The green $P$ gates do not affect circuit operation and are also discarded.]{%
		\input{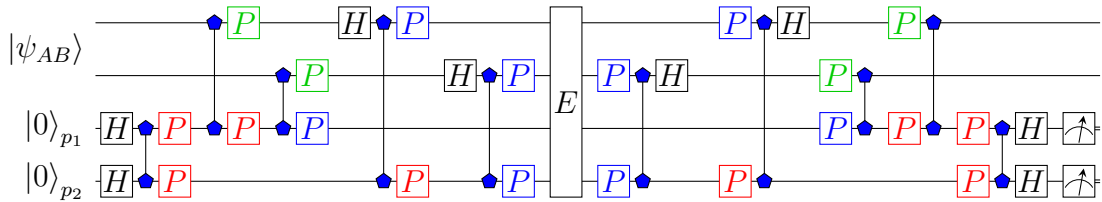}%
		\label{}%
	}

	\subfloat[The compiled \texttt{$[[$}4,2,2\texttt{$]]$} CPC detection code following circuit simplification. Only four single-qubit gates remain in the encoder.]{%
		\input{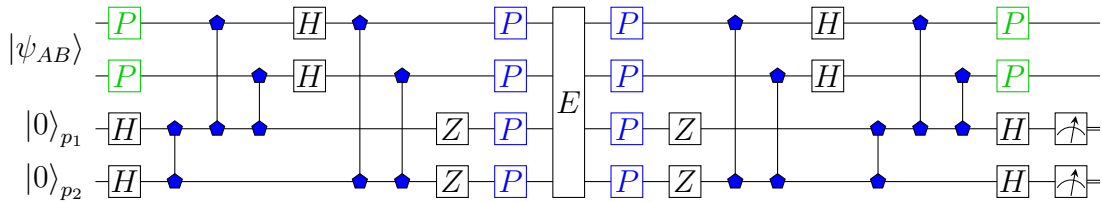}
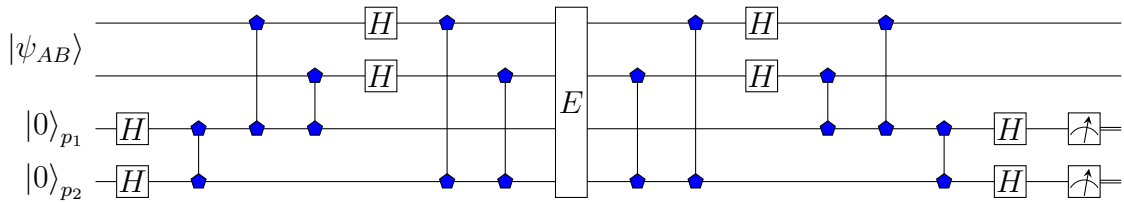%
		\label{}%
	}

	\caption{Compiling the $[[4,2,2]]$ detection code with the SP native gate.}
	
	\label{fig:422_simp}
\end{figure*}

Figure \ref{fig:422_simp} illustrates the steps involved in the compilation and simplification of the $[[4,2,2]]$ CPC code with the SP native gate. The un-compiled circuit, expressed in terms of \cnot and conjugate-propagator gates, is shown in figure \ref{fig:422_simp}a. The first step of compilation involves substituting the \cnot and conjugate-propagator gates with the SP native interaction, via the circuit rewrites rules defined in figure \ref{fig:nat_gate_defs}. The resultant circuit is shown in figure \ref{fig:422_simp}b.

Now that the circuit is written in terms of the native gate, circuit simplifications can be applied to reduce the single-qubit gate count. In figure \ref{fig:422_simp}b, pairs of $H$ gates that cancel to the identity are labelled in red. In the encoder, the $H$ gates labelled in blue are paired with their counterparts from the decoder. We can now exploit the symmetry of the CPC code to further reduce the gate-count. The effect of the blue $H$ gates around the wait-stage is to transform $X$ errors into $Z$ errors and vice-versa, as described by the following matrix transformations
\begin{eqnarray}
H \bigcdot (E=X) \bigcdot H = H \bigcdot X \bigcdot H = Z,\nonumber\\
H \bigcdot (E=Z) \bigcdot H =H \bigcdot Z \bigcdot H= X\rm ,
\end{eqnarray}
where $E$ represents the error that occurs in the wait stage. The $[[4,2,2]]$ code can detect both $X$ and $Z$ errors, as shown in syndrome table \ref{tab:422synd} in section \ref{sec:422}. As a result, the blue $H$ gates do not change the errors into a form that cannot be detected. The blue $H$ gates can therefore be discarded without affecting the operation of the $[[4,2,2]]$ code.

Figure \ref{fig:422_simp}c shows the compiled $[[4,2,2]]$ code following the removal of the unnecessary $H$ gates. Notice that both the $P$ gate and the SP gate are described by diagonal matrices in the computational basis. As a result, we have the freedom to move $P$ gates through the SP native gate as shown in figure \ref{fig:nat_gate_defs}c. Two $P$ gates combine to form a $Z$ gate as follows $P\bigcdot P=Z$. In the circuit in figure \ref{fig:422_simp}c, pairs of $P$ gates are highlighted in red. As $Z$ gates are diagonal in the computational basis, they can also be moved through the SP gates.

In the circuit in figure \ref{fig:422_simp}d, the $Z$ gates and blue $P$ gates have been pushed to the centre of the circuit. \joschka{In the event that no error occurs, these $P$ gates combine to form a $Z$-error via the relation $Z=P\bigcdot P$. However, the locations of these errors are known, and they can therefore be accounted for in post-processing}. \joschka{If an error does occur,} the effect of symmetric $P$ gates about the wait stage, $E$, is to transform $X$-errors into $Y$-errors and vice versa, as described by the following matrix transformation rules
\begin{eqnarray}
P \bigcdot (E=X) \bigcdot P = P \bigcdot X \bigcdot P = (-\text{i})Y,\nonumber\\
P \bigcdot (E=Y) \bigcdot P =P \bigcdot Y \bigcdot P= (-\text{i})X\rm ,
\end{eqnarray}
where the $(-\text{i})$ global phase does not affect the syndrome measurement. These transformations are unproblematic as $[[4,2,2]]$ code can detect both $X$ and $Y$ errors (see syndrome table \ref{tab:422synd}). \joschka{As the effect of the blue $P$ gates can be described in terms of single-qubit Clifford operations on the output, they can be removed from the circuit and accounted for in post-processing}. There are also $P$ gates highlighted in green, located on the register qubits at the beginning and end of each error cycle. These gates occur before the first round of CPC checks, and can therefore be removed from the circuit without affecting the final syndrome readout. Finally, the $Z$ gates located symmetrically about the wait-stage introduce a global phase to the errors. This global phase does not affect the propagation of errors through the circuit, meaning the $Z$ gates can be removed. \joschka{It should be noted that the above simplifications will result in a modified syndrome table. However, the \textit{no-error} case will remain unique meaning the function of the code is maintained.}

The final simplified form of $[[4,2,2]]$ CPC code compiled with the SP native gate is shown in figure \ref{fig:422_simp}e. The single-qubit gate count in the encoder has been lowered from $26$ gates in the original compiled circuit (figure \ref{fig:422_simp}b), to $4$ gates in final circuit (figure \ref{fig:422_simp}e).

\subsection{Requirements for CPC gates}

We have now shown that the $[[4,2,2]]$ code can be efficiently compiled with the SP native gate. Most of the single-qubit corrections can be eliminated, either by direct cancellation between adjacent Hadamards, or by moving $P$ gates through the circuit. We now show that efficient CPC code translation, from the idealised \cnot version to the hardware-compiled version, is possible for a range of native gate types. We begin by outlining the general requirements for two-qubit gates in a CPC circuit.

In a CPC code, the role of two-qubit interaction gates is to distribute error information from the register to the parity qubits. For example, \cnot gates propagate bit-errors from their control to target via the rule in equation (\ref{eq:cnot_prop}). More generally, we require that the two-qubit CPC gate, $\Omega_{q_2}^{q_1}$, has the ability to change the weight of an error operator, $E^{i}_{q_1}\otimes \openone_{q_2}$, such that
\begin{equation} \label{eq:op_rule}
\Omega^{q_1}_{q_2} \bigcdot \left(E^{i}_{q_1}\otimes \openone_{q_2}\right) \bigcdot \left(\Omega^{q_1}_{q_2}\right)^\dagger = \left(E^{j}_{q_1}\otimes E^{k}_{q_2}\right),
\end{equation}
where $q_1$ and $q_2$ are the control and target qubits respectively, and $E^{i,j,k}$ are non-identity elements of the single-qubit Pauli group. \joschka{As both $E^{i}_{q_1}\otimes \openone_{q_2}$ and $E^{j}_{q_1}\otimes E^{k}_{q_2}$ are Pauli group operators, we see that $\Omega_{q_2}^{q_1}$ must be a Clifford gate} (for an overview of the Clifford group see appendix \ref{app:clifford}). CPC quantum memories can be described entirely in terms of Clifford gates, as their operations are restricted to manipulating stabilizer states. This allows for efficient classical simulation. For an example of such a simulation, see the CPC syndrome calculation algorithm we outline in appendix \ref{app:synd_tab}.

Another way of thinking about the CPC interaction gates is in terms of entanglement. In equation (\ref{eq:op_rule}), it can be seen that the general CPC gate de-localises error information from the control to the target, \joschka{suggesting the operation has the potential to entangle states}. Furthermore, we know that elements of the two-qubit stabilizer states are either maximally entangled or separable. Any Clifford entangling gate that maps between these states, and therefore any CPC interaction, is a maximally entangling operation.

We have now established that CPC gates must be maximally entangling Clifford operations. However, many experiments will have native gates that do not satisfy these requirements. For example, several qubit technologies have a native interaction of the form $\sqrt{\swapeq}$ \cite{Loss1998}, which is only partially entangling. In these circumstances, multiple applications of the native  gate, in addition to local operations, are required to synthesise the desired maximally entangling behaviour. \joschka{It is typically the case that quantum computing experiments will have different error rates for single-qubit and two-qubit gates \cite{Linke2017}. Circuit compilation strategies should therefore aim to minimise the gate-type with the highest error rate. In the case of ion traps, for example, the two-qubit gates have lower fidelities than single-qubit gates \cite{Ballance2016,Harty2014}.} 

\subsection{Circuit simplification with any maximally entangling Clifford gate}

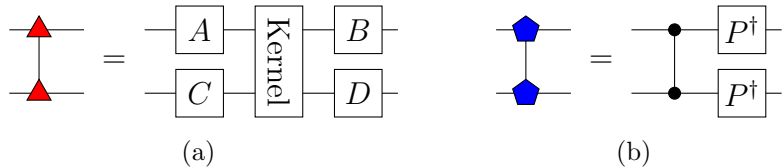
\begin{figure*}[t]

	\subfloat[]{%
		\providecommand{\ket}[1]{\left|#1\right\rangle}
\providecommand{\phase}[1]{e^{2{\pi}i\cdot#1}}
\begin{tikzpicture}[scale=1.500000,x=1pt,y=1pt]
\filldraw[color=white] (0.000000, -8.000000) rectangle (98.000000, 24.000000);
\draw[color=black] (0.000000,16.000000) -- (98.000000,16.000000);
\draw[color=black] (0.000000,0.000000) -- (98.000000,0.000000);
\draw (7.500000,16.000000) -- (7.500000,0.000000);
\begin{scope}
\draw[fill=red] (7.500000, 16.000000) +(-30.000000:3.500000pt) -- +(90.000000:3.500000pt) -- +(210.000000:3.500000pt) -- cycle;
\clip (7.500000, 16.000000) +(-30.000000:3.500000pt) -- +(90.000000:3.500000pt) -- +(210.000000:3.500000pt) -- cycle;
\end{scope}
\begin{scope}
\draw[fill=red] (7.500000, 0.000000) +(-30.000000:3.500000pt) -- +(90.000000:3.500000pt) -- +(210.000000:3.500000pt) -- cycle;
\clip (7.500000, 0.000000) +(-30.000000:3.500000pt) -- +(90.000000:3.500000pt) -- +(210.000000:3.500000pt) -- cycle;
\end{scope}
\draw[fill=white,color=white] (19.000000, -6.000000) rectangle (34.000000, 22.000000);
\draw (26.500000, 8.000000) node {$=$};
\begin{scope}
\draw[fill=white] (48.000000, 16.000000) +(-45.000000:8.485281pt and 8.485281pt) -- +(45.000000:8.485281pt and 8.485281pt) -- +(135.000000:8.485281pt and 8.485281pt) -- +(225.000000:8.485281pt and 8.485281pt) -- cycle;
\clip (48.000000, 16.000000) +(-45.000000:8.485281pt and 8.485281pt) -- +(45.000000:8.485281pt and 8.485281pt) -- +(135.000000:8.485281pt and 8.485281pt) -- +(225.000000:8.485281pt and 8.485281pt) -- cycle;
\draw (48.000000, 16.000000) node {$A$};
\end{scope}
\begin{scope}
\draw[fill=white] (48.000000, -0.000000) +(-45.000000:8.485281pt and 8.485281pt) -- +(45.000000:8.485281pt and 8.485281pt) -- +(135.000000:8.485281pt and 8.485281pt) -- +(225.000000:8.485281pt and 8.485281pt) -- cycle;
\clip (48.000000, -0.000000) +(-45.000000:8.485281pt and 8.485281pt) -- +(45.000000:8.485281pt and 8.485281pt) -- +(135.000000:8.485281pt and 8.485281pt) -- +(225.000000:8.485281pt and 8.485281pt) -- cycle;
\draw (48.000000, -0.000000) node {$C$};
\end{scope}
\draw (68.000000,16.000000) -- (68.000000,0.000000);
\begin{scope}
\draw[fill=white] (68.000000, 8.000000) +(-45.000000:8.485281pt and 19.798990pt) -- +(45.000000:8.485281pt and 19.798990pt) -- +(135.000000:8.485281pt and 19.798990pt) -- +(225.000000:8.485281pt and 19.798990pt) -- cycle;
\clip (68.000000, 8.000000) +(-45.000000:8.485281pt and 19.798990pt) -- +(45.000000:8.485281pt and 19.798990pt) -- +(135.000000:8.485281pt and 19.798990pt) -- +(225.000000:8.485281pt and 19.798990pt) -- cycle;
\draw (68.000000, 8.000000) node {\rotatebox{-90}{Kernel}};
\end{scope}
\begin{scope}
\draw[fill=white] (88.000000, 16.000000) +(-45.000000:8.485281pt and 8.485281pt) -- +(45.000000:8.485281pt and 8.485281pt) -- +(135.000000:8.485281pt and 8.485281pt) -- +(225.000000:8.485281pt and 8.485281pt) -- cycle;
\clip (88.000000, 16.000000) +(-45.000000:8.485281pt and 8.485281pt) -- +(45.000000:8.485281pt and 8.485281pt) -- +(135.000000:8.485281pt and 8.485281pt) -- +(225.000000:8.485281pt and 8.485281pt) -- cycle;
\draw (88.000000, 16.000000) node {$B$};
\end{scope}
\begin{scope}
\draw[fill=white] (88.000000, -0.000000) +(-45.000000:8.485281pt and 8.485281pt) -- +(45.000000:8.485281pt and 8.485281pt) -- +(135.000000:8.485281pt and 8.485281pt) -- +(225.000000:8.485281pt and 8.485281pt) -- cycle;
\clip (88.000000, -0.000000) +(-45.000000:8.485281pt and 8.485281pt) -- +(45.000000:8.485281pt and 8.485281pt) -- +(135.000000:8.485281pt and 8.485281pt) -- +(225.000000:8.485281pt and 8.485281pt) -- cycle;
\draw (88.000000, -0.000000) node {$D$};
\end{scope}
\end{tikzpicture}%
		\label{}%
	}\qquad
	\subfloat[]{%
		\providecommand{\ket}[1]{\left|#1\right\rangle}
\providecommand{\phase}[1]{e^{2{\pi}i\cdot#1}}
\begin{tikzpicture}[scale=1.500000,x=1pt,y=1pt]
\filldraw[color=white] (0.000000, -8.000000) rectangle (72.000000, 24.000000);
\draw[color=black] (0.000000,16.000000) -- (72.000000,16.000000);
\draw[color=black] (0.000000,0.000000) -- (72.000000,0.000000);
\draw (7.500000,16.000000) -- (7.500000,0.000000);
\begin{scope}
\draw[fill=blue] (7.500000, 16.000000) +(-54.000000:3.500000pt) -- +(18.000000:3.500000pt) -- +(90.000000:3.500000pt) -- +(162.000000:3.500000pt) -- +(234.000000:3.500000pt) -- cycle;
\clip (7.500000, 16.000000) +(-54.000000:3.500000pt) -- +(18.000000:3.500000pt) -- +(90.000000:3.500000pt) -- +(162.000000:3.500000pt) -- +(234.000000:3.500000pt) -- cycle;
\end{scope}
\begin{scope}
\draw[fill=blue] (7.500000, 0.000000) +(-54.000000:3.500000pt) -- +(18.000000:3.500000pt) -- +(90.000000:3.500000pt) -- +(162.000000:3.500000pt) -- +(234.000000:3.500000pt) -- cycle;
\clip (7.500000, 0.000000) +(-54.000000:3.500000pt) -- +(18.000000:3.500000pt) -- +(90.000000:3.500000pt) -- +(162.000000:3.500000pt) -- +(234.000000:3.500000pt) -- cycle;
\end{scope}
\draw[fill=white,color=white] (19.000000, -6.000000) rectangle (34.000000, 22.000000);
\draw (26.500000, 8.000000) node {$=$};
\draw (45.000000,16.000000) -- (45.000000,0.000000);
\filldraw (45.000000, 16.000000) circle(1.500000pt);
\filldraw (45.000000, 0.000000) circle(1.500000pt);
\begin{scope}
\draw[fill=white] (62.000000, 16.000000) +(-45.000000:8.485281pt and 8.485281pt) -- +(45.000000:8.485281pt and 8.485281pt) -- +(135.000000:8.485281pt and 8.485281pt) -- +(225.000000:8.485281pt and 8.485281pt) -- cycle;
\clip (62.000000, 16.000000) +(-45.000000:8.485281pt and 8.485281pt) -- +(45.000000:8.485281pt and 8.485281pt) -- +(135.000000:8.485281pt and 8.485281pt) -- +(225.000000:8.485281pt and 8.485281pt) -- cycle;
\draw (62.000000, 16.000000) node {$P^\dagger$};
\end{scope}
\begin{scope}
\draw[fill=white] (62.000000, -0.000000) +(-45.000000:8.485281pt and 8.485281pt) -- +(45.000000:8.485281pt and 8.485281pt) -- +(135.000000:8.485281pt and 8.485281pt) -- +(225.000000:8.485281pt and 8.485281pt) -- cycle;
\clip (62.000000, -0.000000) +(-45.000000:8.485281pt and 8.485281pt) -- +(45.000000:8.485281pt and 8.485281pt) -- +(135.000000:8.485281pt and 8.485281pt) -- +(225.000000:8.485281pt and 8.485281pt) -- cycle;
\draw (62.000000, -0.000000) node {$P^\dagger$};
\end{scope}
\end{tikzpicture}%
		\label{}%
	}
	
	\caption{Left: A general maximally entangling Clifford native gate (red triangles) can be expressed in terms of a kernel supplemented by local corrections on its inputs and outputs. The kernel will always be of the form \CZ or \CZeq-\swap. The local corrections, $\{A, B, C, D\}$, are single-qubit Clifford gates and can expressed as products of $H$ and $P$ gates. Right: The  SP native gate expressed in terms of its \CZ kernel.}
	
	\label{fig:ng_kernel}
\end{figure*}

We will now outline general CPC circuit simplification procedures for maximally entangling Clifford gates. It can be shown that all Clifford entangling gates are local Clifford equivalent to either the \CZ or the \CZeq-\swap interaction. With this knowledge, we can write all maximally entangling Clifford gates in terms of a central kernel, containing either a \CZ or \CZeq-\swap interaction, supplemented by local Clifford gates (see figure \ref{fig:ng_kernel}a).

The single-qubit Clifford group is generated by $P$ and $H$ gates. Any native gate can therefore be constructed from its by kernel via the addition of local gates generated from combinations of $P$ and $H$. The $P$ gates can be trivially pushed through the \CZ kernel. Likewise, it is possible to push $P$ gates through the \CZ-\swap kernels, although the effect of the \swap gate must be taken into account.

In section \ref{sec:ion_comp} we demonstrated the compilation of a CPC code using the SP native gate, which is local Clifford equivalent to a \CZ gate. The exact transformation from \CZ kernel to SP gate is shown in figure \ref{fig:ng_kernel}b. As the local operations in this case consist of $P^\dagger$ gates, we have the freedom move $P$ gates through the SP native gate. Hadamard gates $H$, however, restrict movement, but in many cases will cancel when the native gate is compiled into a CPC circuit.

The general procedure for compiling a CPC code with a given native gate can now be written as follows. First, eliminate any unnecessary $H$ gates by identifying cancellations between adjacent CPC gates.  Second, determine the behaviour when $P$ gates are pushed through the native gate. As all maximally entangling Clifford gates have either a \CZ or \CZeq-\swap kernel, it is often possible to trivially move $P$ gates through each block of the encoder. Once these simplifications rules have been established, they can be applied systematically to substantially reduce the CPC circuit gate count.

\section{The CPC code design process}\label{sec:cpc_code_design}

The first generation of quantum computers will be limited in size to no more than a couple of hundred qubits \cite{nqit, IBM}. In this section, we outline a design process for constructing hardware-optimised quantum codes with the CPC framework. By maximising the encoding density, such bespoke CPC codes will help early quantum computers realise their full potential.

To illustrate our design process, the quantum device we consider is the seven-qubit linear ion trap which was introduced in section \ref{sec:ion_trap_intro}. Our CPC design process is split into three stages. 1) \textbf{CPC code discovery}: numerical search techniques are used to find CPC codes that maximise the quantum encoding density for a seven qubit register. 2) \textbf{Hardware optimisation}: the best CPC codes from the discovered set are identified by analysing which ones have the lowest two-qubit count when implemented on a linear nearest-neighbour architecture. 3) \textbf{Native gate compilation}: further optimisations are made by identifying CPC circuits with efficient translations from the \cnot version of the code to the native gate version, using the circuit simplification strategies outlined in section \ref{sec:any_native_gate}.

\subsection{Stage 1: CPC code discovery}

The idealised ion trap we are considering has seven application qubits. \joschka{We assume that during the wait stage the ion trap qubits are subject to a biased depolarizing noise channel of the form
	\begin{equation}
	\mathcal{E}{\left[\rho\right]}=(1-p_x-p_z-p_xp_z)\rho+p_xX\rho X + 2p_xp_zY\rho Y + p_zZ\rho Z \rm,
	\end{equation}
	where $\rho$ is the single-qubit density matrix, and $p_x$ and $p_z$ are the probabilities of $X$- and $Z$-errors respectively. This error model assumes the ion trap has independent error mechanisms for $X$- and $Z$-errors, but $Y$-errors occur only as a result of successive single-qubit errors of the form $XZ$ and $ZX$ \footnote[1]{Note that the ion trap we consider is an idealised proof-of-concept model, and does not correspond to a specific ion trap experiment.}. Similar error models have recently been considered in \cite{Robertson2017,Tuckett2018}. For the purposes of our ion trap model, we assume that the error probabilities $p_x$ and $p_z$ are low enough that the probability of $Y$-errors becomes negligibly small. The effective error model can then be written as
\begin{equation} \label{eq:error_model}
\mathcal{E}{\left[\rho\right]}\approx(1-p_x-p_z)\rho+p_xX\rho X  + p_zZ\rho Z \rm.
\end{equation}
Under the above error model for the idle ion trap qubits, the CPC quantum memory only needs to correct $X$- and $Z$-errors. We choose this noise model for our proof-of-concept outline of the CPC design process, as it corresponds to the simplest possible non-classical error model. Our aim is to discover non-degenerate quantum codes which produce a unique syndrome for single $X$- and $Z$-errors on any of the seven qubits in the trap.}

The maximum possible encoding density of a non-degenerate quantum error correction code is constrained by the quantum Hamming bound, which states that an $[[n,k,d]]$ code must satisfy the following inequality
\begin{equation}
\sum_{j=0}^{(d-1)/2} \binom{n}{j}|\mathcal{E}|^j \ 2^k \leq 2^n \rm,
\end{equation}

\noindent where $|\mathcal{E}|$ is the size of the single-qubit error set \cite{GottesmanHammingBound}. As we are considering only $X$- and $Z$-errors in the generic ion trap \joschka{under the error model described by equation (\ref{eq:error_model})}, the size of the error set is $|\mathcal{E}|=2$. Under this error model, the quantum Hamming bound tells us that the maximum number of data qubits that can be encoded in $7$ physical qubits is $k_{\text{max}}=3$. The optimal $7$ qubit CPC code will therefore be of the form $[[n=7,k=3,d=3]]$. Note that the code distance is $d=3$, indicating that these $[[7,3,3]]$ codes will be able to correct one error per CPC cycle.

\begin{figure}
\input{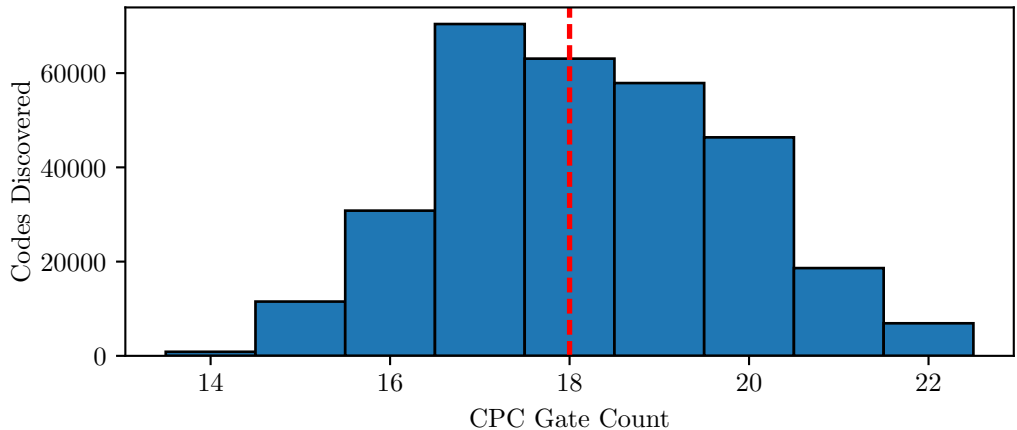}
	\caption{A histogram showing all of the $[[7,3,3]]$ CPC codes discovered in stage 1 of the CPC design process, binned by encoder length. The CPC gate count is the combined total of \cnot and conjugate propagator gates in the encoder. The median length, marked in red, is $18$. In comparison, the shortest $[[7,3,3]]$ circuits have a CPC gate count of $14$. }
	\label{fig:hist1}
\end{figure}

\joschka{An advantage of the CPC framework lies in the fact that new instances of such optimal codes can be discovered numerically}, either using brute-force or more sophisticated optimisation techniques \cite{cpc1}. We now demonstrate these strategies in practice, by showing how optimal $[[7,3,3]]$ CPC codes can be discovered via exhaustive search.

A $[[7,3,3]]$ CPC code has $k=3$ data qubits and $n-k=4$ parity qubits. The adjacency matrices therefore have the form

\begin{equation}\label{eq:733adj}
\begin{split}
m_b=\left(\begin{matrix}
b_{11} & b_{12} & b_{13} & b_{14}\\
b_{21} & b_{22} & b_{23} & b_{24}\\ 
b_{31} & b_{32} & b_{33} & b_{34}
\end{matrix}\right)\rm, \quad 
m_p=\left(\begin{matrix}
h_{11} & h_{12} & h_{13} & h_{14}\\
h_{21} & h_{22} & h_{23} & h_{24}\\ 
h_{31} & h_{32} & h_{33} & h_{34}
\end{matrix}\right)\rm,\\
m_c=\left(\begin{matrix}
0 & c_{12} & c_{13} & c_{14}\\
0 & 0 & c_{23} & c_{24}\\
0 & 0 & 0 & c_{34}\\
0 & 0 & 0 & 0 \\
\end{matrix}\right)\rm,
\end{split}
\end{equation}

\noindent where $b_{xy}$, $h_{xy}$, $c_{xy}$ are binary values. New CPC circuits can be made by generating different instances of these matrices. The single-qubit error syndrome table for each code can be calculated efficiently using a stabilizer simulator or the algorithm we describe in appendix \ref{app:synd_tab}. If the set of syndromes is unique, the respective matrices represent a valid $[[7,3,3]]$ code.

\begin{table}[]
	\centering
	
	\begin{tabular}{|l|l|l|}
		\hline
		\multicolumn{3}{|l|}{\textbf{Encoder gate length (number of CPC gates})} \\ \hline
		\textbf{Mininum} $ \quad \quad \quad \quad \quad \quad  $                 & \textbf{Median}    $ \quad \quad \quad \quad \quad \quad  $         & \textbf{Depth reduction}          \\
		$14$ ($864$ discovered)           & $18$                      & $22\%$                            \\ \hline
		\multicolumn{3}{|l|}{Size of $[[7,3,d=?]]$ search space: $1.07\times 10^9$ circuits}                       \\
		\multicolumn{3}{|l|}{Number of $[[7,3,3]]$ codes discovered: $306,480$ ($0.03\%$ of search space)}                           \\
				\multicolumn{3}{|l|}{Number of symmetry-inequivalent $[[7,3,3]]$ codes: $2190$}                           \\ \hline
	\end{tabular}
	\caption{Summary of the exhaustive search for $[[7,3,3]]$ CPC codes. The number of CPC gates is defined as the combined total of \cnot + conjugate propagator gates in the encoder. The depth reduction is calculated as the percentage decrease in the encoder length of the smallest circuit relative to the median.}
	\label{tab:search_results1}
\end{table}

The number of possible combinations of the adjacency matrices for CPC circuits of type $[[7,3,d=?]]$ is $2^{30}$. By an exhaustive search, we have discovered that $306,480$ of these permutations ($0.03\%$ of the search space) are working $[[7,3,3]]$ codes. These codes have distance $d= 3$, and produce unique syndromes for all single-qubit $X$ and $Z$ errors across the seven qubits. Of the discovered set, there are $2190$ symmetry-inequivalent codes that cannot be transformed from one to another by rearranging the qubit order. However, some symmetry-equivalent code permutations are more amenable to circuit optimisation than others. We will therefore continue to consider the entire set of $306,480$ codes in the CPC design process.

Now that a set of $[[7,3,3]]$ circuits has been found, the next stages in the CPC design process involves analysis to determine which one of the $306,480$ codes is best suited for implementation on the ion trap device. Figure \ref{fig:hist1} shows a histogram of the discovered $[[7,3,3]]$ codes, binned by the combined number of \cnot gates and conjugate propagator gates in their encoder. This quantity will be referred to as the CPC gate count, and can be determined by counting the number of non-zero entries across the three adjacency matrices.

In ion trap hardware, inter-qubit operations are typically the most expensive gate-type in terms of their potential to introduce errors \cite{Ballance2016,Harty2014}. As a result, the CPC circuits with the lowest CPC gate count are most desirable. In the set of $[[7,3,3]]$ codes, the shortest circuit encoders have $14$ CPC gates. This is a $22\%$ reduction in circuit depth compared to the median gate count of $18$. The number of $[[7,3,3]]$ circuits with the minimum encoder depth of $14$ is $864$ of which $245$ are symmetry inequivalent. Further work is therefore necessary to narrow down the code set, and find the optimum quantum memory for the ion trap device.

The encoder length statistics for the $[[7,3,3]]$ CPC codes are summarised in table \ref{tab:search_results1}. Note that in this simple first analysis, we have not accounted for any of the constraints imposed by the ion-trap's nearest-neighbour requirement for two-qubit operations. In the next section, we outline how the $[[7,3,3]]$ codes can be compiled in such a setting through the introduction of additional \swap gates.

The results in this section demonstrate that the CPC framework provides constructive tools for discovery of optimal $[[7,3,3]]$ codes that saturate the quantum Hamming bound. Furthermore, the search was performed using a simple brute-force technique that requires only a basic knowledge of the CPC code structure to implement. The Python script used to perform the code search is approximately $200$ lines long, and required approximately 4 days to run on a CPU clocked at $3.2GHz$ with $8Gb$ of RAM.

\subsection{Stage 2: Hardware optimisation}

\begin{figure*}[]

	\subfloat[]{%
		\includegraphics[width=0.5\textwidth]{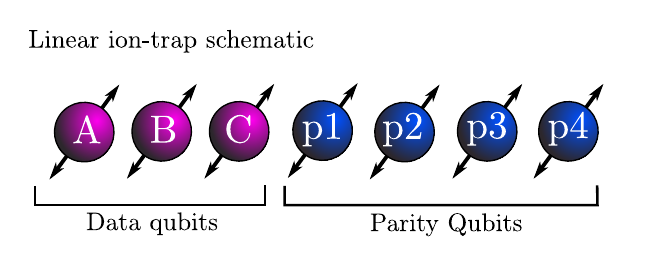}%
		\label{}%
	}\qquad
	\subfloat[]{%
		\usetikzlibrary{decorations.pathreplacing,decorations.pathmorphing}
\providecommand{\ket}[1]{\left|#1\right\rangle}
\providecommand{\phase}[1]{e^{2{\pi}i\cdot#1}}
\begin{tikzpicture}[scale=1.500000,x=1pt,y=1pt]
\filldraw[color=white] (0.000000, -8.000000) rectangle (65.000000, 40.000000);
\draw[color=black] (0.000000,32.000000) -- (65.000000,32.000000);
\draw[color=black] (0.000000,32.000000) node[left] {$\ket{\psi}_B$};
\draw[color=black] (0.000000,16.000000) -- (65.000000,16.000000);
\draw[color=black] (0.000000,16.000000) node[left] {$\ket{\psi}_C$};
\draw[color=black] (0.000000,0.000000) -- (65.000000,0.000000);
\draw[color=black] (0.000000,0.000000) node[left] {$\ket{0}_{p_1}$};
\draw (7.000000,32.000000) -- (7.000000,0.000000);
\filldraw (7.000000, 32.000000) circle(1.500000pt);
\begin{scope}
\draw[fill=white] (7.000000, 0.000000) circle(3.000000pt);
\clip (7.000000, 0.000000) circle(3.000000pt);
\draw (4.000000, 0.000000) -- (10.000000, 0.000000);
\draw (7.000000, -3.000000) -- (7.000000, 3.000000);
\end{scope}
\draw[fill=white,color=white] (18.000000, -6.000000) rectangle (33.000000, 38.000000);
\draw (25.500000, 16.000000) node {$=$};
\draw (44.000000,16.000000) -- (44.000000,0.000000);
\begin{scope}
\draw (41.878680, -2.121320) -- (46.121320, 2.121320);
\draw (41.878680, 2.121320) -- (46.121320, -2.121320);
\end{scope}
\begin{scope}
\draw (41.878680, 13.878680) -- (46.121320, 18.121320);
\draw (41.878680, 18.121320) -- (46.121320, 13.878680);
\end{scope}
\draw (58.000000,32.000000) -- (58.000000,16.000000);
\filldraw (58.000000, 32.000000) circle(1.500000pt);
\begin{scope}
\draw[fill=white] (58.000000, 16.000000) circle(3.000000pt);
\clip (58.000000, 16.000000) circle(3.000000pt);
\draw (55.000000, 16.000000) -- (61.000000, 16.000000);
\draw (58.000000, 13.000000) -- (58.000000, 19.000000);
\end{scope}
\end{tikzpicture}%
		\label{}%
	}
	
	\caption{(a) A schematic of the seven-qubit linear ion trap. Three of the qubits have been labelled as data qubits and four as parity qubits as required by the $[[7,3,3]]$ code. It is assumed that entangling gates can only be performed between nearest-neighbour qubits. (b) A \cnot gate between spatially separated qubits can be implemented using only nearest-neighbour interactions through the addition of \swap gates.}
	
	\label{fig:swaps}
\end{figure*}
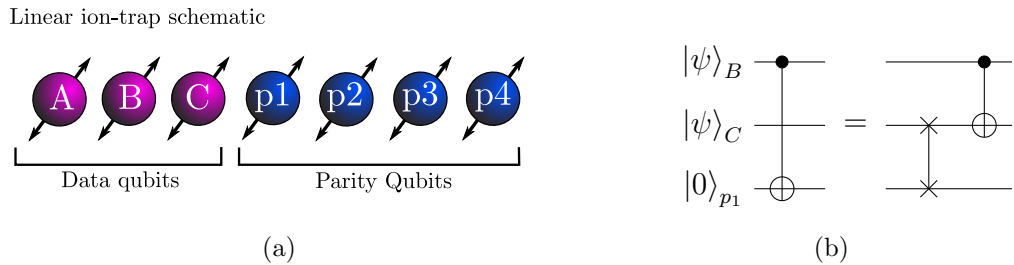

The second stage of the CPC design process involves selecting codes to meet the demands of the chosen quantum hardware and its qubit layout. Figure \ref{fig:swaps}a shows the idealised model ion trap under consideration, labelled with $3$ data qubits and $4$ parity qubits as required by the $[[7,3,3]]$ code. Under the restriction of nearest-neighbour connectivity, interactions between spatially separated qubits can still be realised by performing \swap operations. For example, interacting qubit $B$ with $p_1$ would first require a \swap gate between qubits $B$ and $C$, or qubits $C$ and $p_1$. Circuits with fewer long range interactions will require fewer \swap gates, and will therefore have a reduced two-qubit gate count.

\begin{figure}
	\input{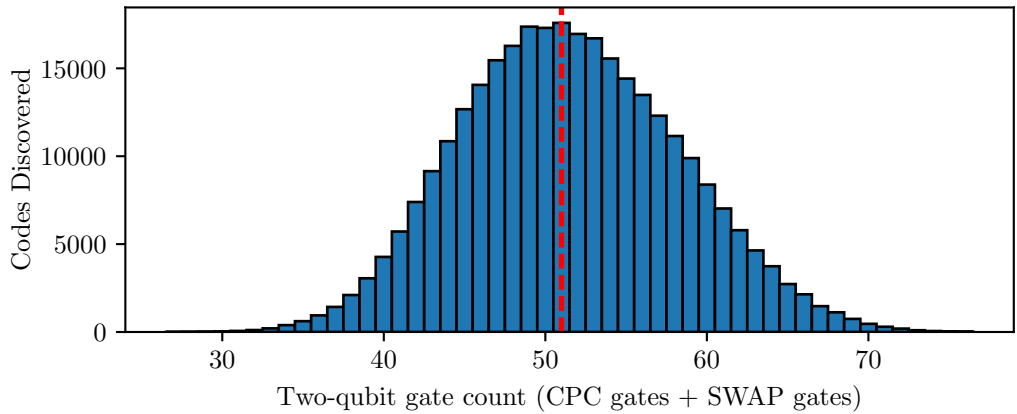}
	
	\caption{The distribution of $[[7,3,3]]$ CPC codes binned by two-qubit gate count after compilation onto a nearest-neighbour architecture. The total gate count is defined as the number of CPC gates + \swap gates in the encode stage of the circuit. }
	\label{fig:733_hist}
\end{figure}

\begin{table}[]
	\centering
	
	\begin{tabular}{|l|l|l|}
		\hline
		\multicolumn{3}{|l|}{\textbf{Encoder gate length (number of two-qubit gates})} \\ \hline
		\textbf{Mininum} $ \quad \quad \quad \quad \quad \quad  $                 & \textbf{Median}    $ \quad \quad \quad \quad \quad \quad  $         & \textbf{Depth reduction}          \\
		$27$ ($1$ discovered)           & $51$                      & $47\%$                            \\ \hline
		\multicolumn{3}{|l|}{Optimum code: }                       \\
		\multicolumn{3}{|l|}{Encoder length=$27$ gates; no. CPC gates=$14$; no. \swap gates=$13$ }                           \\ \hline
	\end{tabular}
	\caption{Summary of the gate-count statistics for the set of $[[7,3,3]]$ codes following the \swap gate compilation. The two-qubit gate count is defined as the number of CPC gates + \swap gates. The depth reduction is the percentage decrease in gate-count of the smallest circuit relative to the median.}
	\label{tab:search_results2}
\end{table}

There are a number of strategies for calculating the sequences of \swap operations required to compile a CPC circuit on a nearest-neighbour architecture. Here we adopt a simple approach in which qubits are always swapped in the upwards direction. As an example of this, in figure \ref{fig:swaps}b, qubit $p_1$ is swapped upwards, instead of qubit $B$ being swapped downwards. More advanced \swap compilation strategies, that combine upwards and downwards moves, can yield circuits with lower \swap counts. However, such analysis is computationally expensive, and can impose a bottleneck in the CPC design process. By restricting our approach to upwards \swap moves only, an exhaustive search of the $[[7,3,3]]$ CPC codes remains possible.      

\begin{figure*}[]
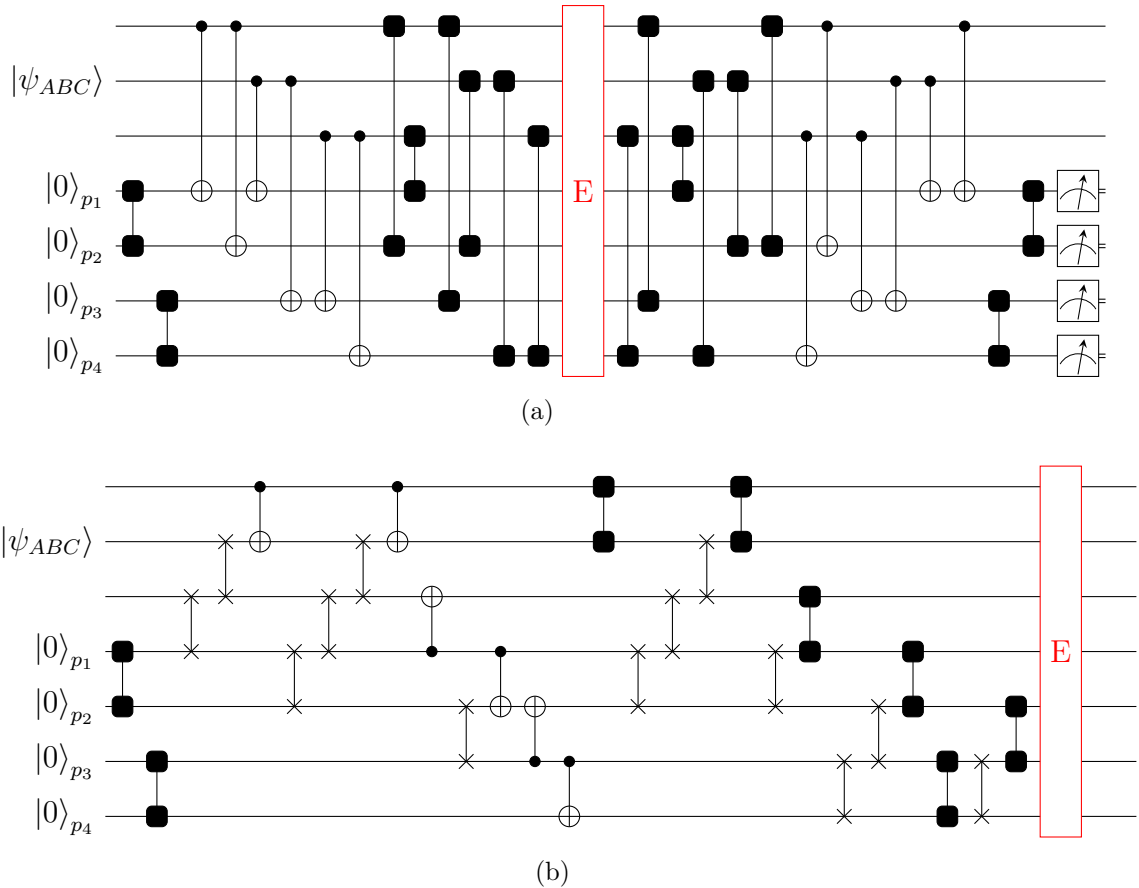

	
	\subfloat[]{
		\input{figures/733_best_encoder.tikz}%
		\label{}%
	}

	\subfloat[]{
		\input{figures/733_best_encoder_swaps.tikz}%
		\label{fsfsdfsd}%
	}

	\caption{Circuit diagrams demonstrating \swap gate compilation for a nearest-neighbour architecture. (a) The $[[7,3,3]]$ code with the smallest two-qubit gate count prior to the addition of \swap gates. (b) The encoder for the same circuit, with \swap gates included.}
	
	\label{fig:733_best_encoder}
\end{figure*}

Figure \ref{fig:733_hist} shows the histogram of the \swap compiled $[[7,3,3]]$ codes distributed by the total two-qubit gate count (CPC gates + \swap gates). The optimum $[[7,3,3]]$ CPC code with the shortest encoder is shown in figure \ref{fig:733_best_encoder}b. The encoder for this circuit includes $14$ CPC gates, and requires an additional $13$ \swap operations to be implemented on a linear, nearest-neighbour architecture. The depth of the encoder, in terms of the number of two qubit gates, is therefore $27$. For comparison, the un-compilied version of this $[[7,3,3]]$ code is shown in figure \ref{fig:733_best_encoder}a.

The results of two-qubit gate count analysis for the $[[7,3,3]]$ codes, following compilation onto the nearest-neighbour hardware, are summarised in table \ref{tab:search_results2}. The optimum circuit has an encoder length of $27$, compared to the median of $51$, a $47\%$ reduction in circuit gate count. Only one CPC code was discovered with the minimum encoder length. The CPC circuit optimisation, with regards to qubit layout, can therefore be considered complete.

\subsection{Stage 3: Native gate compilation}\label{sec:ion_comp}

The ion trap under consideration has a native gate that resembles the symmetrised phase (SP) gate introduced in section \ref{sec:ion_trap_intro}. The final stage of the CPC design process involves systematically applying the SP simplification procedures described in section \ref{sec:gp_nat_gate_simp} to each of the $306,480$ discovered CPC codes. The compilation efficiency of a given code can be quantified by counting the number of local gates that remain in the simplified circuit. The optimal code for the ion trap device is then identified as the circuit with the shortest total encoder length, defined by
\begin{equation}
\mathcal{L}_{\rm CPC}=|\text{\footnotesize CPC}|+|\text{\footnotesize SWAP}|+|\text{\footnotesize LOCAL}|\rm,
\end{equation} 
where $|\text{\footnotesize CPC}|$ is the CPC gate count, $|\text{\footnotesize SWAP}|$ is the \swap gate count and $|\text{\footnotesize LOCAL}|$ is the local gate count.

Table \ref{tab:table4} summarises the simplification statistics for the local gate counts when the $[[7,3,3]]$ CPC codes are compiled with the SP native gate. Without applying any simplifications, the median local gate count is $92$. After applying the simplification routine, the median is $10$, an $89\%$ reduction in gate count. 

\begin{figure}

	\centering
	
	\input{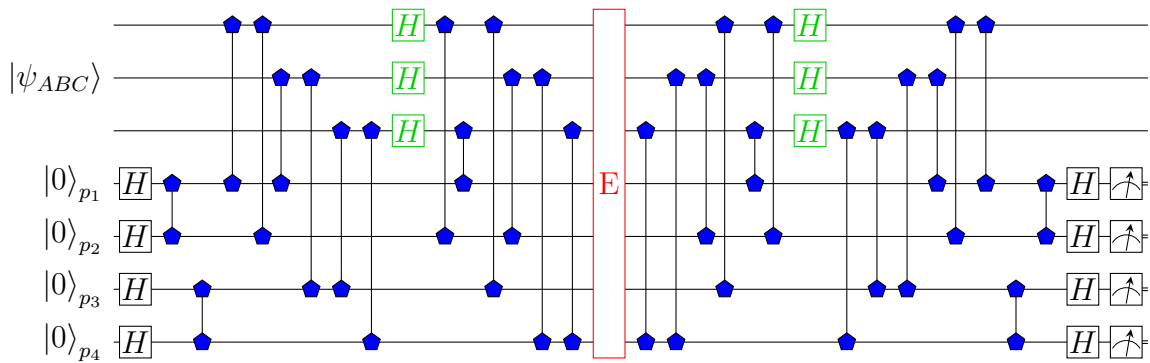}
	
	\caption{The native gate compiled form of the $[[7,3,3]]$ CPC with the lowest total gate count. Note that the \swap operations have been omitted to save space.}
	
	\label{fig:733_best_gate_simp}
\end{figure}

The compiled $[[7,3,3]]$ CPC circuit with the lowest local gate count after simplification is shown in figure \ref{fig:733_best_gate_simp}. This circuit is compiled from the CPC code with the lowest two-qubit gate count, as discovered in the last subsection and depicted in figure \ref{fig:733_best_encoder}. We can therefore identify the compiled $[[7,3,3]]$ code in figure \ref{fig:733_best_gate_simp} as the optimum code for our device with a total gate count of $\mathcal{L}_{\text{CPC}}=34$. For comparison, the median total gate count across all $306,480$ CPC codes was $\mathcal{L}_{\text{CPC}}=61$. The total reduction in circuit depth for the optimised circuit relative to the median is therefore $44\%$. Note that it will not always be the case that the circuit with the lowest two-qubit gate count will also be the circuit that compiles most efficiently. For this reason, the entire discovered set of $[[7,3,3]]$ CPC codes were considered in stage 3 of the design process, rather than restricting the analysis to the single code identified in stage 2.

\begin{table}[]
	
	\begin{tabular}{|l|l|l|}
		\hline
		& \multicolumn{2}{l|}{\textbf{Local gate count (number of single-qubit gates)}} \\ \cline{2-3} 
		& \textbf{Mininum} $\quad \quad\quad \quad\quad \quad$                     & \textbf{Median}          \\ \hline
		\textbf{Before simplification} & $72$ (1 discovered)          & $92$            \\
		\textbf{After simplification}  & $7$ (1 discovered)           & $10$            \\ \hline
		\textbf{$\%$ change}           & $90\%$                       & $89\%$          \\ \hline
	\end{tabular}
	\caption{Summary of the local gate-count for the $[[7,3,3]]$ following compilation with the SP native gate.}
	\label{tab:table4}
\end{table}

\section{Outlook and conclusion} \label{sec:outlook}

In this work, we assessed the real-world functionality of CPC codes by implementing full encode-decode cycles of a $[[4,2,2]]$ quantum error detection code on the IBM 5Q quantum computer.  We then explored the utility of the CPC framework in deriving larger quantum codes. In particular, we illustrated a design process for the automated discovery and optimisation of CPC codes by applying it to a seven qubit ion trap device. In the first stage of the design process, exhaustive code-search methods were used to find $[[7,3,3]]$ CPC codes that saturate the quantum Hamming bound for seven qubits. These circuits were then modified through the addition of \swap gates to allow them to be implemented on a nearest-neighbour architecture. Finally, the circuits were compiled with a SP native gate. At the end of the design process, the optimum hardware-ready code with the lowest gate count of $\mathcal{L}_{\text{CPC}}=34$ was identified.

The design process outlined for ion traps will be adaptable to other qubit technologies. In section \ref{sec:any_native_gate} we demonstrated that the symmetric \textit{encode-error-decode} structure of CPC codes allows for efficient compilation with any realistic maximally entangling Clifford gate. This result means that simplification routines, similar to those seen with the ion trap SP gate, will be possible for a broad range of native gates from different quantum experiments.

The final circuit in the outline of the CPC design process, drawn in figure \ref{fig:733_best_gate_simp}, shows the best CPC code in terms of total gate count. Here it was assumed, however, that each gate type -- CPC, \swap and local -- are equal in terms of the overhead they impose on the code implementation. In practice, however, some types of operations will be more expensive than others. For example, in an ion trap setting, it is typically the case that two-qubit interactions have a lower fidelity than single-qubit operations \cite{Ballance2016, Harty2014}. When implementing the CPC design process, such considerations should be taken into account for choosing the optimum code for the given device. For example, each CPC code could be assigned a weighted total gate count, $\mathcal{R}_{\rm CPC}$, given by
\begin{equation}
\mathcal{R}_{\rm CPC}=\gamma_1 |\text{\footnotesize CPC}|+\gamma_2|\swap|+\gamma_3|\text{\footnotesize LOCAL}|\rm,
\end{equation}
where $|\text{\footnotesize CPC}|$, $|\swapeq|$ and $|\text{\footnotesize LOCAL}|$ are the counts for CPC gates, \swap gates and local gates respectively. The count for each gate-type is weighted by a penalty strength $\gamma$ which is based on the gate count.

In the code discovery stage of the CPC design process for the ion trap device, the aim was to find working $[[7,3,3]]$ codes that saturate the quantum Hamming bound for seven qubits. This involved calculating the code distance for all possible permutations of $[[7,3,d=?]]$ CPC codes, a total of $2^{30}$ circuits. Using the syndrome calculation algorithm outlined in appendix \ref{app:synd_tab}, it was possible to exhaustively analyse all the circuits in less than a week on a desktop computer. In total, the search yielded $306,480$ working $[[7,3,3]]$ codes ($0.03\%$ of the search space).

For a CPC code with $4$ data qubits, the quantum Hamming bound tells us that the optimal CPC code is of type $[[9,4,3]]$. However, there are $2^{50}$ permutations of this circuits of the form $[[9,4,d=?]]$, which is an impractical search space for exhaustive methods. In the original CPC paper, it was shown that $[[9,4,3]]$ codes can be discovered simply by randomised search \cite{cpc1}. In future work, more sophisticated techniques, such as simulated annealing or parallel tempering, could be employed to more efficiently search for CPC codes.

When searching for large CPC codes, the number of circuits in the search space could be reduced by considering hardware constraints in advance. For example, for a nearest-neighbour device, each circuit permutation could be assigned a score on the basis of how many long-range interactions it contains. The code distance would then only be measured for the circuits with fewer long range interactions. \joschka{Another optimisation parameter that could be considered is the weight of the code's stabilizers, a parameter that is useful to minimise when constructing fault tolerant circuits. Exhaustive and random search strategies for quantum code discovery have also been studied in \cite{Grassl06,Brown2013}. The particular strength of the CPC framework is that the symmetric \textit{encode-error-decode} structure ensures the search is constrained to a space of non-disturbing codes. Investigating whether optimised CPC search strategies provide a higher density of good codes compared to other code search techniques would be an interesting area for future research.}

\joschka{Another feature of the CPC framework is that any classical code can be re-purposed for the bit and phase checking stages of the code. If such an approach is adopted, only the space of cross-checks needs to be searched in order to obtain a CPC code with fixed distance.} Owing to the demands of modern high-density communication networks, classical error correction protocols such as low density parity check and turbo codes have been extensively optimised \cite{MacKay1996,Berrou1996_turbo}. At large scales, these codes can be decoded in real time at close to the theoretical maximum rate for information transfer along a noisy channel given by the Shannon limit \cite{mackay}. The tools of the CPC framework could help construct quantum versions of low density parity check and turbo codes. \joschka{A presentation of the CPC framework in terms of classical factor graph notation can be found in \cite{roffe18}.}

\joschka{An important direction for future work is to investigate ways of making CPC circuits fault tolerant. For most quantum computing architectures, it is not realistic to assume that the encode and decode stages will be fault-free, or that errors will only occur within a specified wait-stage. In section \ref{sec:ibm} it was shown that a specific implementation of a $[[4,2,2]]$ CPC detection code can be specially hardened against single-qubit errors occuring after any multi-qubit gate in the encoder or decoder. However, further work is required to develop methods for extending general CPC codes to full fault tolerance. Of particular interest are recent studies into fault tolerant computing using flag checks, which have a similar construction to CPC parity checks \cite{flag,Chamberland2018flagfaulttolerant}. }

The CPC framework lifts many of the restrictions that have hindered the development of traditional QEC codes. In particular, CPC codes have a canonical structure that allows any sequence of parity checks to be performed on a quantum register without risk of decohering the encoded information. The process of deriving CPC codes is therefore reduced to a classical decoding problem, allowing for code discovery via numerical search. This opens up the possibility of constructing custom QEC protocols to meet the hardware and layout demands of a specific quantum computing experiment.

\section*{Acknowledgements}

Joschka Roffe was supported by a Durham Doctoral Studentship (Faculty of Science). Nicholas Chancellor, Dominic Horsman and Viv Kendon were supported by EPSRC (grant ref:grant ref: EP/L022303/1). We acknowledge use of the IBM Quantum Experience for this work. The views expressed are those of the authors and do not reflect the official policy or position of IBM or the IBM Quantum Experience team. The quantum circuits in this paper were drawn using the QPIC package by Thomas Draper and Samuel Kutin \cite{qpic}.

\section*{References}

\bibliographystyle{unsrt}
\bibliography{references/references}

\appendix

\renewcommand\thefigure{\thesubsection\arabic{figure}}

\section*{Appendix}
\renewcommand{\thesubsection}{\Alph{subsection}}
\renewcommand{\theequation}{\thesubsection\arabic{equation}}

\subsection{The Pauli group}\label{app:pauli}

The Pauli group on a single-qubit, $\mathcal{G}_1$, is defined as the set of Pauli operators 
\begin{equation}
\mathcal{G}_1=\{\pm \openone, \pm \text{i} \openone, \pm X, \pm \text{i} X, \pm Y, \pm \text{i} Y, \pm Z, \pm \text{i} Z \}\rm,
\end{equation}
where the $\pm 1$ and $\pm \text{i}$ terms are included to ensure $\mathcal{G}_1$ is closed under multiplication and thus forms a legitimate group \cite{nielsen2010quantum}.
In matrix form, the four Pauli operators are given by
\begin{equation}
\openone = \left(\begin{matrix}
1 & 0\\0& 1
\end{matrix}\right), \quad
X=\left(\begin{matrix}
0 & 1\\1& 0
\end{matrix}\right), \quad
Y=\left(\begin{matrix}
0 & -\text{i}\\ \text{i}& 0
\end{matrix}\right), \quad
Z = \left(\begin{matrix}
1 & 0\\0& -1
\end{matrix}\right)\rm .
\end{equation}
The general Pauli group, $\mathcal{G}$, consists of the set of all operators that are formed from tensor products of the matrices in $\mathcal{G}_1$. For example, the operator
\begin{equation}
\openone \otimes X \otimes \openone \otimes Y \in \mathcal{G}
\end{equation}
is an element of the four-qubit Pauli group. Note that for simplicity, in the context of quantum computing, the above operator would usually be expressed as $X_2Y_4$. The identity operators are omitted, and the remaining elements are subscripted with the label of the qubit they act on.

The elements of the Pauli group have eigenvalues $\{\pm 1, \pm \text{i}\}$. Another useful property of the Pauli group is that its elements either commute or anti-commute with one another.

\subsection{The Clifford group and stabilizer states}\label{app:clifford}
The Clifford group $\mathcal{C}$ is defined as the set of operators that normalise the Pauli group such that
\begin{equation}
U_{\rm C} \bigcdot P_i \bigcdot U_{\rm C}^\dagger=P_j,\quad U_{\rm C}\in \mathcal{C}, \quad \{P_i,P_j\} \in\mathcal{G} \ \forall \ \{i,j\}\rm,
\end{equation}
where $U_{\rm C}\in \mathcal{C}$ is a Clifford operator and $P_k$ are elements of the Pauli group. Clifford gates, $\mathcal{C}$, are generated by the set of three gates $\langle \cnot, H, P \rangle$, such that $\mathcal{C}=\langle \cnot, H, P \rangle$ \cite{Gottesman1998}. Likewise, single-qubit Clifford gates, $\mathcal{C}_1$, are generated by the set $\langle H,P \rangle$, such that $\mathcal{C}_1=\langle H,P \rangle$.

The stabilizer states are all the quantum states that can be reached from a blank register, $\ket{0}^{\otimes N}$, via the application of Clifford gates and computational basis measurements. Quantum circuits consisting only of Clifford gates acting on stabilizer states can be efficiently classically simulated. The proof of this is given by the Gottesman-Knill theorem \cite{GKtheorem}. Although the Clifford group is not a universal quantum gate set, it is sufficient for simulating many QEC circuits and all the quantum memories described in this paper.

\subsection{Efficient calculation of CPC code syndrome table} \label{app:synd_tab}

\begin{figure}
	\centering
	\includegraphics[]{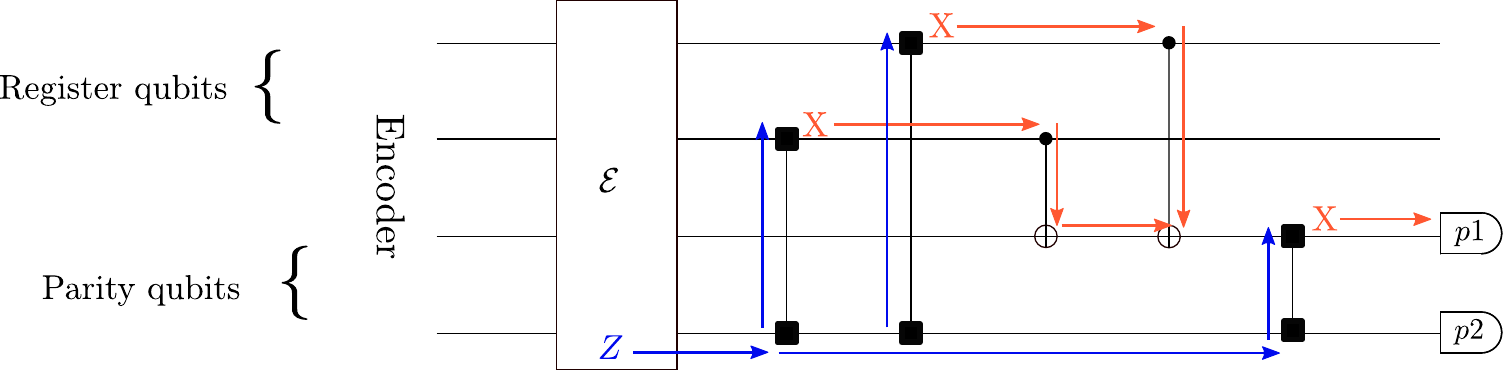}
	
	\caption{The [[4,2,2]] code decoder depicting the different propagation pathways for Z errors on the second parity qubit.}
	\label{422_decoder}
\end{figure} 
In addition to providing a compact way to describe CPC codes, the adjacency matrix representation can be leveraged to create a simple algorithm for calculating syndrome tables, bypassing the need to perform a full stabilizer simulation. We will begin our presentation of this algorithm by considering errors on the data qubits, which are represented in terms of the row vectors $E_{d,x}$ and $E_{d,z}$. For example, in the [[4,2,2]] detection code, depicted in figure \ref{fig:422_circ2}, a bit-flip error on qubit $A$ would have the form $E_{d,x}=\left(1,0\right)$. Likewise, a phase-flip on qubit $B$ would be given by $E_{d,x}=\left(0,1\right)$.

In a CPC code $X$ and $Z$ errors on the data qubits are propagated to the parity qubits via gate sequences described by the adjacency matrices $m_b$ and $m_p$ respectively. The syndromes resulting from this propagation can be calculated by multiplying the error vector by its corresponding adjacency matrix modulo 2. For example, the syndrome for a bit-flip error on qubit $A$ of the [[4,2,2]] code is given by
\begin{equation}\label{eq:dx_prop}
S_{d,x}=E_{d,x}\cdot m_b  =
\left(
\begin{matrix}
1 & 0 \\
\end{matrix}
\right) \left(
\begin{array}{cc}
1 & 0 \\
1 & 0 \\
\end{array}
\right)=\left(
\begin{array}{c}
1 \\
0 \\
\end{array}
\right)\rm .
\end{equation}

\noindent The bit-flip error information is propagated to the parity qubit by a \cnoteq, and the column vector on the right gives the subsequent measurement outcomes of the parity qubits $p1$ and $p2$. Our expression therefore tells us that error on the data qubit $A$ produces the syndrome `$10$', a result in agreement with the values given in table \ref{tab:422synd} in section \ref{sec:422}. Similarly, the syndromes for phase-flip errors on the data qubits can be computed with the expression $S_{d,z}= E_{d,z}\cdot m_p $.

We now need a method for calculating the syndromes for errors occurring on the parity qubits. Again, we represent $X$ and $Z$ errors in terms of two row vectors $ E_{p,x}$ and $ E_{p,z}$. In the case of bit-flip errors, the syndrome is simply given by $S_{p,x}=E_{p,x}$. This is the case as the bit-flip errors commute through the conjugate propagator gates and the \cnot targets, and will therefore propagate directly to the end of the circuit. The final error type to consider are phase-flips on the parity qubits.

Figure \ref{422_decoder} depicts the propagation of such an error through the decoder of the $[[4,2,2]]$ code. To calculate the syndromes, there are two propagation pathways to be considered. Figure \ref{422_decoder} shows that $Z$ errors can be propagated to the register by the phase-check conjugate propagator gates, after which they can be considered as bit-errors. These bit-flip errors are then propagated to the register, as illustrated by the orange arrows in figure \ref{422_decoder}. This propagation pathway can be represented mathematically by the expression $E_{p,z} \cdot m_p^T\cdot m_b$. Note that we have taken the transpose of the phase-check matrix as we are propagating information from the parity bits to register. The second pathway to be considered for phase-flip errors on the parity qubits, is the propagation due to the cross-check operators. As the cross-check operators can act both ways, this pathway is described by the expression $E_{p,z} \cdot \left(m_c+m_c^T\right)$. Combining both error propagation pathways, the syndrome expression for phase-flip errors on the parity qubits is $S_{p,z}=E_{p,z} \cdot m_p^T\cdot m_b \ + \ E_{p,z} \cdot \left(m_c+m_c^T\right)$, where all addition and multiplication is performed modulo 2. The full syndrome equation can now be written by summing the contributions $S_{d,x}$, $S_{d,z}$, $S_{p,x}$ and $S_{p,z}$ to give
\begin{equation}\label{eq:code_test}
S=\left(E_{d,x}\cdot m_b+ E_{d,z}\cdot m_p+E_{p,x}+E_{p,z} \cdot m_p^T\cdot m_b+E_{p,z} \cdot \left(m_c+m_c^T\right)\right) \text{mod} \ 2 \rm.
\end{equation}
The above equation allows the syndromes for a given error circuit to be calculated in time $O(n^2)$. It would be interesting to investigate how this algorithm relates to other efficient stabilizer simulators such as \cite{Aaronson2004} and \cite{Anders2006}.

\subsection{IMBQX4 calibration data} \label{sec:ibm_calib}

The experiment on the IBMQX4 outlined in section \ref{sec:ibm} was run over three days on 25\textsuperscript{th} November 2017, 26\textsuperscript{th} November and 27\textsuperscript{th} November 2017. The calibration data for each of these days can be found below:

\begin{verbatim}
{Date: 11-25-2017,
Single-qubit error rates (10^-3):
{Q0: 0.94, Q1: 0.60, Q2: 1.12, Q3: 1.37, Q4: 1.80},
Readout error rates (10^-2):
{Q0: 4.10, Q1: 5.70, Q2: 4.00, Q3: 3.30, Q4: 5.10},
Two-qubit error rates (10^-2):
{CX1_0: 1.88, CX2_0: 2.09, CX3_2: 1.97, CX2_1: 4.28, CX3_4: 2.15, CX2_4: 3.89}}

{Date: 11-26-2017,
 Single-qubit error rates (10^-3):
 {Q0: 0.86, Q1: 0.69, Q2: 1.12, Q3: 1.89, Q4: 2.06},
 Readout error rates (10^-2):
 {Q0: 4.10, Q1: 4.10, Q2: 4.30, Q3: 5.30, Q4: 7.10},
 Two-qubit error rates (10^-2):
 {CX1_0: 2.31, CX2_0: 2.22, CX3_2: 2.18, CX2_1: 4.80, CX3_4: 2.43, CX2_4: 3.93}}

{Date: 11-27-2017,
Single-qubit error rates (10^-3):
{Q0: 0.77, Q1: 0.43, Q2: 1.20, Q3: 1.72, Q4: 1.89},
Readout error rates (10^-2):
{Q0: 3.70, Q1: 4.90, Q2: 4.20, Q3: 5.30, Q4: 5.80},
Two-qubit error rates (10^-2):
{CX1_0: 2.01, CX2_0: 1.93, CX3_2: 2.75, CX2_1: 4.47, CX3_4: 2.28, CX2_4: 4.13}}
 
\end{verbatim}

\end{document}